\documentclass[]{article}
\usepackage{lmodern}
\usepackage{setspace}
\usepackage{amssymb,amsmath}
\usepackage{ifxetex,ifluatex}
\usepackage{fixltx2e} 
\ifnum 0\ifxetex 1\fi\ifluatex 1\fi=0 
  \usepackage[T1]{fontenc}
  \usepackage[utf8]{inputenc}
\else 
  \ifxetex
    \usepackage{mathspec}
  \else
    \usepackage{fontspec}
  \fi
  \defaultfontfeatures{Ligatures=TeX,Scale=MatchLowercase}
\fi
\IfFileExists{upquote.sty}{\usepackage{upquote}}{}
\IfFileExists{microtype.sty}{%
\usepackage{microtype}
\UseMicrotypeSet[protrusion]{basicmath} 
}{}
\usepackage[margin=1in]{geometry}
\usepackage{hyperref}
\PassOptionsToPackage{usenames,dvipsnames}{color} 
\hypersetup{unicode=true,
            pdftitle={Component response rate variation underlies the stability of highly complex finite systems},
            pdfauthor={A. Bradley Duthie},
            colorlinks=true,
            linkcolor=blue,
            citecolor=Blue,
            urlcolor=Blue,
            breaklinks=true}
\usepackage{color}
\usepackage{fancyvrb}

\DefineVerbatimEnvironment{Highlighting}{Verbatim}{commandchars=\\\{\}}
\usepackage{framed}
\definecolor{shadecolor}{RGB}{248,248,248}
\newenvironment{Shaded}{\begin{snugshade}}{\end{snugshade}}
\newcommand{\KeywordTok}[1]{\textcolor[rgb]{0.13,0.29,0.53}{\textbf{{#1}}}}

\newcommand{\StringTok}[1]{\textcolor[rgb]{0.31,0.60,0.02}{{#1}}}

\newcommand{\NormalTok}[1]{{#1}}
\usepackage{longtable,booktabs}
\usepackage{graphicx,grffile}
\makeatletter
\def\maxwidth{\ifdim\Gin@nat@width>\linewidth\linewidth\else\Gin@nat@width\fi}
\def\maxheight{\ifdim\Gin@nat@height>\textheight\textheight\else\Gin@nat@height\fi}
\makeatother
\setkeys{Gin}{width=\maxwidth,height=\maxheight,keepaspectratio}
\IfFileExists{parskip.sty}{%
\usepackage{parskip}
}{
\setlength{\parindent}{0pt}
\setlength{\parskip}{6pt plus 2pt minus 1pt}
}
\providecommand{\tightlist}{%
  \setlength{\itemsep}{0pt}\setlength{\parskip}{0pt}}
\ifx\paragraph\undefined\else
\let\oldparagraph\paragraph
\renewcommand{\paragraph}[1]{\oldparagraph{#1}\mbox{}}
\fi
\ifx\subparagraph\undefined\else
\let\oldsubparagraph\subparagraph
\renewcommand{\subparagraph}[1]{\oldsubparagraph{#1}\mbox{}}
\fi

\let\rmarkdownfootnote\footnote%
\def\footnote{\protect\rmarkdownfootnote}

\usepackage{titling}

  \title{Component response rate variation underlies the stability of highly
complex finite systems}
  \pretitle{\vspace{\droptitle}\centering\huge}
  \posttitle{\par}
  \author{A. Bradley Duthie (
\href{mailto:alexander.duthie@stir.ac.uk}{\nolinkurl{alexander.duthie@stir.ac.uk}}
)}
  \preauthor{\centering\large\emph}
  \postauthor{\par}
  \predate{\centering\large\emph}
  \postdate{\par}
  \date{Biological and Environmental Sciences, University of Stirling, Stirling,
UK, FK9 4LA}

\usepackage{amsmath}
\usepackage{natbib}
\usepackage[utf8]{inputenc}
\begin{document}
\maketitle


\vspace{10mm}
\hrule
\vspace{2mm}

\textbf{The stability of a complex system generally decreases with increasing
system size and interconnectivity, a counterintuitive result of
widespread importance across the physical, life, and social sciences.
Despite recent interest in the relationship between system properties
and stability, the effect of variation in response rate across system
components remains unconsidered. Here I vary the component response
rates (\(\boldsymbol{\gamma}\)) of randomly generated complex systems. I
use numerical simulations to show that when component response rates
vary, the potential for system stability increases. These results are
robust to common network structures, including small-world and
scale-free networks, and cascade food webs. Variation in
\(\boldsymbol{\gamma}\) is especially important for stability in highly
complex systems, in which the probability of stability would otherwise
be negligible. At such extremes of simulated system complexity, the
largest stable complex systems would be unstable if not for variation in
\(\boldsymbol{\gamma}\). My results therefore reveal a previously
unconsidered aspect of system stability that is likely to be pervasive
across all realistic complex systems.}

\vspace{2mm}
\hrule
\vspace{2mm}

\subsection{Introduction}\label{introduction}

In 1972, May\textsuperscript{\protect\hyperlink{ref-May1972}{1}} first
demonstrated that randomly assembled systems of sufficient complexity
are almost inevitably unstable given infinitesimally small
perturbations. Complexity in this case is defined by the size of the
system (i.e., the number of potentially interacting components; \(S\)),
its connectance (i.e., the probability that one component will interact
with another; \(C\)), and the variance of interaction strengths
(\(\sigma^{2}\))\textsuperscript{\protect\hyperlink{ref-Allesina2012}{2}}.
May's finding that the probability of local stability falls to near zero
given a sufficiently high threshold of \(\sigma\sqrt{SC}\) is broadly
relevant for understanding the dynamics and persistence of systems such
as
ecological\textsuperscript{\protect\hyperlink{ref-May1972}{1}--\protect\hyperlink{ref-Grilli2017}{6}},
neurological\textsuperscript{\protect\hyperlink{ref-Gray2008}{7},\protect\hyperlink{ref-Gray2009}{8}},
biochemical\textsuperscript{\protect\hyperlink{ref-Rosenfeld2009}{9},\protect\hyperlink{ref-MacArthur2010}{10}},
and
socio-economic\textsuperscript{\protect\hyperlink{ref-May2008}{11}--\protect\hyperlink{ref-Bardoscia2017}{14}}
networks. As such, identifying general principles that affect stability
in complex systems is of wide-ranging importance.

Randomly assembled complex systems can be represented as large square
matrices (\(\mathbf{M}\)) with \(S\) components (e.g., networks of
species\textsuperscript{\protect\hyperlink{ref-Allesina2012}{2}} or
banks\textsuperscript{\protect\hyperlink{ref-Haldane2011}{12}}). One
element of such a matrix, \(M_{ij}\), defines how component \(j\)
affects component \(i\) in the system at a point of
equilibrium\textsuperscript{\protect\hyperlink{ref-Allesina2012}{2}}.
Off-diagonal elements (\(i \neq j\)) therefore define interactions
between components, while diagonal elements (\(i = j\)) define component
self-regulation (e.g., carrying capacity in ecological communities).
Traditionally, off-diagonal elements are assigned non-zero values with a
probability \(C\), which are sampled from a distribution with variance
\(\sigma^{2}\); diagonal elements are set to
\(-1\)\textsuperscript{\protect\hyperlink{ref-May1972}{1},\protect\hyperlink{ref-Allesina2012}{2},\protect\hyperlink{ref-Allesina2015}{5}}.
Local system stability is assessed using eigenanalysis on
\(\mathbf{M}\), with the system being stable if the real parts of all
eigenvalues (\(\lambda\)), and therefore the leading eigenvalue
(\(\lambda_{max}\)), are negative
(\(\Re(\lambda_{max}) < 0\))\textsuperscript{\protect\hyperlink{ref-May1972}{1},\protect\hyperlink{ref-Allesina2012}{2}}.
In a large system (high \(S\)), eigenvalues are distributed
uniformly\textsuperscript{\protect\hyperlink{ref-Tao2010}{15}} within a
circle centred at \(\Re = -d\) (\(-d\) is the mean value of diagonal
elements) and \(\Im = 0\), with a radius of
\(\sigma\sqrt{SC}\)\textsuperscript{\protect\hyperlink{ref-May1972}{1},\protect\hyperlink{ref-Allesina2012}{2},\protect\hyperlink{ref-Allesina2015}{5}}
(Fig. 1a). Local stability of randomly assembled systems therefore
becomes increasingly unlikely as \(S\), \(C\), and \(\sigma\) increase.

\clearpage
\hrule
\vspace{2mm}
\textbf{Figure 1: Eigenvalue distributions of random complex systems.}
Each panel shows the real (x-axis) and imaginary (y-axis) parts of
\(S =\) 400 eigenvalues from random \(S \times S\) matrices.
(\(\textbf{a}\)) A system represented by a matrix \(\mathbf{A}\), in
which all elements are sampled from a normal distribution with
\(\mu = 0\) and \(\sigma_{A} = 1/\sqrt{S}\). Points are uniformly
distributed within the blue circle centred at the origin with a radius
of \(\sigma_{A} \sqrt{S} = 1\). (\(\textbf{b}\)) The same system as in 
\(\textbf{a}\) after including variation in the response rates of \(S\)
components, represented by the diagonal matrix \(\gamma\), such that
\(\mathbf{M} = \gamma\mathbf{A}\). Elements of \(\gamma\) are randomly
sampled from a uniform distribution from \(\min = 0\) to \(\max = 2\).
Eigenvalues of \(\mathbf{M}\) are then distributed non-uniformly within
the red circle centred at the origin with a radius of
\(\sqrt{\sigma^{2}_{A}(1 + \sigma^{2}_{\gamma})S} \approx\) 1.15.
(\(\textbf{c}\)) A different random system \(\mathbf{A}\) constructed
from the same parameters as in \(\textbf{a}\), except with diagonal
element values of \(-1\). (\(\textbf{d}\)) The same system
\(\textbf{c}\) after including variation in component response rates,
sampled from \(\mathcal{U}(0, 2)\) as in \(\textbf{b}\).

\includegraphics{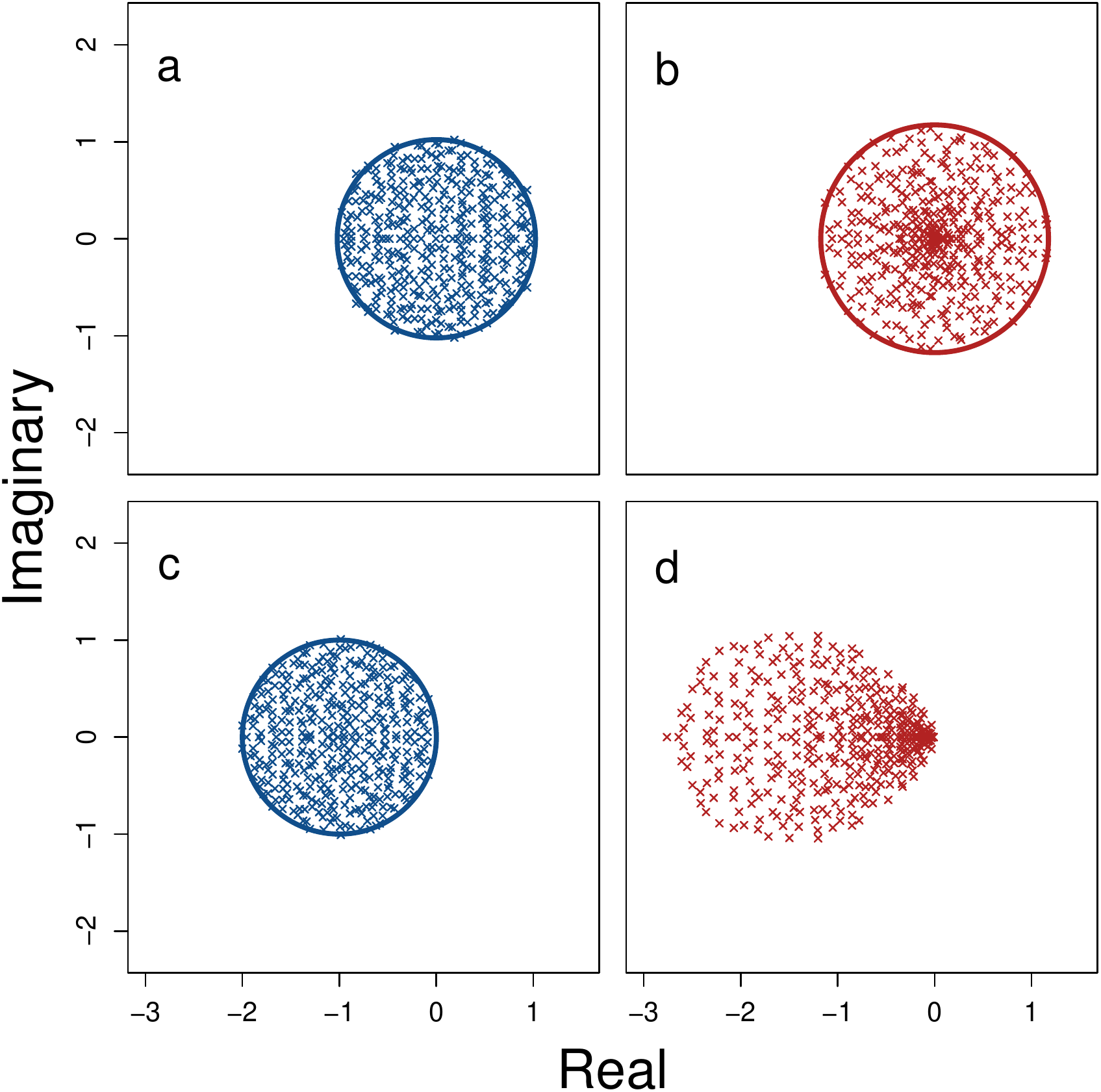}
\vspace{2mm}
\hrule

May's\textsuperscript{\protect\hyperlink{ref-May1972}{1},\protect\hyperlink{ref-Allesina2012}{2}}
stability criterion \(\sigma\sqrt{SC} < d\) assumes that the expected
response rates (\(\gamma\)) of individual components to perturbations of
the system are identical, but this is highly unlikely in any complex
system. In ecological communities, for example, the rate at which
population density changes following perturbation will depend on the
generation time of organisms, which might vary by orders of magnitude
among species. Species with short generation times will respond quickly
(high \(\gamma\)) to perturbations relative to species with long
generation times (low \(\gamma\)). Similarly, the speed at which
individual banks respond to perturbations in financial networks, or
individuals or institutions respond to perturbations in complex social
networks, is likely to vary. The effect of such variance on stability
has not been investigated in complex systems theory. Intuitively,
variation in \(\gamma\) (\(\sigma^{2}_{\gamma}\)) might be expected to
decrease system stability by introducing a new source of variation into
the system and thereby increasing \(\sigma\). Here I show that, despite
higher \(\sigma\), realistic complex systems (in which \(S\) is high but
finite) are actually more likely to be stable if their individual
component response rates vary. My results are robust across commonly
observed network structures, including
random\textsuperscript{\protect\hyperlink{ref-May1972}{1}},
small-world\textsuperscript{\protect\hyperlink{ref-Watts1998}{16}},
scale-free\textsuperscript{\protect\hyperlink{ref-Albert2002}{17}}, and cascade food
web\textsuperscript{\protect\hyperlink{ref-Solow1998}{18},\protect\hyperlink{ref-Williams2000}{19}}
networks.

\subsection{Results}\label{results}

\textbf{Component response rates of random complex systems}. Complex
systems (\(\mathbf{M}\)) are built from two matrices, one modelling
component interactions (\(\mathbf{A}\)), and second modelling component
response rates (\(\boldsymbol{\gamma}\)). Both \(\mathbf{A}\) and
\(\boldsymbol{\gamma}\) are square \(S \times S\) matrices. Rows in
\(\mathbf{A}\) define how a given component \(i\) is affected by each
component \(j\) in the system, including itself (where \(i = j\)).
Off-diagonal elements of \(\mathbf{A}\) are independent and identically
distributed (i.i.d), and diagonal elements are set to \(A_{ii} = -1\) as
in May\textsuperscript{\protect\hyperlink{ref-May1972}{1}}. Diagonal
elements of \(\boldsymbol{\gamma}\) are positive, and off-diagonal
elements are set to zero (i.e., \(\boldsymbol{\gamma}\) is a diagonal
matrix with positive support). The distribution of
\(diag(\boldsymbol{\gamma})\) over \(S\) components thereby models the
distribution of component response rates. The dynamics of the entire
system \(\mathbf{M}\) can be defined as
follows\textsuperscript{\protect\hyperlink{ref-Patel2018}{20}},

\begin{equation} \label{defM}
\mathbf{M} = \boldsymbol{\gamma} \mathbf{A}.
\end{equation}

Equation 1 thereby serves as a null model to investigate how
variation in component response rate (\(\sigma^{2}_{\gamma}\)) affects
complex systems. In the absence of such variation
(\(\sigma^{2}_{\gamma} = 0\)), \(\boldsymbol{\gamma}\) is set to the
identity matrix (diagonal elements all equal 1), and
\(\mathbf{M} = \mathbf{A}\). Under these conditions, eigenvalues of
\(\mathbf{M}\) are distributed
uniformly\textsuperscript{\protect\hyperlink{ref-Tao2010}{15}} in a
circle centred at \((-1, 0)\) with a radius of
\(\sigma \sqrt{SC}\)\textsuperscript{\protect\hyperlink{ref-May1972}{1}}
(Fig. 1a).

\textbf{Effect of \(\mathbf{\sigma^{2}_{\gamma}}\) on \(\mathbf{M}\)
(co)variation}. The value of \(\Re(\lambda_{max})\), and therefore
system stability, can be estimated from five properties of
\(\mathbf{M}\)\textsuperscript{\protect\hyperlink{ref-Tang2014b}{21}}.
These properties include (1) system size (\(S\)), (2) mean
self-regulation of components (\(d\)), (3) mean interaction strength
between components (\(\mu\)), (4) the variance of between component
interaction strengths (hereafter \(\sigma^{2}_{M}\), to distinguish from
\(\sigma^{2}_{A}\) and \(\sigma^{2}_{\gamma}\)), and (5) the correlation
of interaction strengths between components, \(M_{ij}\) and \(M_{ji}\)
(\(\rho\))\textsuperscript{\protect\hyperlink{ref-Sommers1988}{22}}.
Positive \(\sigma^{2}_{\gamma}\) does not change \(S\), nor does it
necessarily change \(E[d]\) or \(E[\mu]\). What \(\sigma^{2}_{\gamma}\)
does change is the total variation in component interaction strengths
(\(\sigma^{2}_{M}\)), and \(\rho\). Introducing variation in \(\gamma\)
increases the total variation in the system. Variation in the
off-diagonal elements of \(\mathbf{M}\) is described by the joint
variation of two random variables,

\begin{equation} \label{var_ref}
\sigma^{2}_{M} = \sigma^{2}_{A}\sigma^{2}_{\gamma} + \sigma^{2}_{A}E[\gamma_{i}]^{2}+\sigma^{2}_{\gamma}E[A_{ij}]^{2}.
\end{equation}

Given \(E[\gamma_{i}] = 1\) and \(E[A_{ij}] = 0\), Eq. 2
can be simplified,

\begin{equation}
\sigma^{2}_{M} = \sigma^{2}_{A}(1 + \sigma^{2}_{\gamma}). \nonumber
\end{equation}

The increase in \(\sigma^{2}_{M}\) caused by \(\sigma^{2}_\gamma\) can
be visualised from the eigenvalue spectra of \(\textbf{A}\) versus
\(\textbf{M} = \boldsymbol{\gamma}\textbf{A}\) (Fig. 1). Given \(d = 0\)
and \(C = 1\), the distribution of eigenvalues of \(\textbf{A}\) and
\(\textbf{M}\) lie within a circle of a radius \(\sigma_{A}\sqrt{S}\)
and \(\sigma_{M}\sqrt{S}\), respectively (Fig. 1a vs.~1b). If
\(d \neq 0\), positive \(\sigma^{2}_\gamma\) changes the distribution of
eigenvalues\textsuperscript{\protect\hyperlink{ref-Ahmadian2015}{23}--\protect\hyperlink{ref-Stone2017}{25}},
potentially affecting stability (Fig. 1c vs.~1d).

Given \(\sigma^{2}_\gamma = 0\), \(\Re(\lambda_{max})\) increases
linearly with \(\rho\) such
that\textsuperscript{\protect\hyperlink{ref-Tang2014c}{26}},

\begin{equation} 
\Re(\lambda_{max}) \approx \sigma_{M}\sqrt{SC}\left(1 + \rho\right). \nonumber
\end{equation}

If \(\rho < 0\), such as when \(\textbf{M}\) models a predator-prey
system in which \(M_{ij}\) and \(M_{ji}\) have opposing signs, stability
increases\textsuperscript{\protect\hyperlink{ref-Allesina2012}{2}}. If
diagonal elements of \(\boldsymbol{\gamma}\) vary independently, the
magnitude of \(\rho\) is decreased because \(\sigma^{2}_{\gamma}\)
increases the variance of \(M_{ij}\) without affecting the expected
covariance between \(M_{ij}\) and \(M_{ji}\) (Figure 2).

\vspace{2mm}
\hrule
\vspace{2mm}

\textbf{Figure 2: Complex system correlation versus stability with and
without variation in component response rates}. Each point represents
10000 replicate numerical simulations of a random complex system
\(\mathbf{M} = \gamma \mathbf{A}\) with a fixed correlation between
off-diagonal elements \(A_{ij}\) and \(A_{ji}\) (\(\rho\), x-axis).
Where real parts of eigenvalues of \(\mathbf{M}\) are negative (y-axis),
\(\mathbf{M}\) is stable (black dotted line). Blue circles show systems
in the absence of variation in component response rates
(\(\sigma^{2}_{\gamma} = 0\)). Red squares show systems in which
\(\sigma^{2}_{\gamma} = 1/3\). Arrows show the range of real parts of
leading eigenvalues observed. Because \(\gamma\) decreases the magnitude
of \(\rho\), purple lines are included to link replicate simulations
before (blue circles) and after (red squares) including \(\gamma\). The
range of \(\rho\) values in which \(\gamma\) decreases the mean real
part of the leading eigenvalue is indicated with grey shading. In all
simulations, system size and connectence were set to \(S = 25\) and
\(C = 1\), respectively. Off-diagonal elements of \(\textbf{A}\) were
randomly sampled from \(A_{ij} \sim \mathcal{N}(0, 0.4^{2})\), and
diagonal elements were set to \(-1\). Elements of \(\gamma\) were
sampled, \(\gamma \sim \mathcal{U}(0, 2)\).

\includegraphics[height=14cm,keepaspectratio]{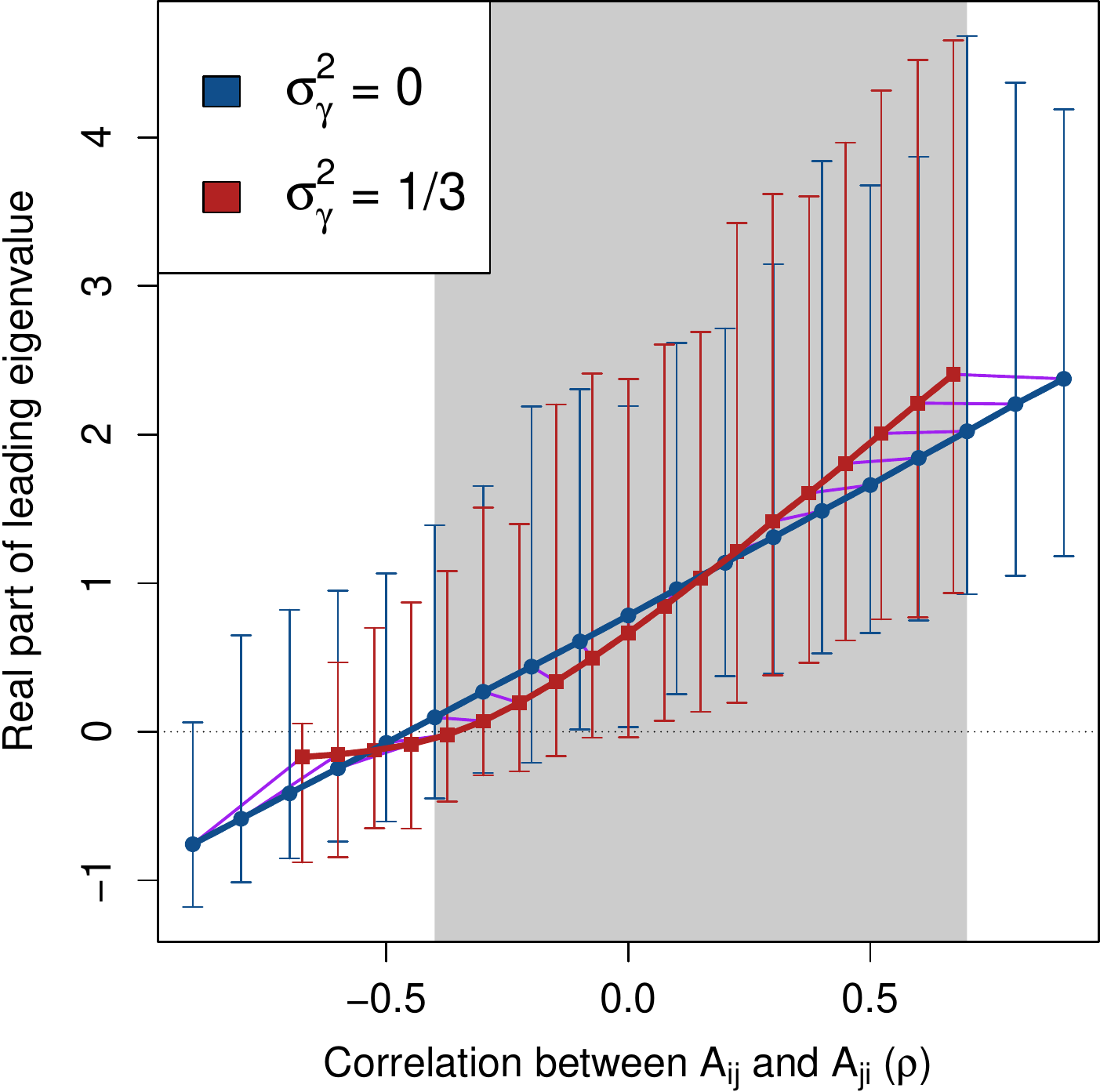}

\vspace{2mm}
\hrule

\textbf{Numerical simulations of random systems with and without
\(\mathbf{\sigma^{2}_{\gamma}}\)}. I used numerical simulations and
eigenanalysis to test how variation in \(\gamma\) affects stability in
random matrices with known properties, comparing the stability of
\(\textbf{A}\) versus \(\mathbf{M} = \gamma\mathbf{A}\). Values of
\(\gamma\) were sampled from a uniform distribution where
\(\gamma \sim \mathcal{U}(0, 2)\) and \(\sigma^{2}_{\gamma} = 1/3\) (see
Supplementary Information for other \(\gamma\) distributions, which gave
similar results). In all simulations, diagonal elements were
standardised to ensure that \(-d\) between individual \(\textbf{A}\) and
\(\textbf{M}\) pairs were identical (also note that
\(E[\gamma_{i}] = 1\)). First I focus on the effect of \(\gamma\) across
values of \(\rho\), then for increasing system sizes (\(S\)) in random
and structured networks. By increasing \(S\), the objective is to
determine the effect of \(\gamma\) as system complexity increases toward
the boundary at which stability is realistic for a finite system.

\textbf{Simulation of random \(\mathbf{M}\) across \(\mathbf{\rho}\)}.
Numerical simulations revealed that \(\sigma^{2}_{\gamma}\) results in a
nonlinear relationship between \(\rho\) and \(\Re(\lambda_{max})\),
which can sometimes increase the stability of the system. Figure 2 shows
a comparison of \(\Re(\lambda_{max})\) across \(\rho\) values for
\(\mathbf{A}\) (\(\sigma^{2}_{\gamma} = 0\)) versus \(\mathbf{M}\)
(\(\sigma^{2}_{\gamma} = 1/3\)) given \(S = 25\), \(C = 1\), and
\(\sigma_{A} = 0.4\). For \(-0.4 \leq \rho \leq 0.7\) (shaded region of
Fig. 2), expected \(\Re(\lambda_{max})\) was lower in \(\mathbf{M}\)
than \(\mathbf{A}\). For \(\rho \geq -0.1\), the lower bound of the
range of \(\Re(\lambda_{max})\) values also decreased given
\(\sigma^{2}_{\gamma}\), resulting in negative \(\Re(\lambda_{max})\) in
\(\mathbf{M}\) for \(\rho = -0.1\) and \(\rho = 0\). Hence, across a
wide range of system correlations, variation in the response rate of
system components had a stabilising effect.

The stabilising effect of \(\sigma^{2}_{\gamma}\) across \(\rho\)
increased with increasing \(S\). Figure 3 shows numerical simulations of
\(\mathbf{M}\) across increasing \(S\) given \(C = 1\) and
\(\sigma_{A} = 0.2\) (\(\sigma_{A}\) has been lowered here to better
illustrate the effect of \(S\); note that now given \(S = 25\),
\(1 = \sigma_{A}\sqrt{SC}\)). For relatively small systems
(\(S \leq 25\)), \(\sigma^{2}_{\gamma}\) never decreased the expected
\(\Re(\lambda_{max})\). But as \(S\) increased, the curvilinear
relationship between \(\rho\) and \(\Re(\lambda_{max})\) decreased
expected \(\Re(\lambda_{max})\) for \(\mathbf{M}\) given low magnitudes
of \(\rho\). In turn, as \(S\) increased, and systems became more
complex, \(\sigma^{2}_{\gamma}\) increased the proportion of numerical
simulations that were observed to be stable (see below).

\textbf{Simulation of random \(\mathbf{M}\) across \(\mathbf{S}\)}. To
investigate the effect of \(\sigma^{2}_{\gamma}\) on stability across
systems of increasing complexity, I simulated random
\(\mathbf{M = \gamma A}\) matrices at \(\sigma_{A} = 0.4\) and \(C = 1\)
across \(S = \{2, 3, ..., 49, 50\}\). One million \(\mathbf{M}\) were
simulated for each \(S\), and the stability of \(\mathbf{A}\) vesus
\(\mathbf{M}\) was assessed given \(\gamma \sim \mathcal{U}(0, 2)\)
(\(\sigma^{2}_{\gamma} = 1/3\)). For all \(S > 10\), I found that the
number of stable random systems was higher in \(\mathbf{M}\) than
\(\mathbf{A}\) (Fig. 4; see Supplementary Information for full table of
results), and that the difference between the probabilities of observing
a stable system increased with an increase in \(S\). In other words, the
potential for \(\sigma^{2}_{\gamma}\) to affect stability increased with
increasing system complexity and was most relevant for systems on the
cusp of being too complex to be realistically stable. For the highest
values of \(S\), nearly all systems that were stable given varying
\(\gamma\) would not have been stable given \(\gamma = 1\).

I also simulated 100000 \(\mathbf{M}\) for three types of random
networks that are typically interpreted as modelling three types of
interspecific ecological
interactions\textsuperscript{\protect\hyperlink{ref-Allesina2012}{2},\protect\hyperlink{ref-Allesina2011}{27}}.
These interaction types are competitive, mutualist, and predator-prey,
as modelled by off-diagonal elements that are constrained to be
negative, positive, or paired such that if \(A_{ij} > 0\) then
\(A_{ji} < 0\),
respectively\textsuperscript{\protect\hyperlink{ref-Allesina2012}{2}}
(but are otherwise identical to the purely random \(\mathbf{A}\)). As
\(S\) increased, a higher number of stable \(\mathbf{M}\) relative to
\(\mathbf{A}\) was observed for competitor and predator-prey, but not
mutualist, systems. A higher number of stable systems was observed
whenever \(S > 12\) and \(S > 40\) for competitive and predator-prey
systems, respectively (note that \(\rho < 0\) for predator-prey systems,
making stability more likely overall). The stability of mutualist
systems was never affected by \(\sigma^{2}_{\gamma}\).

The effect of \(\sigma^{2}_{\gamma}\) on stability did not change
qualitatively across values of \(C\), \(\sigma_{A}\), or for different
distributions of \(\gamma\) (see Supporting Information).

\textbf{Simulation of structured \(\mathbf{M}\) across \(\mathbf{S}\)}.
To investigate how \(\sigma^{2}_{\gamma}\) affects the stability of
commonly observed network structures, I simulated one million
\(\mathbf{M = \gamma A}\) for
small-world\textsuperscript{\protect\hyperlink{ref-Watts1998}{16}},
scale-free\textsuperscript{\protect\hyperlink{ref-Albert2002}{17}}, and
cascade food
web\textsuperscript{\protect\hyperlink{ref-Solow1998}{18},\protect\hyperlink{ref-Williams2000}{19}}
networks. In all of these networks, rules determining the presence or
absence of an interaction between components \(i\) and \(j\) constrain
the overall structure of the network. In small-world networks,
interactions between components are constrained so that the expected
degree of separation between any two components increases in proportion
to \(\log(S)\)\textsuperscript{\protect\hyperlink{ref-Watts1998}{16}}.
In scale-free networks, the distribution of the number of components
with which a focal component interacts follows a power law; a few
components have many interactions while most components have few
interactions\textsuperscript{\protect\hyperlink{ref-Albert2002}{17}}. In
cascade food webs, species are ranked and interactions are constrained
such that a species \(i\) can only feed on \(j\) if the rank of
\(i > j\).

\clearpage

\hrule
\vspace{2mm}

\textbf{Figure 3: System correlation versus stability across different
system sizes}. In each panel, 10000 random complex systems
\(\mathbf{M} = \gamma \mathbf{A}\) are simulated for each correlation
\(\rho = \{-0.90, -0.85, ..., 0.85, 0.90 \}\) between off-diagonal
elements \(A_{ij}\) and \(A_{ji}\). Lines show the expected real part of
the leading eigenvalues of \(\mathbf{M}\) (red squares;
\(\sigma^{2}_{\gamma} = 1/3\)) versus \(\mathbf{A}\) (blue circles;
\(\sigma^{2}_{\gamma} = 0\)) across \(\rho\), where negative values
(below the dotted black line) indicate system stability. Differences
between lines thereby show the effect of component response rate
variation (\(\gamma\)) on system stability across system correlations
and sizes (\(S\)). For all simulations, system connectance was
\(C = 1\). Off-diagonal elements of \(\textbf{A}\) were randomly sampled
from \(A_{ij} \sim \mathcal{N}(0, 0.2^{2})\), and diagonal elements were
set to \(-1\). Elements of \(\gamma\) were sampled such that
\(\gamma \sim \mathcal{U}(0, 2)\), so \(\sigma^{2}_{\gamma} = 1/3\).

\includegraphics{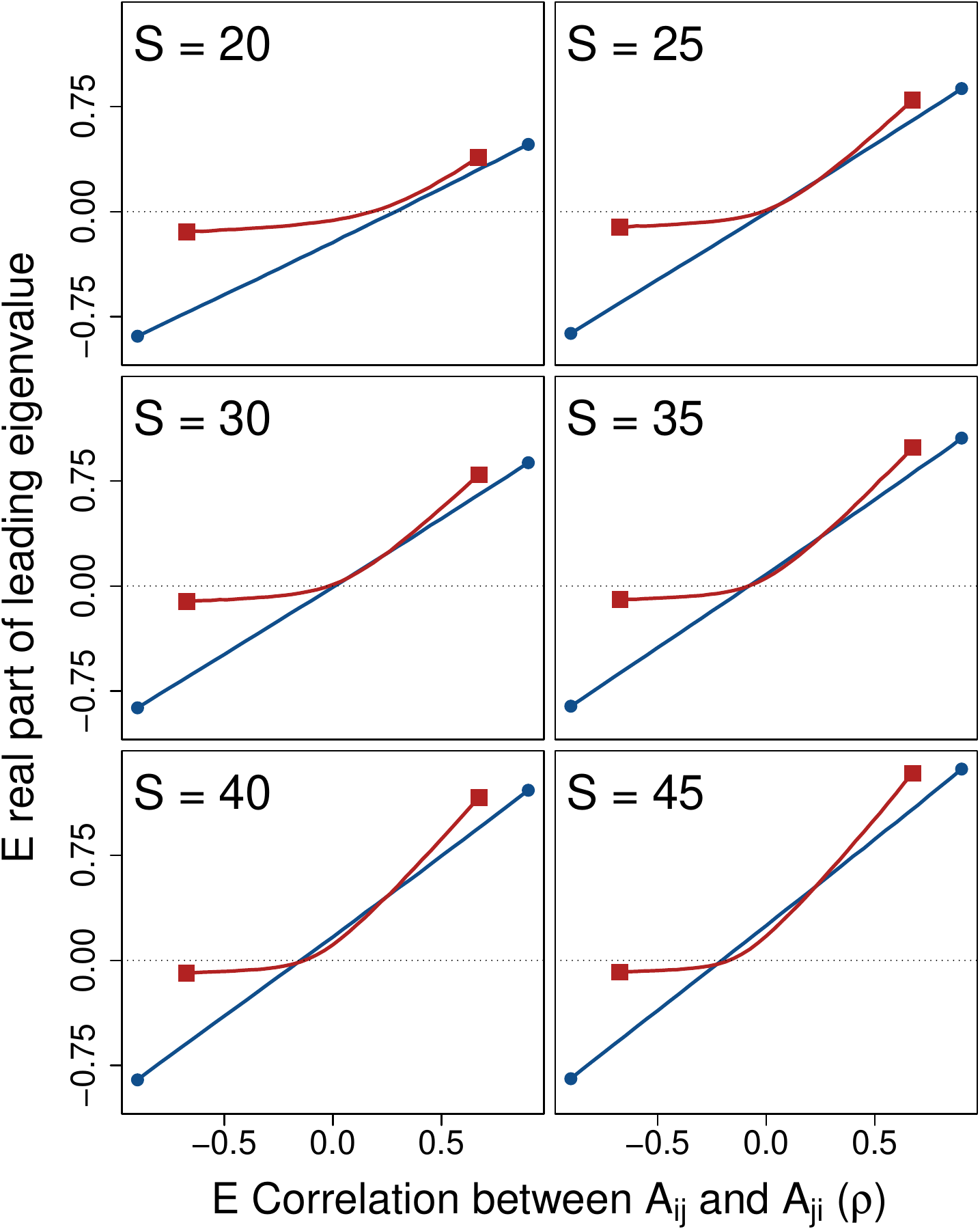}

\vspace{2mm}
\hrule

Network structure did not strongly modulate the effect that
\(\sigma^{2}_{\gamma}\) had on stability. For comparable magnitudes of
complexity, structured networks still had a higher number of stable
\(\mathbf{M}\) than \(\mathbf{A}\). For random networks,
\(\sigma^{2}_{\gamma}\) increased stability given \(S > 10\)
(\(\sigma_{A} = 0.4\) and \(C = 1\)), and therefore complexity
\(\sigma_{A} \sqrt{SC} \gtrapprox 1.26\). This threshold of complexity,
above which more \(\mathbf{M}\) than \(\mathbf{A}\) were stable, was
comparable for small-world networks, and slightly lower for scale-free
networks (note that algorithms for generating small-world and scale-free
networks necessarily led to varying \(C\); see methods). Varying
\(\gamma\) increased stability in cascade food webs for \(S > 27\), and
therefore at a relatively low complexity magnitudes compared to random
predator-prey networks (\(S > 40\)). Overall, network structure did not
greatly change the effect that \(\sigma^{2}_{\gamma}\) had on increasing
the upper bound of complexity within which stability might reasonably be
observed.

\vspace{2mm}
\hrule
\vspace{2mm}

\textbf{Figure 4: Stability of large complex systems with and without
variation in component response rate (\(\boldsymbol{\gamma}\)).} The
\(\log\) number of systems that are stable across different system sizes
(\(S = \{2, 3, ..., 49, 50 \}\)) given \(C = 1\), and the proportion of
systems for which variation in \(\gamma\) is critical for system
stability. For each \(S\), 1 million complex systems are randomly
generated. Stability of each complex system is tested given variation in
\(\gamma\) by randomly sampling \(\gamma \sim \mathcal{U}(0, 2)\).
Stability given \(\sigma^{2}_{\gamma}>0\) is then compared to stability
in an otherwise identical system in which
\(\gamma_{i} = E[\mathcal{U}(0, 2)]\) for all components. Blue and red
bars show the number of stable systems in the absence and presence of
\(\sigma^{2}_{\gamma}\), respectively. The black line shows the
proportion of systems that are stable when \(\sigma^{2}_{\gamma}>0\),
but would be unstable if \(\sigma^{2}_{\gamma}=0\) (i.e., the
conditional probability that \(\mathbf{A}\) is unstable given that
\(\mathbf{M}\) is stable).

\includegraphics{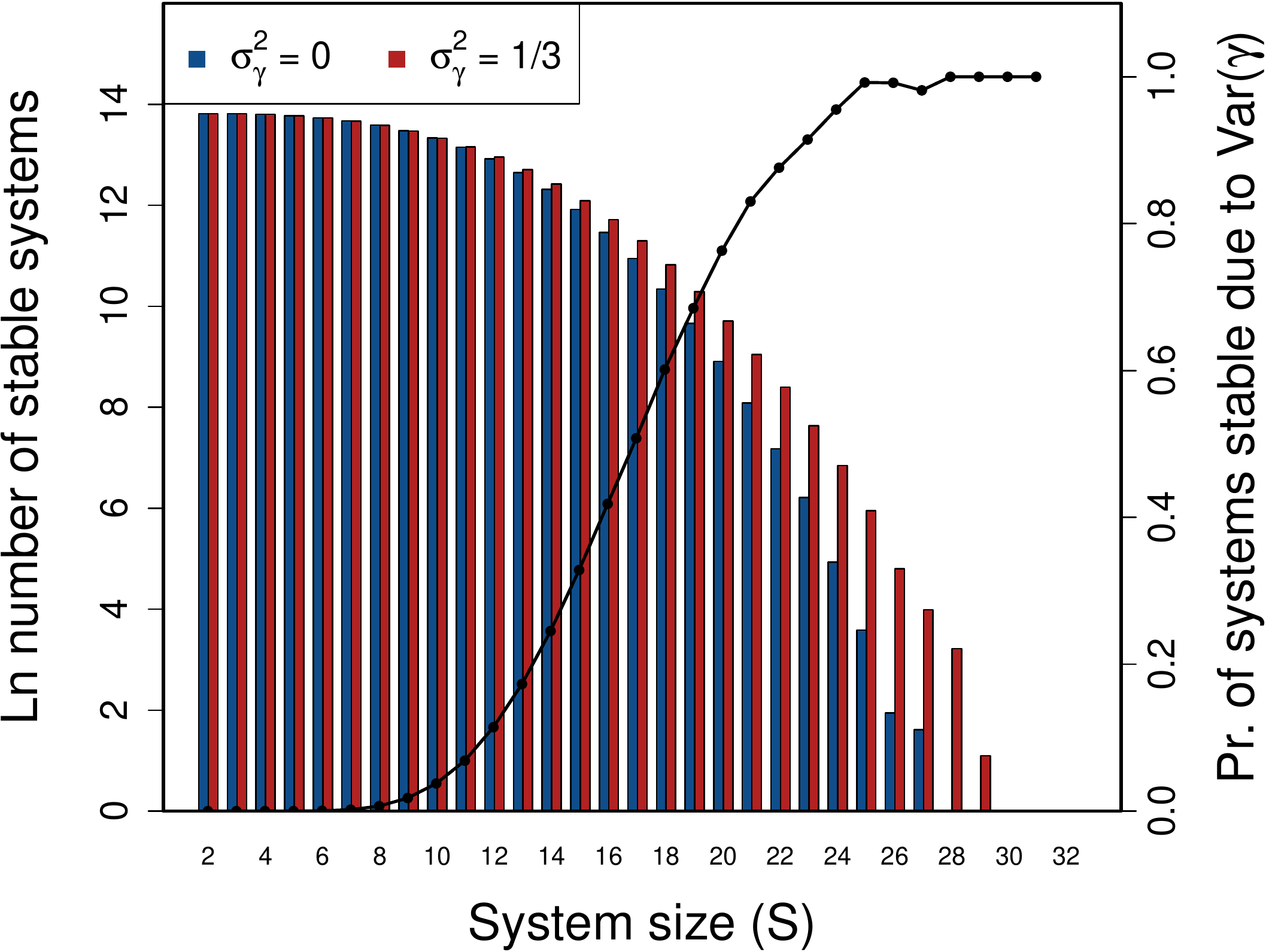}

\vspace{2mm}
\hrule

\textbf{System feasibility given \(\mathbf{\sigma^{2}_{\gamma}}\)} For
complex systems in which individual system components represent the
density of some tangible quantity, it is relevant to consider the
feasibility of the system. Feasibility assumes that values of all
components are positive at
equilibrium\textsuperscript{\protect\hyperlink{ref-Grilli2017}{6},\protect\hyperlink{ref-Dougoud2018}{28},\protect\hyperlink{ref-Song2018}{29}}.
This is of particular interest for ecological communities because
population density (\(n\)) cannot take negative values, meaning that
ecological systems need to be feasible for stability to be biologically
realistic\textsuperscript{\protect\hyperlink{ref-Dougoud2018}{28}}.
While my results are intended to be general to all complex systems, and
not restricted to species networks, I have also performed a feasibility
analysis on all matrices tested for stability. I emphasise that
\(\gamma\) is not interpreted as population density in this analysis,
but instead as a fundamental property of species life history such as
expected generation time. Feasibility was unaffected by
\(\sigma^{2}_{\gamma}\) and instead occurred with a fixed probability of
\(1/2^{S}\), consistent with a recent proof by Serván et
al.\textsuperscript{\protect\hyperlink{ref-Servan2018}{30}} (see
Supplementary Information). Hence, for pure interacting species
networks, variation in component response rate (i.e., species generation
time) does not affect stability at biologically realistic species
densities.

\textbf{Targeted manipulation of \(\boldsymbol{\gamma}\)}. To further
investigate the potential of \(\sigma^{2}_{\gamma}\) to be stabilising,
I used a genetic algorithm. Genetic algorithms are heuristic tools that
mimic evolution by natural selection, and are useful when the space of
potential solutions (in this case, possible combinations of \(\gamma\)
values leading to stability in a complex system) is too large to search
exhaustively\textsuperscript{\protect\hyperlink{ref-Hamblin2013}{31}}.
Generations of selection on \(\gamma\) value combinations to minimise
\(\Re(\lambda_{max})\) demonstrated the potential for
\(\sigma^{2}_{\gamma}\) to increase system stability. Across
\(S = \{2, 3, ..., 39, 40\}\), sets of \(\gamma\) values were found that
resulted in stable systems with probabilities that were up to four
orders of magnitude higher than when \(\gamma = 1\) (see Supplementary
Information), meaning that stability could often be achieved by
manipulating \(S\) \(\gamma\) values rather than \(S \times S\)
\(\mathbf{M}\) elements (i.e., by manipulating component response rates
rather than interactions between components).

\subsection{Discussion}\label{discussion}

I have shown that the stability of complex systems might often be
contigent upon variation in the response rates of their individual
components, meaning that factors such as rate of trait evolution (in
biological networks), transaction speed (in economic networks), or
communication speed (in social networks) need to be considered when
investigating the stability of complex systems. Variation in component
response rate is more likely to be critical for stability in systems
that are especially complex, and it can ultimately increase the
probability that system stability is observed above that predicted by
May's\textsuperscript{\protect\hyperlink{ref-May1972}{1}} classically
derived \(\sigma \sqrt{SC}\) criterion. The logic outlined here is
general, and potentially applies to any complex system in which
individual system components can vary in their reaction rates to system
perturbation.

It is important to recognise that variation in component response rate
is not stabilising per se; that is, adding variation in component
response rates to a particular system does not increase the probability
that the system will be stable. Rather, highly complex systems that are
observed to be stable are more likely to have varying component response
rates, and for this variation to be critical to their stability (Fig.
4). This is caused by the shift to a non-uniform distribution of
eigenvalues that occurs by introducing variation in \(\gamma\) (Fig. 1),
which can sometimes cause all of the real components of the eigenvalues
of the system matrix to become negative, but might also increase the
real components of eigenvalues.

My focus here is distinct from Gibbs et
al.\textsuperscript{\protect\hyperlink{ref-Gibbs2017}{24}}, who applied
the same mathematical framework to investigate how a diagonal matrix
\(\mathbf{X}\) (equivalent to \(\boldsymbol{\gamma}\) in my model)
affects the stability of a community matrix \(\mathbf{M}\) given an
interaction matrix \(\mathbf{A}\) within a generalised Lotka-Volterra
model, where \(\mathbf{M} = \mathbf{XA}\). Gibbs et
al.\textsuperscript{\protect\hyperlink{ref-Gibbs2017}{24}} analytically
demonstrated that the effect of \(\mathbf{X}\) on system stability
decreases exponentially as system size becomes arbitrarily large
(\(S \to \infty\)) for a given magnitude of complexity
\(\sigma\sqrt{SC}\). My numerical results do not contradict this
prediction because I did not scale \(\sigma = 1 / \sqrt{S}\), but
instead fixed \(\sigma\) and increased \(S\) to thereby increase total
system complexity (see Supplemental Information for results simulated
across \(\sigma\) and \(C\)). Overall, I show that component response
rate variation increases the upper bound of complexity at which
stability can be realistically observed, meaning that highly complex
systems are more likely than not to vary in their component response
rates, and for this variation to be critical for system stability.

Interestingly, while complex systems were more likely to be stable given
variation in component response rate, they were not more likely to be
feasible, meaning that stability was not increased when component values
were also restricted to being positive at equilibrium. Feasibility is
important to consider, particularly for the study of ecological networks
of
species\textsuperscript{\protect\hyperlink{ref-Grilli2017}{6},\protect\hyperlink{ref-Stone2017}{25},\protect\hyperlink{ref-Dougoud2018}{28},\protect\hyperlink{ref-Servan2018}{30}}
because population densities cannot realistically be negative. My
results therefore suggest that variation in the rate of population
responses to perturbation (e.g., due to differences in generation time
among species) is unlikely to be critical to the stability of purely
multi-species interaction networks (see also Supplementary Information).
Nevertheless, ecological interactions do not exist in isolation in
empirical
systems\textsuperscript{\protect\hyperlink{ref-Patel2018}{20}}, but
instead interact with evolutionary, abiotic, or social-economic systems.
The relevance of component response rate for complex system stability
should therefore not be ignored in the broader context of ecological
communities.

The potential importance of component response rate variation was most
evident from the results of simulations in which the genetic algorithm
was used in attempt to maximise the probability of system stability. The
probability that some combination of component response rates could be
found to stabilise the system was shown to be up to four orders of
magnitude higher than the background probabilities of stability in the
absence of any component response rate variation. Instead of
manipulating the \(S \times S\) interactions between system components,
it might therefore be possible to manipulate only the \(S\) response
rates of individual system components to achieve stability. Hence,
managing the response rates of system components in a targeted way could
potentially facilitate the stabilisation of complex systems through a
reduction in dimensionality.

A general mathematical framework encompassing shifts in eigenvalue
distributions caused by a diagonal matrix \(\boldsymbol{\gamma}\) has
been
investigated\textsuperscript{\protect\hyperlink{ref-Ahmadian2015}{23}}
and recently applied to questions concerning species density and
feasibility\textsuperscript{\protect\hyperlink{ref-Gibbs2017}{24},\protect\hyperlink{ref-Stone2017}{25}},
but \(\boldsymbol{\gamma}\) has not been interpreted as rates of
response of individual system components to perturbation. My model
focuses on component response rates for systems of a finite size, in
which complexity is high but not yet high enough to make the probability
of stability unrealistically low for actual empirical systems. For this
upper range of system size, randomly assembled complex systems are more
likely to be stable if their component response rates vary (e.g.,
\(10 < S < 30\) for parameter values in Fig. 4). Variation in component
response rate might therefore be critical for maintaining stability in
many highly complex empirical systems. These results are broadly
applicable for understanding the stability of complex networks across
the physical, life, and social sciences.

\subsection{Methods}\label{methods}

\textbf{Component response rate (\(\boldsymbol{\gamma}\)) variation}. In
a synthesis of eco-evolutionary feedbacks on community stability, Patel
et al.\textsuperscript{\protect\hyperlink{ref-Patel2018}{20}} model a system that includes a vector of potentially changing
species densities (\(\mathbf{n}\)) and a vector of potentially evolving
traits
(\(\mathbf{x}\)).
For any species \(i\) or trait \(j\), change in species density
(\(n_{i}\)) or trait value (\(x_{j}\)) with time (\(t\)) is a function
of the vectors \(\mathbf{n}\) and \(\mathbf{x}\),

\[\frac{dn_{i}}{dt} = n_{i}f_{i}(\mathbf{n}, \mathbf{x}),\]

\[\frac{dx_{j}}{dt} = \epsilon g_{j}(\mathbf{n}, \mathbf{x}).\]

In the above, \(f_{i}\) and \(g_{j}\) are functions that define the
effects of all species densities and trait values on the density of a
species \(i\) and the value of trait \(j\), respectively. 
Patel et al.\textsuperscript{\protect\hyperlink{ref-Patel2018}{20}}
were interested in stability when the evolution of traits was relatively
slow or fast in comparison with the change in species
densities, and
this is modulated in the above by the scalar \(\epsilon\). The value of
\(\epsilon\) thereby determines the timescale separation between ecology
and evolution, with high \(\epsilon\) modelling relatively fast
evolution and low \(\epsilon\) modelling relatively slow
evolution\textsuperscript{\protect\hyperlink{ref-Patel2018}{20}}.

I use the same principle that Patel et al.\textsuperscript{\protect\hyperlink{ref-Patel2018}{20}} 
use to modulate the relative
rate of evolution to modulate rates of component responses for \(S\)
components. Following
May\textsuperscript{\protect\hyperlink{ref-May1972}{1},\protect\hyperlink{ref-May1973}{32}},
the value of a component \(i\) at time \(t\) (\(v_{i}(t)\)) is affected
by the value of \(j\) (\(v_{j}(t)\)) and \(j\)'s marginal effect on
\(i\) (\(a_{ij}\)), and by \(i\)'s response rate (\(\gamma_{i}\)),

\[\frac{dv_{i}(t)}{dt} = \gamma_{i} \sum_{j=1}^{S}a_{ij}v_{j}(t).\]

In matrix notation\textsuperscript{\protect\hyperlink{ref-May1973}{32}},

\[\frac{d\mathbf{v}(t)}{dt} = \boldsymbol{\gamma} \mathbf{A}\mathbf{v}(t).\]

In the above, \(\boldsymbol{\gamma}\) is a diagonal matrix in which
elements correspond to individual component response rates. Therefore,
\(\mathbf{M} = \boldsymbol{\gamma} \mathbf{A}\) defines the change in
values of system components and can be analysed using the techniques of
May\textsuperscript{\protect\hyperlink{ref-May1972}{1},\protect\hyperlink{ref-Ahmadian2015}{23},\protect\hyperlink{ref-May1973}{32}}.
In these analyses, row means of \(\mathbf{A}\) are expected to be
identical, but variation around this expectation will naturally arise
due to random sampling of \(\mathbf{A}\) off-diagonal elements and
finite \(S\). In simulations, the total variation in \(\mathbf{M}\) row
means that is attributable to \(\mathbf{A}\) is small relative to that
attributable to \(\boldsymbol{\gamma}\), especially at high \(S\).
Variation in \(\boldsymbol{\gamma}\) specifically isolates the effects
of differing component response rates, hence causing differences in
expected \(\mathbf{M}\) row means.

\textbf{Construction of random and structured networks}. I used the R
programming language for all numerical simulations and
analyses\textsuperscript{\protect\hyperlink{ref-Rproject}{33}}. Purely
random networks were generated by sampling off-diagonal elements from an
i.i.d \(A_{ij} \sim \mathcal{N}(0, 0.4^{2})\) with a probability \(C\)
(unsampled elements were set to zero). Diagonal elements \(A_{ii}\) were
set to \(-1\). Elements of \(\boldsymbol{\gamma}\) were simulated i.i.d.
from a distribution with positive support (typically
\(\gamma \sim \mathcal{U}(0, 2)\)). Random \(\mathbf{A}\) matrices with
correlated elements \(A_{ij}\) and \(A_{ji}\) were built using Cholesky
decomposition. Competitor networks in which off-diagonal elements
\(A_{ij} \leq 0\) were constructed by first building a random
\(\mathbf{A}\), then flipping the sign of any elements in which
\(A_{ij} > 0\). Similarly, mutualist networks were constructed by
building a random \(\mathbf{A}\), then flipping the sign of elements
where \(A_{ij} < 0\). Predator-prey networks were constructed by first
building a random \(\mathbf{A}\), then flipping the sign of either
\(A_{ij}\) or \(A_{ji}\) if \(A_{ij} \times A_{ji} > 0\).

Small-world networks were constructed using the method of Watts and
Strogatz\textsuperscript{\protect\hyperlink{ref-Watts1998}{16}}. First,
a regular network\textsuperscript{\protect\hyperlink{ref-Watts1998}{16}}
was created such that components were arranged in a circle. Each
component was initially set to interact with its \(k/2\) closest
neighbouring components on each side, where \(k\) was an even natural
number (e.g., for \(k = 2\) the regular network forms a ring in which
each component interacts with its two adjacent neighbours; see
Supplemental Material for examples). Each interaction between a focal
component and its neighbour was then removed and replaced with with a
probability of \(\beta\). In replacement, a new component was randomly
selected to interact with the focal component; selection was done with
equal probability among all but the focal component. The resulting
small-world network was represented by a square \(S \times S\) binary
matrix \(\mathbf{B}\) in which 1s represented interactions between
components and 0s represented the absence of an interaction. A new
random matrix \(\mathbf{J}\) was then generated with elements \(J_{ij}\)
sampled i.i.d. from \(\mathcal{N}(0, 0.4^{2})\). To build the
interaction matrix \(\mathbf{A}\), I used element-wise multiplication
\(\mathbf{A} = \mathbf{J} \odot \mathbf{B}\), then set
\(diag(\mathbf{A}) = -1\). I set \(k = S/12\) and simulated small-world
networks across all combinations of
\(S = \{24, 48, 72, 96, 120, 144, 168\}\) and
\(\beta = \{0, 0.01, 0.1, 0.25, 1\}\).

Scale-free networks were constructed using the method of Albert and
Barabási\textsuperscript{\protect\hyperlink{ref-Albert2002}{17}}. First,
a saturated network (all components interact with each other) of size
\(m \leq S\) was created. New components were then added sequentially to
the network; each newly added component was set to interact with \(m\)
randomly selected existing components. When the system size reached
\(S\), the distribution of the number of total interactions that
components had followed a power-law
tail\textsuperscript{\protect\hyperlink{ref-Albert2002}{17}}. The
resulting network was represented by an \(S \times S\) binary matrix
\(\mathbf{G}\), where 1s and 0s represent the presence and absence of an
interaction, respectively. As with small-world networks, a random matrix
\(\mathbf{J}\) was generated, and
\(\mathbf{A} = \mathbf{J} \odot \mathbf{G}\). Diagonal elements were set
to \(-1\). I simulated scale-free networks across all combinations of
\(S = \{24, 48, 72, 96, 120\}\) and \(m = \{2, 3, ..., 11, 12\}\).

Cascade food webs were constructed following Solow and
Beet\textsuperscript{\protect\hyperlink{ref-Solow1998}{18}}. First, a
random matrix \(\mathbf{A}\) was generated with off-diagonal elements
sampled i.i.d so that \(A_{ij} \sim \mathcal{N}(0, 0.4^2)\). Each
component in the system was ranked from \(1\) to \(S\). If component
\(i\) had a higher rank than component \(j\) and \(A_{ij} < 0\), then
\(A_{ij}\) was multiplied by \(-1\). If \(i\) had a lower rank than
\(j\) and \(A_{ji} < 0\), then \(A_{ji}\) was multiplied by \(-1\). In
practice, this resulted in a matrix \(\mathbf{A}\) with negative and
positive values in the lower and upper triangles, respectively. Diagonal
elements of \(\mathbf{A}\) were set to \(-1\) and \(C = 1\). I simulated
cascade food webs for \(S = \{2, 3, ..., 59, 60\}\).

\textbf{System feasibility}. Dougoud et
al.\textsuperscript{\protect\hyperlink{ref-Dougoud2018}{28}} identify
the following feasibility criteria for ecological systems characterised
by \(S\) interacting species with varying densities in a generalised
Lotka-Volterra model,

\[\mathbf{n^{*}} = -\left(\theta \mathbf{I} + (CS)^{-\delta}\mathbf{J} \right)^{-1}\mathbf{r}.\]

In the above, \(\mathbf{n^{*}}\) is the vector of species densities at
equilibrium. Feasibility is satisfied if all elements in
\(\mathbf{n^{*}}\) are positive. The matrix \(\mathbf{I}\) is the
identity matrix, and the value \(\theta\) is the strength of
intraspecific competition (diagonal elements). Diagonal values are set
to \(-1\), so \(\theta = -1\). The variable \(\delta\) is a
normalisation parameter that modulates the strength of interactions
(\(\sigma\)) for \(\mathbf{J}\). Implicitly, here \(\delta = 0\)
underlying strong interactions. Hence, \((CS)^{-\delta} = 1\), so in the
above, a diagonal matrix of -1s (\(\theta \mathbf{I}\)) is added to
\(\mathbf{J}\), which has a diagonal of all zeros and an off-diagonal
affecting species interactions (i.e., the expression \((CS)^{-\delta}\)
relates to May's\textsuperscript{\protect\hyperlink{ref-May1972}{1}}
stability
criterion\textsuperscript{\protect\hyperlink{ref-Dougoud2018}{28}} by
\(\frac{\sigma}{(CS)^{-\delta}}\sqrt{SC} < 1\), and hence for my
purposes \((CS)^{-\delta} = 1\)). Given
\(\mathbf{A} = \theta\mathbf{I + J}\), the above criteria is therefore
reduced to the below (see also Serván et
al.\textsuperscript{\protect\hyperlink{ref-Servan2018}{30}}),

\[\mathbf{n^{*} = -A^{-1}r}.\]

To check the feasibility criteria for \(\mathbf{M = \gamma A}\), I
therefore evaluated \(\mathbf{-M^{-1}r}\) (\(\mathbf{r}\) elements were
sampled i.i.d. from \(r \sim \mathcal{N}(0, 0.4^{2})\)). Feasibility is
satisfied if all of the elements of the resulting vector are positive.

\textbf{Genetic algorithm}. Ideally, to investigate the potential of
\(\sigma^{2}_{\gamma}\) for increasing the proportion of stable complex
systems, the search space of all possible \(diag(\boldsymbol{\gamma})\)
vectors would be evaluated for each unique \(\mathbf{M = \gamma A}\).
This is technically impossible because \(\gamma_{i}\) can take any real
value between 0-2, but even rounding \(\gamma_{i}\) to reasonable values
would result in a search space too large to practically explore. Under
these conditions, genetic algorithms are highly useful tools for finding
practical solutions by mimicking the process of biological
evolution\textsuperscript{\protect\hyperlink{ref-Hamblin2013}{31}}. In
this case, the practical solution is finding vectors of
\(diag(\boldsymbol{\gamma})\) that decrease the most positive real
eigenvalue of \(\mathbf{M}\). The genetic algorithm used achieves this
by initialising a large population of 1000 different potential
\(diag(\boldsymbol{\gamma})\) vectors and allowing this population to
evolve through a process of mutation, crossover (swaping \(\gamma_{i}\)
values between vectors), selection, and reproduction until either a
\(diag(\boldsymbol{\gamma})\) vector is found where all
\(\Re(\lambda) < 0\) or some ``giving up'' critiera is met.

For each \(S = \{2, 3, ..., 39, 40\}\), the genetic algorithm was run
for 100000 random \(\mathbf{M = \gamma A}\) (\(\sigma_{A} = 0.4\),
\(C = 1\)). The genetic algorithm was initialised with a population of
1000 different \(diag(\boldsymbol{\gamma})\) vectors with elements
sampled i.i.d from \(\gamma \sim \mathcal{U}(0, 2)\). Eigenanalysis was
performed on the \(\mathbf{M}\) resulting from each
\(\boldsymbol{\gamma}\), and the 20 \(diag(\boldsymbol{\gamma})\)
vectors resulting in \(\mathbf{M}\) with the lowest
\(\Re(\lambda_{max})\) each produced 50 clonal offspring with subsequent
random mutation and crossover between the resulting new generation of
1000 \(diag(\boldsymbol{\gamma})\) vectors. Mutation of each
\(\gamma_{i}\) in a \(diag(\boldsymbol{\gamma})\) vector occurred with a
probability of 0.2, resulting in a mutation effect of size
\(\mathcal{N}(0, 0.02^{2})\) being added to generate the newly mutated
\(\gamma_{i}\) (any \(\gamma_{i}\) values that mutated below zero were
multiplied by \(-1\), and any values that mutated above 2 were set to
2). Crossover occurred between two sets of 100
\(diag(\boldsymbol{\gamma})\) vectors paired in each generation; vectors
were randomly sampled with replacement among but not within sets. Vector
pairs selected for crossover swapped all elements between and including
two \(\gamma_{i}\) randomly selected with replacement (this allowed for
reversal of vector element positions during crossover; e.g.,
\(\{\gamma_{4}, \gamma_{5}, \gamma_{6}, \gamma_{7}\} \to \{\gamma_{7}, \gamma_{6}, \gamma_{5}, \gamma_{4}\}\)
). The genetic algorithm terminated if a stable \(\mathbf{M}\) was
found, 20 generations occurred, or if the mean \(\boldsymbol{\gamma}\)
fitness increase between generations was less than 0.01 (where fitness
was defined as \(W_{\gamma} = -\Re(\lambda_{max})\) for \(\mathbf{M}\)).

\textbf{Acknowledgements:} I am supported by a Leverhulme Trust Early
Career Fellowship (ECF-2016-376). Conversations with L. Bussière and N.
Bunnefeld, and comments from J. J. Cusack and I. L. Jones, improved the
quality of this work.

\textbf{Supplementary Information:} Full tables of stability results for
simulations across different system size (\(S\)) values, ecological
community types, connectance (\(C\)) values, interaction strengths
(\(\sigma\)), and \(\gamma\) distributions are provided as supplementary
material. An additional table also shows results for how feasibility
changes across \(S\). All code and simulation outputs are publicly
available as part of the RandomMatrixStability package on GitHub
(\url{https://github.com/bradduthie/RandomMatrixStability}).

\textbf{Additional Information:} The author declares no competing
interests. All work was carried out by A. Bradley Duthie, and all code
and data are accessible on
\href{https://github.com/bradduthie/RandomMatrixStability}{GitHub}.

\textbf{References}

\hypertarget{refs}{}
\hypertarget{ref-May1972}{}
1. May, R. M. Will a large complex system be stable? \emph{Nature}
\textbf{238,} 413--414 (1972).

\hypertarget{ref-Allesina2012}{}
2. Allesina, S. \& Tang, S. Stability criteria for complex ecosystems.
\emph{Nature} \textbf{483,} 205--208 (2012).

\hypertarget{ref-Townsend2010a}{}
3. Townsend, S. E., Haydon, D. T. \& Matthews, L. On the generality of
stability-complexity relationships in Lotka-Volterra ecosystems.
\emph{Journal of Theoretical Biology} \textbf{267,} 243--251 (2010).

\hypertarget{ref-Mougi2012}{}
4. Mougi, A. \& Kondoh, M. Diversity of interaction types and ecological
community stability. \emph{Science} \textbf{337,} 349--351 (2012).

\hypertarget{ref-Allesina2015}{}
5. Allesina, S. \emph{et al.} Predicting the stability of large
structured food webs. \emph{Nature Communications} \textbf{6,} 7842
(2015).

\hypertarget{ref-Grilli2017}{}
6. Grilli, J. \emph{et al.} Feasibility and coexistence of large
ecological communities. \emph{Nature Communications} \textbf{8,} (2017).

\hypertarget{ref-Gray2008}{}
7. Gray, R. T. \& Robinson, P. A. Stability and synchronization of
random brain networks with a distribution of connection strengths.
\emph{Neurocomputing} \textbf{71,} 1373--1387 (2008).

\hypertarget{ref-Gray2009}{}
8. Gray, R. T. \& Robinson, P. A. Stability of random brain networks
with excitatory and inhibitory connections. \emph{Neurocomputing}
\textbf{72,} 1849--1858 (2009).

\hypertarget{ref-Rosenfeld2009}{}
9. Rosenfeld, S. Patterns of stochastic behavior in dynamically unstable
high-dimensional biochemical networks. \emph{Gene Regulation and Systems
Biology} \textbf{3,} 1--10 (2009).

\hypertarget{ref-MacArthur2010}{}
10. MacArthur, B. D., Sanchez-Garcia, R. J. \& Ma'ayan, A. Microdynamics
and criticality of adaptive regulatory networks. \emph{Physics Review
Letters} \textbf{104,} 168701 (2010).

\hypertarget{ref-May2008}{}
11. May, R. M., Levin, S. A. \& Sugihara, G. Complex systems: Ecology
for bankers. \emph{Nature} \textbf{451,} 893--895 (2008).

\hypertarget{ref-Haldane2011}{}
12. Haldane, A. G. \& May, R. M. Systemic risk in banking ecosystems.
\emph{Nature} \textbf{469,} 351--355 (2011).

\hypertarget{ref-Suweis2014}{}
13. Suweis, S. \& D'Odorico, P. Early warning signs in social-ecological
networks. \emph{PLoS ONE} \textbf{9,} (2014).

\hypertarget{ref-Bardoscia2017}{}
14. Bardoscia, M., Battiston, S., Caccioli, F. \& Caldarelli, G.
Pathways towards instability in financial networks. \emph{Nature
Communications} \textbf{8,} 1--7 (2017).

\hypertarget{ref-Tao2010}{}
15. Tao, T. \& Vu, V. Random matrices: Universality of ESDs and the
circular law. \emph{Annals of Probability} \textbf{38,} 2023--2065
(2010).

\hypertarget{ref-Watts1998}{}
16. Watts, D. J. \& Strogatz, S. H. Collective dynamics of 'small world'
networks. \emph{Nature} \textbf{393,} 440--442 (1998).

\hypertarget{ref-Albert2002}{}
17. Albert, R. \& Barabási, A. L. Statistical mechanics of complex
networks. \emph{Reviews of Modern Physics} \textbf{74,} 47--97 (2002).

\hypertarget{ref-Solow1998}{}
18. Solow, A. R. \& Beet, A. R. On lumping species in food webs.
\emph{Ecology} \textbf{79,} 2013--2018 (1998).

\hypertarget{ref-Williams2000}{}
19. Williams, R. J. \& Martinez, N. D. Simple rules yield complex food
webs. \emph{Nature} \textbf{404,} 180--183 (2000).

\hypertarget{ref-Patel2018}{}
20. Patel, S., Cortez, M. H. \& Schreiber, S. J. Partitioning the
effects of eco-evolutionary feedbacks on community stability.
\emph{American Naturalist} \textbf{191,} 1--29 (2018).

\hypertarget{ref-Tang2014b}{}
21. Tang, S. \& Allesina, S. Reactivity and stability of large
ecosystems. \emph{Frontiers in Ecology and Evolution} \textbf{2,} 1--8
(2014).

\hypertarget{ref-Sommers1988}{}
22. Sommers, H. J., Crisanti, A., Sompolinsky, H. \& Stein, Y. Spectrum
of large random asymmetric matrices. \emph{Physical Review Letters}
\textbf{60,} 1895--1898 (1988).

\hypertarget{ref-Ahmadian2015}{}
23. Ahmadian, Y., Fumarola, F. \& Miller, K. D. Properties of networks
with partially structured and partially random connectivity.
\emph{Physical Review E - Statistical, Nonlinear, and Soft Matter
Physics} \textbf{91,} 012820 (2015).

\hypertarget{ref-Gibbs2017}{}
24. Gibbs, T., Grilli, J., Rogers, T. \& Allesina, S. The effect of
population abundances on the stability of large random ecosystems.
\emph{Physical Review E - Statistical, Nonlinear, and Soft Matter
Physics} \textbf{98,} 022410 (2018).

\hypertarget{ref-Stone2017}{}
25. Stone, L. The feasibility and stability of large complex biological
networks: a random matrix approach. \emph{Scientific Reports}
\textbf{8,} 8246 (2018).

\hypertarget{ref-Tang2014c}{}
26. Tang, S., Pawar, S. \& Allesina, S. Correlation between interaction
strengths drives stability in large ecological networks. \textbf{17,}
1094--1100 (2014).

\hypertarget{ref-Allesina2011}{}
27. Allesina, S. \& Levine, J. M. A competitive network theory of
species diversity. \emph{Proceedings of the National Academy of Sciences
of the United States of America} \textbf{108,} 5638--5642 (2011).

\hypertarget{ref-Dougoud2018}{}
28. Dougoud, M., Vinckenbosch, L., Rohr, R., Bersier, L.-F. \& Mazza, C.
The feasibility of equilibria in large ecosystems: a primary but
neglected concept in the complexity-stability debate. \emph{PLOS
Computational Biology} \textbf{14,} e1005988 (2018).

\hypertarget{ref-Song2018}{}
29. Song, C. \& Saavedra, S. Will a small randomly assembled community
be feasible and stable? \emph{Ecology} \textbf{99,} 743--751 (2018).

\hypertarget{ref-Servan2018}{}
30. Serván, C. A., Capitán, J. A., Grilli, J., Morrison, K. E. \&
Allesina, S. Coexistence of many species in random ecosystems.
\emph{Nature Ecology and Evolution} \textbf{2,} 1237--1242 (2018).

\hypertarget{ref-Hamblin2013}{}
31. Hamblin, S. On the practical usage of genetic algorithms in ecology
and evolution. \emph{Methods in Ecology and Evolution} \textbf{4,}
184--194 (2013).

\hypertarget{ref-May1973}{}
32. May, R. M. Qualitative stability in model ecosystems. \emph{Ecology}
\textbf{54,} 638--641 (1973).

\hypertarget{ref-Rproject}{}
33. R Core Team. \emph{R: A language and environment for statistical
computing}. (R Foundation for Statistical Computing, 2018).

\clearpage

\section{Supplemental Information}

\vspace{2mm}
\hrule
\vspace{2mm}

\textbf{This supplemental information supports the manuscript
``Component response rate variation underlies the stability of complex
systems'' with additional analyses to support its conclusions. All text,
code, and data underlying this manuscript are publicly available on
\href{https://github.com/bradduthie/RandomMatrixStability}{GitHub} as
part of the RandomMatrixStability R package.}

The
\href{https://github.com/bradduthie/RandomMatrixStability}{RandomMatrixStability
package} includes all functions and tools for recreating the text, this
supplemental information, and running all code; additional documentation
is also provided for package functions. The RandomMatrixStability
package is available on
\href{https://github.com/bradduthie/RandomMatrixStability}{GitHub}; to
download it, the
\href{https://cran.r-project.org/web/packages/devtools/index.html}{\texttt{devtools}
library} is needed.

\begin{Shaded}
\begin{Highlighting}[]
\KeywordTok{install.packages}\NormalTok{(}\StringTok{"devtools"}\NormalTok{);}
\KeywordTok{library}\NormalTok{(devtools);}
\end{Highlighting}
\end{Shaded}

The code below installs the RandomMatrixStability package using
devtools.

\begin{Shaded}
\begin{Highlighting}[]
\KeywordTok{install_github}\NormalTok{(}\StringTok{"bradduthie/RandomMatrixStability"}\NormalTok{);}
\end{Highlighting}
\end{Shaded}

\vspace{2mm}
\hrule
\vspace{2mm}

\section{Supplemental Information table of
contents}\label{supplemental-information-table-of-contents}

\begin{itemize}
\tightlist
\item
  \protect\hyperlink{IncrS}{Stability across increasing \(S\)}
\item
  \protect\hyperlink{ecological}{Stability of random ecological
  networks}

  \begin{itemize}
  \tightlist
  \item
    \protect\hyperlink{competition}{Competitor networks}
  \item
    \protect\hyperlink{mutualism}{Mutualist networks}
  \item
    \protect\hyperlink{pred-prey}{Predator-prey networks}
  \end{itemize}
\item
  \protect\hyperlink{connectance}{Sensitivity of connectance (C) values}

  \begin{itemize}
  \tightlist
  \item
    \protect\hyperlink{connect3}{C = 0.3}
  \item
    \protect\hyperlink{connect5}{C = 0.5}
  \item
    \protect\hyperlink{connect7}{C = 0.7}
  \item
    \protect\hyperlink{connect9}{C = 0.9}
  \end{itemize}
\item
  \protect\hyperlink{sigma}{Large networks of \(C = 0.05\) across \(S\)
  and \(\sigma\)}

  \begin{itemize}
  \tightlist
  \item
    \protect\hyperlink{sigma3}{\(\sigma\) = 0.3}
  \item
    \protect\hyperlink{sigma4}{\(\sigma\) = 0.4}
  \item
    \protect\hyperlink{sigma5}{\(\sigma\) = 0.5}
  \item
    \protect\hyperlink{sigma6}{\(\sigma\) = 0.6}
  \end{itemize}
\item
  \protect\hyperlink{gam_dist}{Sensitivity of distribution of
  \(\gamma\)}
\item
  \protect\hyperlink{structured}{Stability of structured networks}
\item
  \protect\hyperlink{Feasibility}{Feasibility of complex systems}
\item
  \protect\hyperlink{ga}{Stability given targeted manipulation of
  \(\gamma\) (genetic algorithm)}
\item
  \protect\hyperlink{Gibbs}{Consistency with Gibbs et al. (2018)}
\item
  \protect\hyperlink{repr}{Reproducing simulation results}
\item
  \protect\hyperlink{ref}{Literature cited}
\end{itemize}

\clearpage

\hypertarget{IncrS}{\section{\texorpdfstring{Stability across increasing
\(S\)}{Stability across increasing S}}\label{IncrS}}

Figure 4 of the main text reports the number of stable random complex
systems found over 1 million iterations. The table below shows the
results for all simulations of random \(\mathbf{M}\) matrices at
\(\sigma_{A} = 0.4\) and \(C = 1\) given a range of
\(S = \{2, 3, ..., 49, 50\}\). In this table, the \texttt{A} refers to
\(\mathbf{A}\) matrices where \(\gamma = 1\), while \texttt{M} refers to
\(\mathbf{M}\) matrices after \(\sigma^{2}_{\gamma}\) is added and
\(\gamma \sim \mathcal{U}(0, 2)\). Each row summarises data for a given
\(S\) over 1 million randomly simulated \(\mathbf{M}\). The column
\texttt{A\_unstable} shows the number of \(\mathbf{A}\) matrices that
are unstable, and the column \texttt{A\_stable} shows the number of
\(\mathbf{A}\) matrices that are stable (these two columns sum to 1
million). Similarly, the column \texttt{M\_unstable} shows the number of
\(\mathbf{M}\) matrices that are unstable and \texttt{M\_stable} shows
the number that are stable. The columns \texttt{A\_stabilised} and
\texttt{A\_destabilised} show how many \(\mathbf{M}\) matrices were
stabilised or destabilised, respectively, by \(\sigma^{2}_{\gamma}\).

\begin{longtable}[]{@{}rrrrrrr@{}}
\toprule
S & A\_unstable & A\_stable & M\_unstable & M\_stable & A\_stabilised &
A\_destabilised\tabularnewline
\midrule
\endhead
2 & 293 & 999707 & 293 & 999707 & 0 & 0\tabularnewline
3 & 3602 & 996398 & 3609 & 996391 & 0 & 7\tabularnewline
4 & 14937 & 985063 & 15008 & 984992 & 0 & 71\tabularnewline
5 & 39289 & 960711 & 39783 & 960217 & 36 & 530\tabularnewline
6 & 78845 & 921155 & 80207 & 919793 & 389 & 1751\tabularnewline
7 & 133764 & 866236 & 136904 & 863096 & 1679 & 4819\tabularnewline
8 & 204112 & 795888 & 208241 & 791759 & 5391 & 9520\tabularnewline
9 & 288041 & 711959 & 291775 & 708225 & 12619 & 16353\tabularnewline
10 & 384024 & 615976 & 384931 & 615069 & 23153 & 24060\tabularnewline
11 & 485975 & 514025 & 481019 & 518981 & 35681 & 30725\tabularnewline
12 & 590453 & 409547 & 577439 & 422561 & 48302 & 35288\tabularnewline
13 & 689643 & 310357 & 669440 & 330560 & 57194 & 36991\tabularnewline
14 & 777496 & 222504 & 751433 & 248567 & 60959 & 34896\tabularnewline
15 & 850159 & 149841 & 821613 & 178387 & 58567 & 30021\tabularnewline
16 & 905057 & 94943 & 877481 & 122519 & 51255 & 23679\tabularnewline
17 & 943192 & 56808 & 919536 & 80464 & 40854 & 17198\tabularnewline
18 & 969018 & 30982 & 949944 & 50056 & 30102 & 11028\tabularnewline
19 & 984301 & 15699 & 970703 & 29297 & 20065 & 6467\tabularnewline
20 & 992601 & 7399 & 983507 & 16493 & 12587 & 3493\tabularnewline
21 & 996765 & 3235 & 991532 & 8468 & 7030 & 1797\tabularnewline
22 & 998693 & 1307 & 995567 & 4433 & 3884 & 758\tabularnewline
23 & 999503 & 497 & 997941 & 2059 & 1883 & 321\tabularnewline
24 & 999861 & 139 & 999059 & 941 & 899 & 97\tabularnewline
25 & 999964 & 36 & 999617 & 383 & 380 & 33\tabularnewline
26 & 999993 & 7 & 999878 & 122 & 121 & 6\tabularnewline
27 & 999995 & 5 & 999946 & 54 & 53 & 4\tabularnewline
28 & 1000000 & 0 & 999975 & 25 & 25 & 0\tabularnewline
29 & 1000000 & 0 & 999997 & 3 & 3 & 0\tabularnewline
30 & 1000000 & 0 & 999999 & 1 & 1 & 0\tabularnewline
31 & 1000000 & 0 & 999999 & 1 & 1 & 0\tabularnewline
32 & 1000000 & 0 & 1000000 & 0 & 0 & 0\tabularnewline
33 & 1000000 & 0 & 1000000 & 0 & 0 & 0\tabularnewline
34 & 1000000 & 0 & 1000000 & 0 & 0 & 0\tabularnewline
35 & 1000000 & 0 & 1000000 & 0 & 0 & 0\tabularnewline
36 & 1000000 & 0 & 1000000 & 0 & 0 & 0\tabularnewline
37 & 1000000 & 0 & 1000000 & 0 & 0 & 0\tabularnewline
38 & 1000000 & 0 & 1000000 & 0 & 0 & 0\tabularnewline
39 & 1000000 & 0 & 1000000 & 0 & 0 & 0\tabularnewline
40 & 1000000 & 0 & 1000000 & 0 & 0 & 0\tabularnewline
41 & 1000000 & 0 & 1000000 & 0 & 0 & 0\tabularnewline
42 & 1000000 & 0 & 1000000 & 0 & 0 & 0\tabularnewline
43 & 1000000 & 0 & 1000000 & 0 & 0 & 0\tabularnewline
44 & 1000000 & 0 & 1000000 & 0 & 0 & 0\tabularnewline
45 & 1000000 & 0 & 1000000 & 0 & 0 & 0\tabularnewline
46 & 1000000 & 0 & 1000000 & 0 & 0 & 0\tabularnewline
47 & 1000000 & 0 & 1000000 & 0 & 0 & 0\tabularnewline
48 & 1000000 & 0 & 1000000 & 0 & 0 & 0\tabularnewline
49 & 1000000 & 0 & 1000000 & 0 & 0 & 0\tabularnewline
50 & 1000000 & 0 & 1000000 & 0 & 0 & 0\tabularnewline
\bottomrule
\end{longtable}

Overall, the ratio of stable \(\mathbf{A}\) matrices to stable
\(\mathbf{M}\) matrices found is greater than 1 whenever \(S > 10\)
(compare column 3 to column 5), and this ratio increases with increasing
\(S\) (column 1). Hence, more randomly created complex systems
(\(\mathbf{M}\)) are stable given variation in \(\gamma\) than when
\(\gamma = 1\). Note that feasibility results were omitted for the table
above, but are \protect\hyperlink{Feasibility}{reported below}.

\hypertarget{ecological}{\section{Stability of random ecological
networks}\label{ecological}}

While the foundational work of
May\textsuperscript{\protect\hyperlink{ref-May1972}{1}} applies broadly
to complex networks, much attention has been given specifically to
ecological networks of interacting species. In these networks, the
matrix \(\mathbf{A}\) is interpreted as a community matrix and each row
and column is interpreted as a single species. The per capita effect
that the density of any species \(i\) has on the population dynamics of
species \(j\) is found in \(A_{ij}\), meaning that \(\mathbf{A}\) holds
the effects of pair-wise interactions between \(S\)
species\textsuperscript{\protect\hyperlink{ref-Allesina2012}{2},\protect\hyperlink{ref-Allesina2015}{3}}.
While May's original
work\textsuperscript{\protect\hyperlink{ref-May1972}{1}} considered only
randomly assembled communities, recent work has specifically looked at
more restricted ecological communities including competitive networks
(all off-diagonal elements of \(\mathbf{A}\) are negative), mutualist
networks (all off-diagonal elements of \(\mathbf{A}\) are positive), and
predator-prey networks (for any pair of \(i\) and \(j\), the effect of
\(i\) on \(j\) is negative and \(j\) on \(i\) is positive, or vice
versa)\textsuperscript{\protect\hyperlink{ref-Allesina2012}{2},\protect\hyperlink{ref-Allesina2015}{3}}.
In general, competitor and mutualist networks tend to be unstable, while
predator-prey networks tend to be highly
stabilising\textsuperscript{\protect\hyperlink{ref-Allesina2012}{2}}.

I investigated competitor, mutualist, and predator-prey networks
following Allesina et
al.\textsuperscript{\protect\hyperlink{ref-Allesina2012}{2}}. To create
these networks, I first generated a random matrix \(\mathbf{A}\), then
changed the elements of \(\mathbf{A}\) accordingly. If \(\mathbf{A}\)
was a competitive network, then the sign of any positive off-diagonal
elements was reversed to be negative. If \(\mathbf{A}\) was a mutualist
network, then the sign of any positive off-diagonal elements was
reversed to be positive. And if \(\mathbf{A}\) was a predator-prey
network, then all \(i\) and \(j\) pairs of elements were checked; any
pairs of the same sign were changed so that one was negative and the
other was positive.

The number of stable \(\mathbf{M = \gamma A}\) systems was calculated
\protect\hyperlink{IncrS}{exactly as it was} for random matrices for
values of \(S\) from 2 to 50 (100 in the case of the relatively more
stable predator-prey interactions), except that only 100000 random
\(\mathbf{M}\) were generated instead of 1 million.

The following tables for restricted ecological communities can therefore
be compared with the random \(\mathbf{M}\)
\protect\hyperlink{IncrS}{results above} (but note that counts from
systems with comparable probabilities of stability will be an order of
magnitude lower in the tables below due to the smaller number of
\(\mathbf{M}\) matrices generated). As with the
\protect\hyperlink{IncrS}{results above}, in the tables below,
\texttt{A} refers to matrices \(\mathbf{A}\) when \(\gamma = 1\) and
\texttt{M} refers to matrices after \(\sigma^{2}_{\gamma}\) is added.
The column \texttt{A\_unstable} shows the number of \(\mathbf{A}\)
matrices that are unstable, and the column \texttt{A\_stable} shows the
number of \(\mathbf{A}\) matrices that are stable (these two columns sum
to 100000). Similarly, the column \texttt{M\_unstable} shows the number
of \(\mathbf{M}\) matrices that are unstable and \texttt{M\_stable}
shows the number that are stable. The columns \texttt{A\_stabilised} and
\texttt{A\_destabilised} show how many \(\mathbf{A}\) matrices were
stabilised or destabilised, respectively, by \(\sigma^{2}_{\gamma}\).

\textbf{Competition}

Results for competitor interaction networks are shown below

\begin{longtable}[]{@{}lllllll@{}}
\toprule
S & A\_unstable & A\_stable & M\_unstable & M\_stable & A\_stabilised &
A\_destabilised\tabularnewline
\midrule
\endhead
2 & 48 & 99952 & 48 & 99952 & 0 & 0\tabularnewline
3 & 229 & 99771 & 231 & 99769 & 0 & 2\tabularnewline
4 & 701 & 99299 & 704 & 99296 & 0 & 3\tabularnewline
5 & 1579 & 98421 & 1587 & 98413 & 0 & 8\tabularnewline
6 & 3218 & 96782 & 3253 & 96747 & 6 & 41\tabularnewline
7 & 5519 & 94481 & 5619 & 94381 & 23 & 123\tabularnewline
8 & 9062 & 90938 & 9237 & 90763 & 77 & 252\tabularnewline
9 & 13436 & 86564 & 13729 & 86271 & 230 & 523\tabularnewline
10 & 18911 & 81089 & 19303 & 80697 & 505 & 897\tabularnewline
11 & 25594 & 74406 & 25961 & 74039 & 1011 & 1378\tabularnewline
12 & 33207 & 66793 & 33382 & 66618 & 1724 & 1899\tabularnewline
13 & 41160 & 58840 & 41089 & 58911 & 2655 & 2584\tabularnewline
14 & 50575 & 49425 & 49894 & 50106 & 3777 & 3096\tabularnewline
15 & 59250 & 40750 & 57892 & 42108 & 4824 & 3466\tabularnewline
16 & 67811 & 32189 & 65740 & 34260 & 5634 & 3563\tabularnewline
17 & 75483 & 24517 & 73056 & 26944 & 5943 & 3516\tabularnewline
18 & 82551 & 17449 & 79878 & 20122 & 5780 & 3107\tabularnewline
19 & 88030 & 11970 & 85204 & 14796 & 5417 & 2591\tabularnewline
20 & 92254 & 7746 & 89766 & 10234 & 4544 & 2056\tabularnewline
21 & 95233 & 4767 & 93002 & 6998 & 3695 & 1464\tabularnewline
22 & 97317 & 2683 & 95451 & 4549 & 2803 & 937\tabularnewline
23 & 98508 & 1492 & 97122 & 2878 & 1991 & 605\tabularnewline
24 & 99240 & 760 & 98407 & 1593 & 1216 & 383\tabularnewline
25 & 99669 & 331 & 99082 & 918 & 739 & 152\tabularnewline
26 & 99871 & 129 & 99490 & 510 & 452 & 71\tabularnewline
27 & 99938 & 62 & 99732 & 268 & 240 & 34\tabularnewline
28 & 99985 & 15 & 99888 & 112 & 108 & 11\tabularnewline
29 & 99990 & 10 & 99951 & 49 & 46 & 7\tabularnewline
30 & 100000 & 0 & 99981 & 19 & 19 & 0\tabularnewline
31 & 100000 & 0 & 99993 & 7 & 7 & 0\tabularnewline
32 & 100000 & 0 & 99996 & 4 & 4 & 0\tabularnewline
33 & 100000 & 0 & 99998 & 2 & 2 & 0\tabularnewline
34 & 100000 & 0 & 100000 & 0 & 0 & 0\tabularnewline
\ldots{} & \ldots{} & \ldots{} & \ldots{} & \ldots{} & \ldots{} &
\ldots{}\tabularnewline
50 & 100000 & 0 & 100000 & 0 & 0 & 0\tabularnewline
\bottomrule
\end{longtable}

\textbf{Mutualism}

Results for mutualist interaction networks are shown below

\begin{longtable}[]{@{}lllllll@{}}
\toprule
S & A\_unstable & A\_stable & M\_unstable & M\_stable & A\_stabilised &
A\_destabilised\tabularnewline
\midrule
\endhead
2 & 56 & 99944 & 56 & 99944 & 0 & 0\tabularnewline
3 & 3301 & 96699 & 3301 & 96699 & 0 & 0\tabularnewline
4 & 34446 & 65554 & 34446 & 65554 & 0 & 0\tabularnewline
5 & 86520 & 13480 & 86520 & 13480 & 0 & 0\tabularnewline
6 & 99683 & 317 & 99683 & 317 & 0 & 0\tabularnewline
7 & 99998 & 2 & 99998 & 2 & 0 & 0\tabularnewline
8 & 100000 & 0 & 100000 & 0 & 0 & 0\tabularnewline
9 & 100000 & 0 & 100000 & 0 & 0 & 0\tabularnewline
10 & 100000 & 0 & 100000 & 0 & 0 & 0\tabularnewline
11 & 100000 & 0 & 100000 & 0 & 0 & 0\tabularnewline
12 & 100000 & 0 & 100000 & 0 & 0 & 0\tabularnewline
\ldots{} & \ldots{} & \ldots{} & \ldots{} & \ldots{} & \ldots{} &
\ldots{}\tabularnewline
50 & 100000 & 0 & 100000 & 0 & 0 & 0\tabularnewline
\bottomrule
\end{longtable}

\textbf{Predator-prey}

Results for predator-prey interaction networks are shown below

\begin{longtable}[]{@{}rrrrrrr@{}}
\toprule
S & A\_unstable & A\_stable & M\_unstable & M\_stable & A\_stabilised &
A\_destabilised\tabularnewline
\midrule
\endhead
2 & 0 & 100000 & 0 & 100000 & 0 & 0\tabularnewline
3 & 0 & 100000 & 0 & 100000 & 0 & 0\tabularnewline
4 & 0 & 100000 & 0 & 100000 & 0 & 0\tabularnewline
5 & 1 & 99999 & 1 & 99999 & 0 & 0\tabularnewline
6 & 4 & 99996 & 4 & 99996 & 0 & 0\tabularnewline
7 & 2 & 99998 & 2 & 99998 & 0 & 0\tabularnewline
8 & 5 & 99995 & 5 & 99995 & 0 & 0\tabularnewline
9 & 20 & 99980 & 21 & 99979 & 0 & 1\tabularnewline
10 & 20 & 99980 & 22 & 99978 & 0 & 2\tabularnewline
11 & 38 & 99962 & 39 & 99961 & 0 & 1\tabularnewline
12 & 64 & 99936 & 66 & 99934 & 0 & 2\tabularnewline
13 & 87 & 99913 & 91 & 99909 & 0 & 4\tabularnewline
14 & 157 & 99843 & 159 & 99841 & 0 & 2\tabularnewline
15 & 215 & 99785 & 227 & 99773 & 0 & 12\tabularnewline
16 & 293 & 99707 & 310 & 99690 & 0 & 17\tabularnewline
17 & 383 & 99617 & 408 & 99592 & 0 & 25\tabularnewline
18 & 443 & 99557 & 473 & 99527 & 3 & 33\tabularnewline
19 & 642 & 99358 & 675 & 99325 & 4 & 37\tabularnewline
20 & 836 & 99164 & 887 & 99113 & 7 & 58\tabularnewline
21 & 1006 & 98994 & 1058 & 98942 & 10 & 62\tabularnewline
22 & 1153 & 98847 & 1228 & 98772 & 20 & 95\tabularnewline
23 & 1501 & 98499 & 1593 & 98407 & 30 & 122\tabularnewline
24 & 1841 & 98159 & 1996 & 98004 & 40 & 195\tabularnewline
25 & 2146 & 97854 & 2316 & 97684 & 58 & 228\tabularnewline
26 & 2643 & 97357 & 2809 & 97191 & 119 & 285\tabularnewline
27 & 3034 & 96966 & 3258 & 96742 & 158 & 382\tabularnewline
28 & 3690 & 96310 & 3928 & 96072 & 201 & 439\tabularnewline
29 & 4257 & 95743 & 4532 & 95468 & 290 & 565\tabularnewline
30 & 4964 & 95036 & 5221 & 94779 & 424 & 681\tabularnewline
31 & 5627 & 94373 & 5978 & 94022 & 452 & 803\tabularnewline
32 & 6543 & 93457 & 6891 & 93109 & 666 & 1014\tabularnewline
33 & 7425 & 92575 & 7777 & 92223 & 818 & 1170\tabularnewline
34 & 8540 & 91460 & 8841 & 91159 & 1071 & 1372\tabularnewline
35 & 9526 & 90474 & 9842 & 90158 & 1337 & 1653\tabularnewline
36 & 10617 & 89383 & 10891 & 89109 & 1624 & 1898\tabularnewline
37 & 12344 & 87656 & 12508 & 87492 & 2021 & 2185\tabularnewline
38 & 13675 & 86325 & 13877 & 86123 & 2442 & 2644\tabularnewline
39 & 15264 & 84736 & 15349 & 84651 & 2870 & 2955\tabularnewline
40 & 17026 & 82974 & 17053 & 82947 & 3363 & 3390\tabularnewline
41 & 18768 & 81232 & 18614 & 81386 & 3905 & 3751\tabularnewline
42 & 20791 & 79209 & 20470 & 79530 & 4579 & 4258\tabularnewline
43 & 23150 & 76850 & 22754 & 77246 & 5217 & 4821\tabularnewline
44 & 25449 & 74551 & 24184 & 75816 & 6285 & 5020\tabularnewline
45 & 27702 & 72298 & 26464 & 73536 & 6754 & 5516\tabularnewline
46 & 30525 & 69475 & 28966 & 71034 & 7646 & 6087\tabularnewline
47 & 32832 & 67168 & 31125 & 68875 & 8487 & 6780\tabularnewline
48 & 36152 & 63848 & 33865 & 66135 & 9479 & 7192\tabularnewline
49 & 38714 & 61286 & 36242 & 63758 & 10125 & 7653\tabularnewline
50 & 41628 & 58372 & 38508 & 61492 & 11036 & 7916\tabularnewline
51 & 44483 & 55517 & 41023 & 58977 & 11704 & 8244\tabularnewline
52 & 48134 & 51866 & 44287 & 55713 & 12573 & 8726\tabularnewline
53 & 51138 & 48862 & 46721 & 53279 & 13223 & 8806\tabularnewline
54 & 54261 & 45739 & 49559 & 50441 & 13757 & 9055\tabularnewline
55 & 57647 & 42353 & 52403 & 47597 & 14324 & 9080\tabularnewline
56 & 60630 & 39370 & 55293 & 44707 & 14669 & 9332\tabularnewline
57 & 63647 & 36353 & 57787 & 42213 & 15103 & 9243\tabularnewline
58 & 66961 & 33039 & 60439 & 39561 & 15450 & 8928\tabularnewline
59 & 69968 & 30032 & 63708 & 36292 & 15246 & 8986\tabularnewline
60 & 72838 & 27162 & 66270 & 33730 & 15177 & 8609\tabularnewline
61 & 75609 & 24391 & 68873 & 31127 & 15006 & 8270\tabularnewline
62 & 77999 & 22001 & 71318 & 28682 & 14538 & 7857\tabularnewline
63 & 80616 & 19384 & 73517 & 26483 & 14510 & 7411\tabularnewline
64 & 83089 & 16911 & 76209 & 23791 & 13784 & 6904\tabularnewline
65 & 85150 & 14850 & 78086 & 21914 & 13412 & 6348\tabularnewline
66 & 86908 & 13092 & 80437 & 19563 & 12477 & 6006\tabularnewline
67 & 88671 & 11329 & 82379 & 17621 & 11718 & 5426\tabularnewline
68 & 90537 & 9463 & 84483 & 15517 & 10878 & 4824\tabularnewline
69 & 91969 & 8031 & 86233 & 13767 & 10033 & 4297\tabularnewline
70 & 93181 & 6819 & 87914 & 12086 & 9070 & 3803\tabularnewline
71 & 94330 & 5670 & 89200 & 10800 & 8401 & 3271\tabularnewline
72 & 95324 & 4676 & 90833 & 9167 & 7359 & 2868\tabularnewline
73 & 96143 & 3857 & 91805 & 8195 & 6726 & 2388\tabularnewline
74 & 96959 & 3041 & 93065 & 6935 & 5900 & 2006\tabularnewline
75 & 97543 & 2457 & 93987 & 6013 & 5222 & 1666\tabularnewline
76 & 97969 & 2031 & 94900 & 5100 & 4481 & 1412\tabularnewline
77 & 98497 & 1503 & 95756 & 4244 & 3809 & 1068\tabularnewline
78 & 98744 & 1256 & 96442 & 3558 & 3269 & 967\tabularnewline
79 & 99045 & 955 & 96942 & 3058 & 2837 & 734\tabularnewline
80 & 99276 & 724 & 97528 & 2472 & 2329 & 581\tabularnewline
81 & 99481 & 519 & 97996 & 2004 & 1894 & 409\tabularnewline
82 & 99556 & 444 & 98321 & 1679 & 1597 & 362\tabularnewline
83 & 99691 & 309 & 98722 & 1278 & 1227 & 258\tabularnewline
84 & 99752 & 248 & 98943 & 1057 & 1015 & 206\tabularnewline
85 & 99833 & 167 & 99144 & 856 & 837 & 148\tabularnewline
86 & 99895 & 105 & 99346 & 654 & 642 & 93\tabularnewline
87 & 99925 & 75 & 99461 & 539 & 530 & 66\tabularnewline
88 & 99945 & 55 & 99566 & 434 & 428 & 49\tabularnewline
89 & 99976 & 24 & 99675 & 325 & 324 & 23\tabularnewline
90 & 99977 & 23 & 99756 & 244 & 243 & 22\tabularnewline
91 & 99982 & 18 & 99839 & 161 & 155 & 12\tabularnewline
92 & 99988 & 12 & 99865 & 135 & 135 & 12\tabularnewline
93 & 99994 & 6 & 99885 & 115 & 115 & 6\tabularnewline
94 & 99993 & 7 & 99911 & 89 & 88 & 6\tabularnewline
95 & 99998 & 2 & 99953 & 47 & 47 & 2\tabularnewline
96 & 99999 & 1 & 99965 & 35 & 35 & 1\tabularnewline
97 & 99999 & 1 & 99979 & 21 & 21 & 1\tabularnewline
98 & 100000 & 0 & 99973 & 27 & 27 & 0\tabularnewline
99 & 100000 & 0 & 99984 & 16 & 16 & 0\tabularnewline
100 & 100000 & 0 & 99989 & 11 & 11 & 0\tabularnewline
\bottomrule
\end{longtable}

Overall, as
expected\textsuperscript{\protect\hyperlink{ref-Allesina2012}{2}},
predator-prey communities are relatively stable while mutualist
communties are highly unstable. But interestingly, while
\(\sigma^{2}_{\gamma}\) stabilises predator-prey and competitor
communities, it does not stabilise mutualist communities. This is
unsurprising because purely mutualist communities are characterised by a
very positive\textsuperscript{\protect\hyperlink{ref-Allesina2012}{2}}
leading \(\Re(\lambda)\), and it is highly unlikely that
\(\sigma^{2}_{\gamma}\) alone will shift all real parts of eigenvalues
to negative values.

\hypertarget{connectance}{\section{Sensitivity of connectance (C)
values}\label{connectance}}

In the main text, for simplicity, I assumed connectance values of
\(C = 1\), meaning that all off-diagonal elements of a matrix
\(\mathbf{M}\) were potentially nonzero and sampled from a normal
distribution \(\mathcal{N}(0, \sigma^{2}_{A})\) where
\(\sigma_{A} = 0.4\). Here I present four tables showing the number of
stable communities given \(C = \{0.3, 0. 5, 0.7, 0.9 \}\). In all cases,
uniform variation in component response rate
(\(\gamma \sim \mathcal{U}(0, 2)\)) led to a higher number of stable
communities than when \(\gamma\) did not vary (\(\gamma = 1\)). In
contrast to the main text, 100000 rather than 1 million \(\mathbf{M}\)
were simulated. As with the results on
\protect\hyperlink{IncrS}{stability with increasing \(S\)} shown above,
in the tables below \texttt{A} refers to \(\mathbf{A}\) matrices when
\(\gamma = 1\), and \texttt{M} refers to \(\mathbf{M}\) matrices after
\(\sigma^{2}_{\gamma}\) is added. The column \texttt{A\_unstable} shows
the number of \(\mathbf{A}\) matrices that are unstable, and the column
\texttt{A\_stable} shows the number of \(\mathbf{A}\) matrices that are
stable (these two columns sum to 100000). Similarly, the column
\texttt{M\_unstable} shows the number of \(\mathbf{M}\) matrices that
are unstable and \texttt{M\_stable} shows the number that are stable.
The columns \texttt{A\_stabilised} and \texttt{A\_destabilised} show how
many \(\mathbf{A}\) matrices were stabilised or destabilised,
respectively, by \(\sigma^{2}_{\gamma}\).

\textbf{Connectance \(\mathbf{C = 0.3}\)}

\begin{longtable}[]{@{}lllllll@{}}
\toprule
S & A\_unstable & A\_stable & M\_unstable & M\_stable & A\_stabilised &
A\_destabilised\tabularnewline
\midrule
\endhead
2 & 5 & 99995 & 5 & 99995 & 0 & 0\tabularnewline
3 & 6 & 99994 & 6 & 99994 & 0 & 0\tabularnewline
4 & 24 & 99976 & 24 & 99976 & 0 & 0\tabularnewline
5 & 59 & 99941 & 59 & 99941 & 0 & 0\tabularnewline
6 & 98 & 99902 & 98 & 99902 & 0 & 0\tabularnewline
7 & 160 & 99840 & 161 & 99839 & 0 & 1\tabularnewline
8 & 290 & 99710 & 293 & 99707 & 0 & 3\tabularnewline
9 & 430 & 99570 & 434 & 99566 & 0 & 4\tabularnewline
10 & 648 & 99352 & 653 & 99347 & 1 & 6\tabularnewline
11 & 946 & 99054 & 957 & 99043 & 0 & 11\tabularnewline
12 & 1392 & 98608 & 1415 & 98585 & 4 & 27\tabularnewline
13 & 2032 & 97968 & 2065 & 97935 & 5 & 38\tabularnewline
14 & 2627 & 97373 & 2688 & 97312 & 10 & 71\tabularnewline
15 & 3588 & 96412 & 3647 & 96353 & 35 & 94\tabularnewline
16 & 5019 & 94981 & 5124 & 94876 & 51 & 156\tabularnewline
17 & 6512 & 93488 & 6673 & 93327 & 79 & 240\tabularnewline
18 & 8444 & 91556 & 8600 & 91400 & 165 & 321\tabularnewline
19 & 10416 & 89584 & 10667 & 89333 & 244 & 495\tabularnewline
20 & 13254 & 86746 & 13477 & 86523 & 425 & 648\tabularnewline
21 & 16248 & 83752 & 16481 & 83519 & 642 & 875\tabularnewline
22 & 19497 & 80503 & 19719 & 80281 & 929 & 1151\tabularnewline
23 & 23654 & 76346 & 23776 & 76224 & 1368 & 1490\tabularnewline
24 & 28485 & 71515 & 28389 & 71611 & 1914 & 1818\tabularnewline
25 & 32774 & 67226 & 32483 & 67517 & 2428 & 2137\tabularnewline
26 & 38126 & 61874 & 37411 & 62589 & 3221 & 2506\tabularnewline
27 & 43435 & 56565 & 42418 & 57582 & 3828 & 2811\tabularnewline
28 & 49333 & 50667 & 47840 & 52160 & 4565 & 3072\tabularnewline
29 & 55389 & 44611 & 53381 & 46619 & 5329 & 3321\tabularnewline
30 & 60826 & 39174 & 58388 & 41612 & 5918 & 3480\tabularnewline
31 & 66820 & 33180 & 64043 & 35957 & 6345 & 3568\tabularnewline
32 & 72190 & 27810 & 69036 & 30964 & 6685 & 3531\tabularnewline
33 & 77053 & 22947 & 73587 & 26413 & 6826 & 3360\tabularnewline
34 & 81816 & 18184 & 78157 & 21843 & 6673 & 3014\tabularnewline
35 & 85651 & 14349 & 82041 & 17959 & 6383 & 2773\tabularnewline
36 & 88985 & 11015 & 85657 & 14343 & 5721 & 2393\tabularnewline
37 & 92072 & 7928 & 88805 & 11195 & 5180 & 1913\tabularnewline
38 & 94329 & 5671 & 91444 & 8556 & 4451 & 1566\tabularnewline
39 & 95912 & 4088 & 93295 & 6705 & 3804 & 1187\tabularnewline
40 & 97232 & 2768 & 95201 & 4799 & 2967 & 936\tabularnewline
41 & 98179 & 1821 & 96506 & 3494 & 2356 & 683\tabularnewline
42 & 98826 & 1174 & 97489 & 2511 & 1786 & 449\tabularnewline
43 & 99275 & 725 & 98312 & 1688 & 1251 & 288\tabularnewline
44 & 99583 & 417 & 98872 & 1128 & 903 & 192\tabularnewline
45 & 99776 & 224 & 99339 & 661 & 576 & 139\tabularnewline
46 & 99865 & 135 & 99518 & 482 & 413 & 66\tabularnewline
47 & 99938 & 62 & 99744 & 256 & 226 & 32\tabularnewline
48 & 99956 & 44 & 99824 & 176 & 151 & 19\tabularnewline
49 & 99980 & 20 & 99914 & 86 & 85 & 19\tabularnewline
50 & 99993 & 7 & 99950 & 50 & 46 & 3\tabularnewline
51 & 99998 & 2 & 99971 & 29 & 28 & 1\tabularnewline
52 & 99998 & 2 & 99986 & 14 & 14 & 2\tabularnewline
53 & 99999 & 1 & 99992 & 8 & 7 & 0\tabularnewline
54 & 100000 & 0 & 99997 & 3 & 3 & 0\tabularnewline
55 & 100000 & 0 & 99999 & 1 & 1 & 0\tabularnewline
56 & 100000 & 0 & 99998 & 2 & 2 & 0\tabularnewline
57 & 100000 & 0 & 99999 & 1 & 1 & 0\tabularnewline
58 & 100000 & 0 & 100000 & 0 & 0 & 0\tabularnewline
\ldots{} & \ldots{} & \ldots{} & \ldots{} & \ldots{} & \ldots{} &
\ldots{}\tabularnewline
100 & 100000 & 0 & 100000 & 0 & 0 & 0\tabularnewline
\bottomrule
\end{longtable}

\textbf{Connectance \(\mathbf{C = 0.5}\)}

\begin{longtable}[]{@{}lllllll@{}}
\toprule
S & A\_unstable & A\_stable & M\_unstable & M\_stable & A\_stabilised &
A\_destabilised\tabularnewline
\midrule
\endhead
2 & 7 & 99993 & 7 & 99993 & 0 & 0\tabularnewline
3 & 32 & 99968 & 32 & 99968 & 0 & 0\tabularnewline
4 & 122 & 99878 & 122 & 99878 & 0 & 0\tabularnewline
5 & 320 & 99680 & 321 & 99679 & 0 & 1\tabularnewline
6 & 667 & 99333 & 673 & 99327 & 0 & 6\tabularnewline
7 & 1233 & 98767 & 1252 & 98748 & 0 & 19\tabularnewline
8 & 2123 & 97877 & 2156 & 97844 & 3 & 36\tabularnewline
9 & 3415 & 96585 & 3471 & 96529 & 16 & 72\tabularnewline
10 & 5349 & 94651 & 5450 & 94550 & 30 & 131\tabularnewline
11 & 7990 & 92010 & 8185 & 91815 & 81 & 276\tabularnewline
12 & 11073 & 88927 & 11301 & 88699 & 219 & 447\tabularnewline
13 & 14971 & 85029 & 15204 & 84796 & 445 & 678\tabularnewline
14 & 19754 & 80246 & 19992 & 80008 & 764 & 1002\tabularnewline
15 & 25020 & 74980 & 25239 & 74761 & 1185 & 1404\tabularnewline
16 & 30860 & 69140 & 30938 & 69062 & 1902 & 1980\tabularnewline
17 & 37844 & 62156 & 37562 & 62438 & 2758 & 2476\tabularnewline
18 & 44909 & 55091 & 44251 & 55749 & 3595 & 2937\tabularnewline
19 & 52322 & 47678 & 51011 & 48989 & 4573 & 3262\tabularnewline
20 & 60150 & 39850 & 58295 & 41705 & 5382 & 3527\tabularnewline
21 & 67147 & 32853 & 64895 & 35105 & 5925 & 3673\tabularnewline
22 & 74177 & 25823 & 71358 & 28642 & 6310 & 3491\tabularnewline
23 & 80297 & 19703 & 77034 & 22966 & 6507 & 3244\tabularnewline
24 & 85372 & 14628 & 82039 & 17961 & 6209 & 2876\tabularnewline
25 & 89719 & 10281 & 86539 & 13461 & 5562 & 2382\tabularnewline
26 & 92947 & 7053 & 90141 & 9859 & 4707 & 1901\tabularnewline
27 & 95436 & 4564 & 92950 & 7050 & 3844 & 1358\tabularnewline
28 & 97196 & 2804 & 95171 & 4829 & 2999 & 974\tabularnewline
29 & 98300 & 1700 & 96842 & 3158 & 2115 & 657\tabularnewline
30 & 99103 & 897 & 98033 & 1967 & 1466 & 396\tabularnewline
31 & 99502 & 498 & 98665 & 1335 & 1068 & 231\tabularnewline
32 & 99745 & 255 & 99185 & 815 & 696 & 136\tabularnewline
33 & 99881 & 119 & 99572 & 428 & 375 & 66\tabularnewline
34 & 99955 & 45 & 99788 & 212 & 191 & 24\tabularnewline
35 & 99979 & 21 & 99900 & 100 & 95 & 16\tabularnewline
36 & 99995 & 5 & 99950 & 50 & 50 & 5\tabularnewline
37 & 99997 & 3 & 99970 & 30 & 28 & 1\tabularnewline
38 & 99998 & 2 & 99986 & 14 & 13 & 1\tabularnewline
39 & 99999 & 1 & 99991 & 9 & 9 & 1\tabularnewline
40 & 100000 & 0 & 100000 & 0 & 0 & 0\tabularnewline
41 & 100000 & 0 & 99999 & 1 & 1 & 0\tabularnewline
42 & 100000 & 0 & 99999 & 1 & 1 & 0\tabularnewline
43 & 100000 & 0 & 100000 & 0 & 0 & 0\tabularnewline
\ldots{} & \ldots{} & \ldots{} & \ldots{} & \ldots{} & \ldots{} &
\ldots{}\tabularnewline
50 & 100000 & 0 & 100000 & 0 & 0 & 0\tabularnewline
\bottomrule
\end{longtable}

\textbf{Connectance \(\mathbf{C = 0.7}\)}

\begin{longtable}[]{@{}lllllll@{}}
\toprule
S & A\_unstable & A\_stable & M\_unstable & M\_stable & A\_stabilised &
A\_destabilised\tabularnewline
\midrule
\endhead
2 & 7 & 99993 & 7 & 99993 & 0 & 0\tabularnewline
3 & 106 & 99894 & 106 & 99894 & 0 & 0\tabularnewline
4 & 395 & 99605 & 397 & 99603 & 0 & 2\tabularnewline
5 & 1117 & 98883 & 1123 & 98877 & 0 & 6\tabularnewline
6 & 2346 & 97654 & 2367 & 97633 & 6 & 27\tabularnewline
7 & 4314 & 95686 & 4388 & 95612 & 16 & 90\tabularnewline
8 & 7327 & 92673 & 7456 & 92544 & 61 & 190\tabularnewline
9 & 11514 & 88486 & 11792 & 88208 & 150 & 428\tabularnewline
10 & 16247 & 83753 & 16584 & 83416 & 415 & 752\tabularnewline
11 & 22481 & 77519 & 22759 & 77241 & 884 & 1162\tabularnewline
12 & 29459 & 70541 & 29729 & 70271 & 1548 & 1818\tabularnewline
13 & 37631 & 62369 & 37567 & 62433 & 2419 & 2355\tabularnewline
14 & 46317 & 53683 & 45696 & 54304 & 3548 & 2927\tabularnewline
15 & 54945 & 45055 & 53695 & 46305 & 4671 & 3421\tabularnewline
16 & 63683 & 36317 & 61643 & 38357 & 5567 & 3527\tabularnewline
17 & 72004 & 27996 & 69375 & 30625 & 6124 & 3495\tabularnewline
18 & 79220 & 20780 & 76158 & 23842 & 6413 & 3351\tabularnewline
19 & 85286 & 14714 & 82283 & 17717 & 5982 & 2979\tabularnewline
20 & 90240 & 9760 & 87181 & 12819 & 5398 & 2339\tabularnewline
21 & 93676 & 6324 & 91077 & 8923 & 4468 & 1869\tabularnewline
22 & 96203 & 3797 & 94045 & 5955 & 3425 & 1267\tabularnewline
23 & 97866 & 2134 & 96161 & 3839 & 2496 & 791\tabularnewline
24 & 98842 & 1158 & 97633 & 2367 & 1713 & 504\tabularnewline
25 & 99433 & 567 & 98630 & 1370 & 1079 & 276\tabularnewline
26 & 99760 & 240 & 99259 & 741 & 655 & 154\tabularnewline
27 & 99895 & 105 & 99576 & 424 & 377 & 58\tabularnewline
28 & 99950 & 50 & 99790 & 210 & 194 & 34\tabularnewline
29 & 99981 & 19 & 99915 & 85 & 80 & 14\tabularnewline
30 & 99994 & 6 & 99952 & 48 & 47 & 5\tabularnewline
31 & 99998 & 2 & 99972 & 28 & 28 & 2\tabularnewline
32 & 99999 & 1 & 99992 & 8 & 8 & 1\tabularnewline
33 & 100000 & 0 & 99997 & 3 & 3 & 0\tabularnewline
34 & 100000 & 0 & 99999 & 1 & 1 & 0\tabularnewline
35 & 100000 & 0 & 100000 & 0 & 0 & 0\tabularnewline
\ldots{} & \ldots{} & \ldots{} & \ldots{} & \ldots{} & \ldots{} &
\ldots{}\tabularnewline
50 & 100000 & 0 & 100000 & 0 & 0 & 0\tabularnewline
\bottomrule
\end{longtable}

\textbf{Connectance \(\mathbf{C = 0.9}\)}

\begin{longtable}[]{@{}lllllll@{}}
\toprule
S & A\_unstable & A\_stable & M\_unstable & M\_stable & A\_stabilised &
A\_destabilised\tabularnewline
\midrule
\endhead
2 & 14 & 99986 & 14 & 99986 & 0 & 0\tabularnewline
3 & 240 & 99760 & 240 & 99760 & 0 & 0\tabularnewline
4 & 1008 & 98992 & 1016 & 98984 & 0 & 8\tabularnewline
5 & 2708 & 97292 & 2729 & 97271 & 2 & 23\tabularnewline
6 & 5669 & 94331 & 5755 & 94245 & 13 & 99\tabularnewline
7 & 9848 & 90152 & 10057 & 89943 & 91 & 300\tabularnewline
8 & 15903 & 84097 & 16201 & 83799 & 336 & 634\tabularnewline
9 & 22707 & 77293 & 23110 & 76890 & 765 & 1168\tabularnewline
10 & 30796 & 69204 & 31122 & 68878 & 1526 & 1852\tabularnewline
11 & 40224 & 59776 & 40082 & 59918 & 2649 & 2507\tabularnewline
12 & 49934 & 50066 & 49288 & 50712 & 3773 & 3127\tabularnewline
13 & 60138 & 39862 & 58803 & 41197 & 4984 & 3649\tabularnewline
14 & 69100 & 30900 & 67110 & 32890 & 5755 & 3765\tabularnewline
15 & 77607 & 22393 & 74884 & 25116 & 6273 & 3550\tabularnewline
16 & 84663 & 15337 & 81780 & 18220 & 5975 & 3092\tabularnewline
17 & 90075 & 9925 & 87290 & 12710 & 5209 & 2424\tabularnewline
18 & 93944 & 6056 & 91419 & 8581 & 4271 & 1746\tabularnewline
19 & 96650 & 3350 & 94530 & 5470 & 3287 & 1167\tabularnewline
20 & 98160 & 1840 & 96698 & 3302 & 2191 & 729\tabularnewline
21 & 99111 & 889 & 98133 & 1867 & 1389 & 411\tabularnewline
22 & 99588 & 412 & 98905 & 1095 & 903 & 220\tabularnewline
23 & 99837 & 163 & 99480 & 520 & 452 & 95\tabularnewline
24 & 99932 & 68 & 99744 & 256 & 228 & 40\tabularnewline
25 & 99976 & 24 & 99863 & 137 & 133 & 20\tabularnewline
26 & 99995 & 5 & 99950 & 50 & 49 & 4\tabularnewline
27 & 99996 & 4 & 99986 & 14 & 13 & 3\tabularnewline
28 & 100000 & 0 & 99993 & 7 & 7 & 0\tabularnewline
29 & 100000 & 0 & 99996 & 4 & 4 & 0\tabularnewline
30 & 100000 & 0 & 99998 & 2 & 2 & 0\tabularnewline
31 & 100000 & 0 & 100000 & 0 & 0 & 0\tabularnewline
\ldots{} & \ldots{} & \ldots{} & \ldots{} & \ldots{} & \ldots{} &
\ldots{}\tabularnewline
50 & 100000 & 0 & 100000 & 0 & 0 & 0\tabularnewline
\bottomrule
\end{longtable}

\hypertarget{sigma}{\section{\texorpdfstring{Sensitivity of interaction
strength (\(\sigma_{A}\))
values}{Sensitivity of interaction strength (\textbackslash{}sigma\_\{A\}) values}}\label{sigma}}

Results below show stability results given varying interaction strengths
(\(\sigma_{A}\)) for \(C = 0.05\) (note that system size \(S\) values
are larger and increase by 10 with increasing rows). In the tables below
(as \protect\hyperlink{IncrS}{above}), \texttt{A} and \texttt{M} refers
to matrices for \(\gamma = 1\) and \(\sigma^{2}_{\gamma}\),
respectively.

\textbf{Interaction strength \(\mathbf{\sigma_{A} = 0.3}\)}

\begin{longtable}[]{@{}rrrrrrr@{}}
\toprule
S & A\_unstable & A\_stable & M\_unstable & M\_stable & A\_stabilised &
A\_destabilised\tabularnewline
\midrule
\endhead
10 & 0 & 100000 & 0 & 100000 & 0 & 0\tabularnewline
20 & 0 & 100000 & 0 & 100000 & 0 & 0\tabularnewline
30 & 0 & 100000 & 0 & 100000 & 0 & 0\tabularnewline
40 & 0 & 100000 & 0 & 100000 & 0 & 0\tabularnewline
50 & 0 & 100000 & 0 & 100000 & 0 & 0\tabularnewline
60 & 2 & 99998 & 2 & 99998 & 0 & 0\tabularnewline
70 & 4 & 99996 & 4 & 99996 & 0 & 0\tabularnewline
80 & 6 & 99994 & 6 & 99994 & 0 & 0\tabularnewline
90 & 5 & 99995 & 5 & 99995 & 0 & 0\tabularnewline
100 & 11 & 99989 & 11 & 99989 & 0 & 0\tabularnewline
110 & 12 & 99988 & 13 & 99987 & 0 & 1\tabularnewline
120 & 23 & 99977 & 23 & 99977 & 0 & 0\tabularnewline
130 & 40 & 99960 & 40 & 99960 & 0 & 0\tabularnewline
140 & 62 & 99938 & 65 & 99935 & 0 & 3\tabularnewline
150 & 162 & 99838 & 165 & 99835 & 0 & 3\tabularnewline
160 & 325 & 99675 & 329 & 99671 & 2 & 6\tabularnewline
170 & 829 & 99171 & 851 & 99149 & 6 & 28\tabularnewline
180 & 1817 & 98183 & 1860 & 98140 & 31 & 74\tabularnewline
190 & 3927 & 96073 & 3989 & 96011 & 143 & 205\tabularnewline
200 & 8084 & 91916 & 8048 & 91952 & 557 & 521\tabularnewline
210 & 15558 & 84442 & 15147 & 84853 & 1534 & 1123\tabularnewline
220 & 26848 & 73152 & 25342 & 74658 & 3625 & 2119\tabularnewline
230 & 43386 & 56614 & 39535 & 60465 & 6992 & 3141\tabularnewline
240 & 62734 & 37266 & 56684 & 43316 & 9815 & 3765\tabularnewline
250 & 80128 & 19872 & 73080 & 26920 & 10128 & 3080\tabularnewline
260 & 92206 & 7794 & 86619 & 13381 & 7490 & 1903\tabularnewline
270 & 97946 & 2054 & 94824 & 5176 & 3797 & 675\tabularnewline
280 & 99659 & 341 & 98534 & 1466 & 1265 & 140\tabularnewline
290 & 99962 & 38 & 99696 & 304 & 281 & 15\tabularnewline
300 & 99994 & 6 & 99964 & 36 & 34 & 4\tabularnewline
\bottomrule
\end{longtable}

\textbf{Interaction strength \(\mathbf{\sigma_{A} = 0.4}\)}

\begin{longtable}[]{@{}lllllll@{}}
\toprule
S & A\_unstable & A\_stable & M\_unstable & M\_stable & A\_stabilised &
A\_destabilised\tabularnewline
\midrule
\endhead
10 & 3 & 99997 & 3 & 99997 & 0 & 0\tabularnewline
20 & 15 & 99985 & 15 & 99985 & 0 & 0\tabularnewline
30 & 48 & 99952 & 48 & 99952 & 0 & 0\tabularnewline
40 & 85 & 99915 & 85 & 99915 & 0 & 0\tabularnewline
50 & 163 & 99837 & 163 & 99837 & 0 & 0\tabularnewline
60 & 280 & 99720 & 282 & 99718 & 0 & 2\tabularnewline
70 & 561 & 99439 & 566 & 99434 & 3 & 8\tabularnewline
80 & 1009 & 98991 & 1029 & 98971 & 6 & 26\tabularnewline
90 & 2126 & 97874 & 2175 & 97825 & 31 & 80\tabularnewline
100 & 4580 & 95420 & 4653 & 95347 & 142 & 215\tabularnewline
110 & 9540 & 90460 & 9632 & 90368 & 465 & 557\tabularnewline
120 & 19090 & 80910 & 18668 & 81332 & 1676 & 1254\tabularnewline
130 & 35047 & 64953 & 33220 & 66780 & 4172 & 2345\tabularnewline
140 & 56411 & 43589 & 52439 & 47561 & 7297 & 3325\tabularnewline
150 & 78003 & 21997 & 72574 & 27426 & 8477 & 3048\tabularnewline
160 & 92678 & 7322 & 88438 & 11562 & 5901 & 1661\tabularnewline
170 & 98614 & 1386 & 96670 & 3330 & 2397 & 453\tabularnewline
180 & 99839 & 161 & 99418 & 582 & 499 & 78\tabularnewline
190 & 99990 & 10 & 99945 & 55 & 52 & 7\tabularnewline
200 & 100000 & 0 & 99995 & 5 & 5 & 0\tabularnewline
210 & 100000 & 0 & 100000 & 0 & 0 & 0\tabularnewline
\ldots{} & \ldots{} & \ldots{} & \ldots{} & \ldots{} & \ldots{} &
\ldots{}\tabularnewline
300 & 100000 & 0 & 100000 & 0 & 0 & 0\tabularnewline
\bottomrule
\end{longtable}

\textbf{Interaction strength \(\mathbf{\sigma_{A} = 0.5}\)}

\begin{longtable}[]{@{}lllllll@{}}
\toprule
S & A\_unstable & A\_stable & M\_unstable & M\_stable & A\_stabilised &
A\_destabilised\tabularnewline
\midrule
\endhead
10 & 36 & 99964 & 36 & 99964 & 0 & 0\tabularnewline
20 & 195 & 99805 & 195 & 99805 & 0 & 0\tabularnewline
30 & 519 & 99481 & 523 & 99477 & 0 & 4\tabularnewline
40 & 1096 & 98904 & 1101 & 98899 & 2 & 7\tabularnewline
50 & 2375 & 97625 & 2397 & 97603 & 9 & 31\tabularnewline
60 & 4898 & 95102 & 4968 & 95032 & 83 & 153\tabularnewline
70 & 10841 & 89159 & 10916 & 89084 & 432 & 507\tabularnewline
80 & 22281 & 77719 & 21988 & 78012 & 1622 & 1329\tabularnewline
90 & 42010 & 57990 & 39998 & 60002 & 4458 & 2446\tabularnewline
100 & 67289 & 32711 & 63098 & 36902 & 7153 & 2962\tabularnewline
110 & 88137 & 11863 & 84023 & 15977 & 6108 & 1994\tabularnewline
120 & 97678 & 2322 & 95557 & 4443 & 2740 & 619\tabularnewline
130 & 99795 & 205 & 99304 & 696 & 578 & 87\tabularnewline
140 & 99989 & 11 & 99948 & 52 & 49 & 8\tabularnewline
150 & 100000 & 0 & 100000 & 0 & 0 & 0\tabularnewline
\ldots{} & \ldots{} & \ldots{} & \ldots{} & \ldots{} & \ldots{} &
\ldots{}\tabularnewline
300 & 100000 & 0 & 100000 & 0 & 0 & 0\tabularnewline
\bottomrule
\end{longtable}

\textbf{Interaction strength \(\mathbf{\sigma_{A} = 0.6}\)}

\begin{longtable}[]{@{}lllllll@{}}
\toprule
S & A\_unstable & A\_stable & M\_unstable & M\_stable & A\_stabilised &
A\_destabilised\tabularnewline
\midrule
\endhead
10 & 162 & 99838 & 162 & 99838 & 0 & 0\tabularnewline
20 & 798 & 99202 & 799 & 99201 & 0 & 1\tabularnewline
30 & 2273 & 97727 & 2289 & 97711 & 6 & 22\tabularnewline
40 & 5259 & 94741 & 5298 & 94702 & 70 & 109\tabularnewline
50 & 12084 & 87916 & 12054 & 87946 & 446 & 416\tabularnewline
60 & 26072 & 73928 & 25511 & 74489 & 1810 & 1249\tabularnewline
70 & 50121 & 49879 & 47747 & 52253 & 4748 & 2374\tabularnewline
80 & 77806 & 22194 & 73810 & 26190 & 6421 & 2425\tabularnewline
90 & 94862 & 5138 & 92069 & 7931 & 3842 & 1049\tabularnewline
100 & 99527 & 473 & 98822 & 1178 & 870 & 165\tabularnewline
110 & 99984 & 16 & 99912 & 88 & 80 & 8\tabularnewline
120 & 100000 & 0 & 99998 & 2 & 2 & 0\tabularnewline
130 & 100000 & 0 & 100000 & 0 & 0 & 0\tabularnewline
\ldots{} & \ldots{} & \ldots{} & \ldots{} & \ldots{} & \ldots{} &
\ldots{}\tabularnewline
300 & 100000 & 0 & 100000 & 0 & 0 & 0\tabularnewline
\bottomrule
\end{longtable}

\hypertarget{gam_dist}{\section{\texorpdfstring{Sensitivity of
distribution of
\(\gamma\)}{Sensitivity of distribution of \textbackslash{}gamma}}\label{gam_dist}}

In the main text, I considered a uniform distribution of component
response rates \(\gamma \sim \mathcal{U}(0, 2)\). The number of unstable
and stable \(\mathbf{M}\) matrices are reported in
\protect\hyperlink{IncrS}{a table above} across different values of
\(S\). Here I show complementary results for three different
distributions including an exponential, beta, and gamma distribution of
\(\gamma\) values. The shape of these distributions is shown in the
figure below.

\begin{center}\rule{0.5\linewidth}{\linethickness}\end{center}

\textbf{Distributions of component response rate
(\(\boldsymbol{\gamma}\)) values in complex systems.} The stabilities of
simulated complex systems with these \(\gamma\) distributions are
compared to identical systems in which \(\gamma = 1\) across different
system sizes (\(S\); i.e., component numbers) given a unit \(\gamma\)
standard deviation (\(\sigma_{\gamma} = 1\)) for b-d. Distributions are
as follows: (a) uniform, (b) exponential, (c) beta (\(\alpha = 0.5\) and
\(\beta = 0.5\)), and (d) gamma (\(k = 2\) and \(\theta = 2\)). Each
panel shows 1 million randomly generated \(\gamma\) values.

\begin{center}\includegraphics{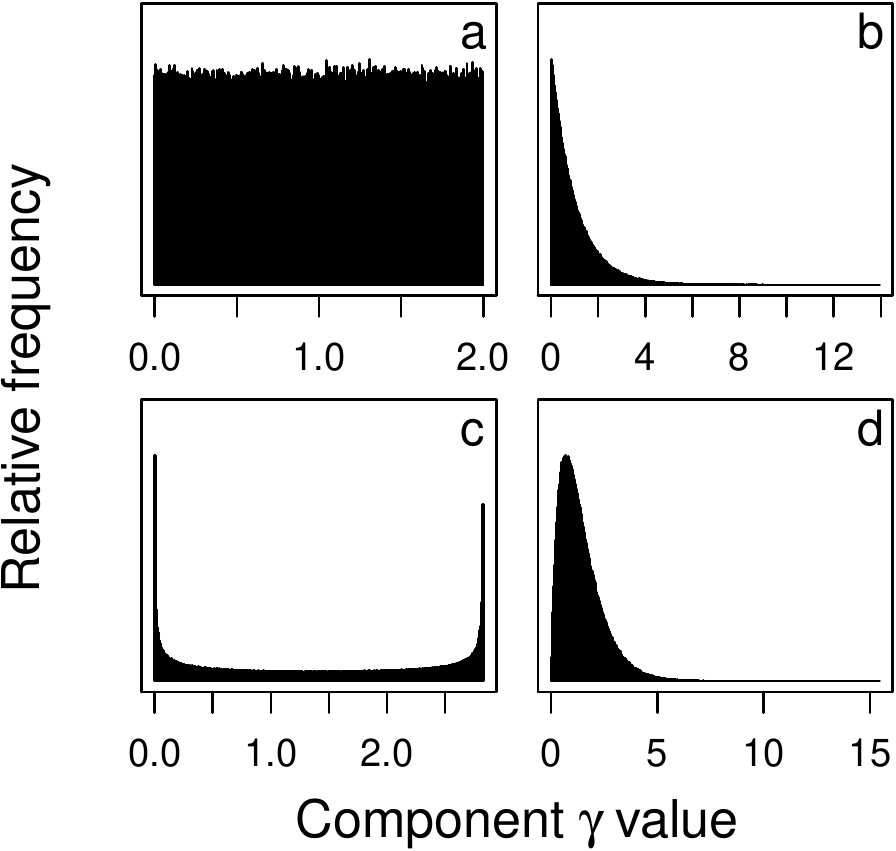} \end{center}

\begin{center}\rule{0.5\linewidth}{\linethickness}\end{center}

The stability of \(\mathbf{A}\) versus \(\mathbf{M}\) was investigated
for each of the distributions of \(\gamma\) shown in panels b-d above.
The table below shows the number of \(\mathbf{A}\) versus \(\mathbf{M}\)
that were stable for the exponential (exp), beta, and gamma
distributions.

\begin{longtable}[]{@{}lllllll@{}}
\toprule
S & exp\_A & exp\_M & beta\_A & beta\_M & gamma\_A &
gamma\_M\tabularnewline
\midrule
\endhead
2 & 99965 & 99965 & 99974 & 99974 & 99977 & 99977\tabularnewline
3 & 99636 & 99635 & 99650 & 99648 & 99628 & 99628\tabularnewline
4 & 98576 & 98564 & 98482 & 98470 & 98508 & 98492\tabularnewline
5 & 96053 & 95971 & 96156 & 96096 & 96068 & 96004\tabularnewline
6 & 92036 & 91867 & 92104 & 91927 & 92233 & 92029\tabularnewline
7 & 86667 & 86333 & 86456 & 86070 & 86604 & 86161\tabularnewline
8 & 79670 & 79153 & 79392 & 78822 & 79393 & 78771\tabularnewline
9 & 71389 & 70911 & 70998 & 70529 & 71070 & 70548\tabularnewline
10 & 61674 & 61609 & 61794 & 61586 & 61265 & 61093\tabularnewline
11 & 51150 & 51935 & 51352 & 51924 & 51313 & 51951\tabularnewline
12 & 41209 & 42925 & 40954 & 42670 & 40708 & 42183\tabularnewline
13 & 30827 & 33462 & 30969 & 33770 & 31046 & 33522\tabularnewline
14 & 22203 & 25767 & 22208 & 25629 & 22342 & 25435\tabularnewline
15 & 15003 & 18877 & 15206 & 18913 & 15025 & 18464\tabularnewline
16 & 9613 & 13372 & 9504 & 13357 & 9418 & 12737\tabularnewline
17 & 5579 & 8967 & 5570 & 8976 & 5719 & 8487\tabularnewline
18 & 3104 & 5833 & 3048 & 5853 & 3060 & 5447\tabularnewline
19 & 1516 & 3578 & 1553 & 3633 & 1600 & 3185\tabularnewline
20 & 717 & 2067 & 799 & 2179 & 769 & 1862\tabularnewline
21 & 312 & 1196 & 310 & 1200 & 331 & 1039\tabularnewline
22 & 129 & 643 & 128 & 654 & 135 & 510\tabularnewline
23 & 48 & 321 & 48 & 359 & 57 & 242\tabularnewline
24 & 11 & 161 & 19 & 159 & 20 & 120\tabularnewline
25 & 1 & 59 & 5 & 81 & 7 & 45\tabularnewline
26 & 0 & 30 & 0 & 48 & 0 & 22\tabularnewline
27 & 0 & 10 & 0 & 16 & 0 & 6\tabularnewline
28 & 1 & 3 & 2 & 2 & 0 & 3\tabularnewline
29 & 0 & 2 & 0 & 0 & 0 & 0\tabularnewline
30 & 0 & 0 & 0 & 1 & 0 & 0\tabularnewline
31 & 0 & 0 & 0 & 1 & 0 & 0\tabularnewline
32 & 0 & 0 & 0 & 0 & 0 & 0\tabularnewline
\ldots{} & \ldots{} & \ldots{} & \ldots{} & \ldots{} & \ldots{} &
\ldots{}\tabularnewline
50 & 0 & 0 & 0 & 0 & 0 & 0\tabularnewline
\bottomrule
\end{longtable}

In comparison to the uniform distribution (a), proportionally fewer
random systems are found with the exponential distribution (b), while
more are found with the beta (c) and gamma (d) distributions.

\hypertarget{structured}{\section{Stability of structured
networks}\label{structured}}

I tested the stability of one million random, small-world, scale-free,
and cascade food web networks for different network parameters. Each of
these networks is structured differently. In the main text, the random
networks and cascade food webs that I built were saturated (\(C = 1\)),
meaning that every component was connected to, and interacted with,
every other component (see immediately below).

\vspace{2mm}

\begin{center}\includegraphics{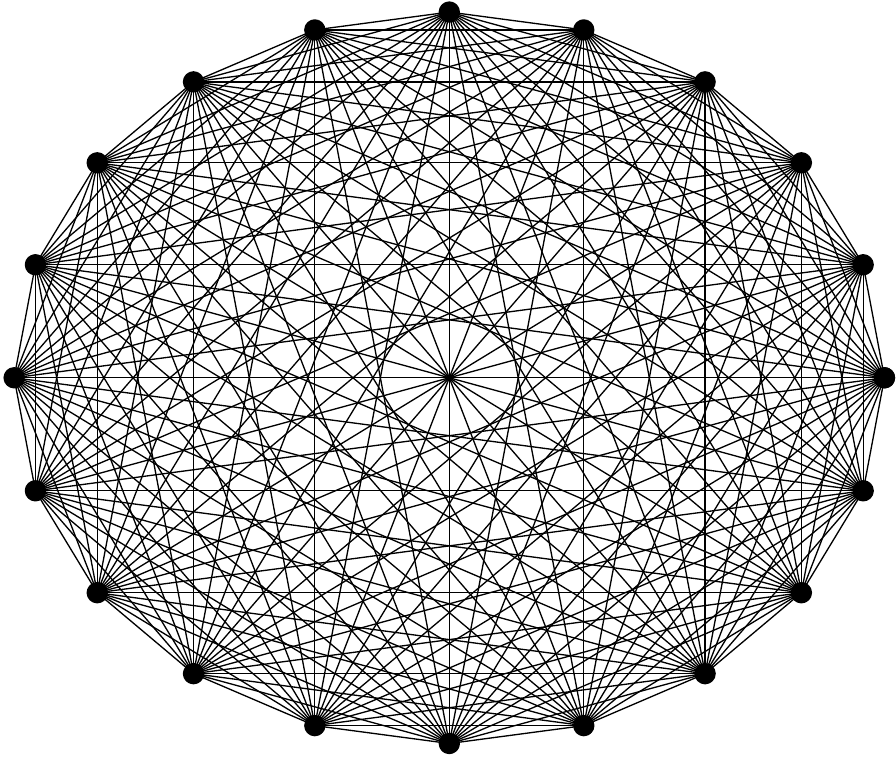} \end{center}

\vspace{2mm}

Small-world networks, in contrast, are not saturated. They are instead
defined by components that interact mostly with other closely
neighbouring components, but have a proportion of interactions
(\(\beta\)) that are instead between
non-neighbours\textsuperscript{\protect\hyperlink{ref-Watts1998}{4}}.
Two small-world networks are shown below.

\vspace{2mm}

\begin{center}\includegraphics{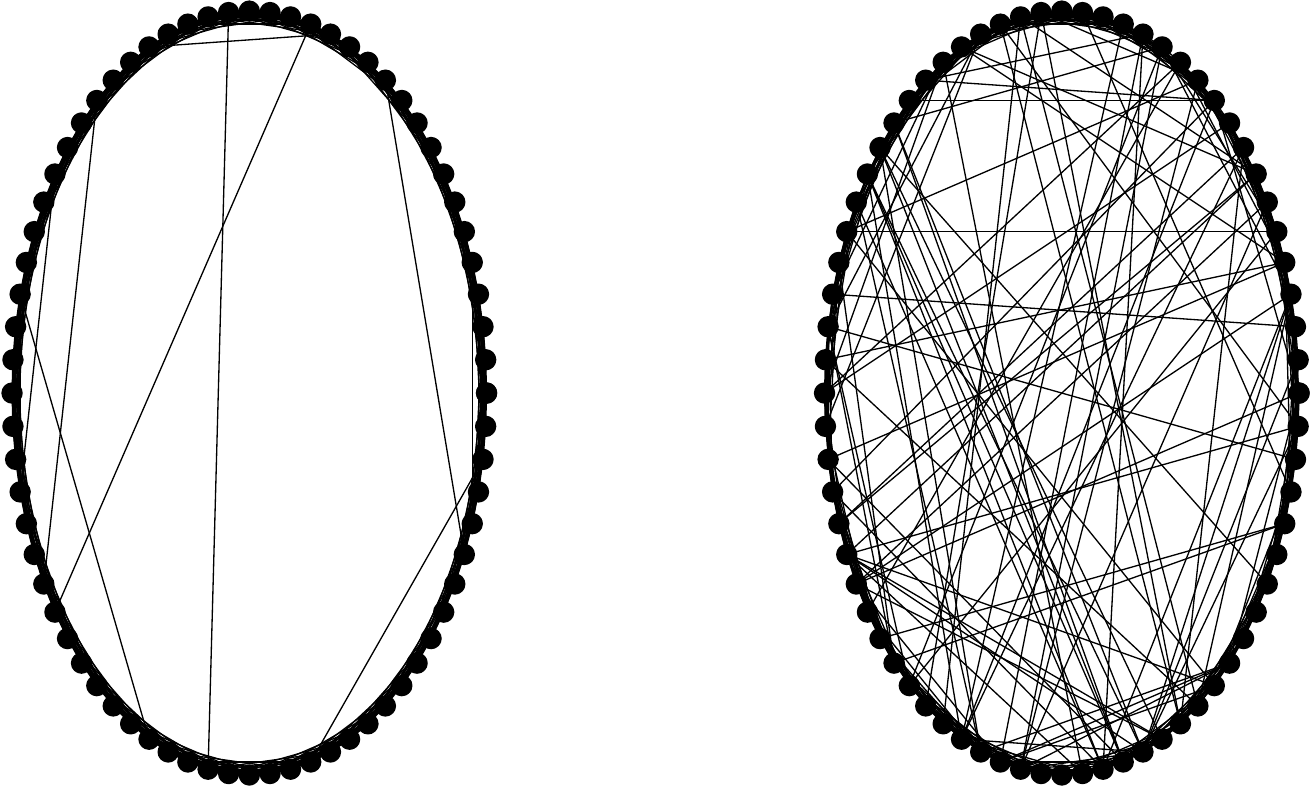} \end{center}

\vspace{2mm}

The small-world network on the left shows a system in which
\(\beta = 0.01\), while the small-world network on the right shows one
in which \(\beta = 0.1\). At the extremes of \(\beta = 0\) and
\(\beta = 1\), networks are regular and random, respectively. The table
below shows how \(\sigma^{2}_\gamma\) affects stability in small world
networks across different values of \(S\) and \(\beta\).

\begin{longtable}[]{@{}rrrrrrrrr@{}}
\toprule
beta & S & A\_unstable & A\_stable & M\_unstable & M\_stable &
complex\_A & complex\_M & C\tabularnewline
\midrule
\endhead
0.00 & 24 & 17388 & 982612 & 17446 & 982554 & 0.5748066 & 0.6582632 &
0.1304348\tabularnewline
0.00 & 48 & 258024 & 741976 & 260579 & 739421 & 0.8073918 & 0.9294192 &
0.1063830\tabularnewline
0.00 & 72 & 715036 & 284964 & 722639 & 277361 & 0.9860840 & 1.1364805 &
0.0985915\tabularnewline
0.00 & 96 & 961434 & 38566 & 962788 & 37212 & 1.1369395 & 1.3110263 &
0.0947368\tabularnewline
0.00 & 120 & 999008 & 992 & 998857 & 1143 & 1.2700387 & 1.4649832 &
0.0924370\tabularnewline
0.00 & 144 & 999997 & 3 & 999994 & 6 & 1.3903192 & 1.6041216 &
0.0909091\tabularnewline
0.00 & 168 & 1000000 & 0 & 1000000 & 0 & 1.5010334 & 1.7320676 &
0.0898204\tabularnewline
0.01 & 24 & 17673 & 982327 & 17720 & 982280 & 0.5747156 & 0.6581503 &
0.1304319\tabularnewline
0.01 & 48 & 255038 & 744962 & 257647 & 742353 & 0.8073388 & 0.9292952 &
0.1063800\tabularnewline
0.01 & 72 & 708892 & 291108 & 716829 & 283171 & 0.9859457 & 1.1363940 &
0.0985884\tabularnewline
0.01 & 96 & 960635 & 39365 & 961876 & 38124 & 1.1370640 & 1.3112193 &
0.0947337\tabularnewline
0.01 & 120 & 999040 & 960 & 998794 & 1206 & 1.2698715 & 1.4648280 &
0.0924338\tabularnewline
0.01 & 144 & 999997 & 3 & 999994 & 6 & 1.3901601 & 1.6039285 &
0.0909060\tabularnewline
0.01 & 168 & 1000000 & 0 & 1000000 & 0 & 1.5009490 & 1.7319739 &
0.0898173\tabularnewline
0.10 & 24 & 20382 & 979618 & 20455 & 979545 & 0.5742520 & 0.6573563 &
0.1302974\tabularnewline
0.10 & 48 & 237747 & 762253 & 240370 & 759630 & 0.8066604 & 0.9284434 &
0.1062311\tabularnewline
0.10 & 72 & 679874 & 320126 & 685575 & 314425 & 0.9849695 & 1.1352553 &
0.0984349\tabularnewline
0.10 & 96 & 961984 & 38016 & 960128 & 39872 & 1.1358912 & 1.3097957 &
0.0945788\tabularnewline
0.10 & 120 & 999546 & 454 & 999275 & 725 & 1.2687142 & 1.4634587 &
0.0922779\tabularnewline
0.10 & 144 & 1000000 & 0 & 1000000 & 0 & 1.3890356 & 1.6025900 &
0.0907489\tabularnewline
0.10 & 168 & 1000000 & 0 & 1000000 & 0 & 1.4994818 & 1.7302649 &
0.0896598\tabularnewline
0.25 & 24 & 23654 & 976346 & 23775 & 976225 & 0.5722185 & 0.6546853 &
0.1296712\tabularnewline
0.25 & 48 & 228318 & 771682 & 231208 & 768792 & 0.8033257 & 0.9244966 &
0.1055259\tabularnewline
0.25 & 72 & 666982 & 333018 & 669104 & 330896 & 0.9808676 & 1.1304109 &
0.0977066\tabularnewline
0.25 & 96 & 966456 & 33544 & 961545 & 38455 & 1.1307841 & 1.3039452 &
0.0938392\tabularnewline
0.25 & 120 & 999749 & 251 & 999507 & 493 & 1.2632327 & 1.4571506 &
0.0915316\tabularnewline
0.25 & 144 & 1000000 & 0 & 1000000 & 0 & 1.3827642 & 1.5953248 &
0.0899987\tabularnewline
0.25 & 168 & 1000000 & 0 & 1000000 & 0 & 1.4926700 & 1.7224506 &
0.0889064\tabularnewline
1.00 & 24 & 26331 & 973669 & 26478 & 973522 & 0.5561013 & 0.6356655 &
0.1249651\tabularnewline
1.00 & 48 & 211199 & 788801 & 214154 & 785846 & 0.7720342 & 0.8881302 &
0.0991370\tabularnewline
1.00 & 72 & 613621 & 386379 & 615771 & 384229 & 0.9394912 & 1.0825566 &
0.0908153\tabularnewline
1.00 & 96 & 943191 & 56809 & 936396 & 63604 & 1.0812364 & 1.2466510 &
0.0867047\tabularnewline
1.00 & 120 & 999157 & 843 & 998396 & 1604 & 1.2065026 & 1.3916458 &
0.0842561\tabularnewline
1.00 & 144 & 1000000 & 0 & 999997 & 3 & 1.3199179 & 1.5227509 &
0.0826325\tabularnewline
1.00 & 168 & 1000000 & 0 & 1000000 & 0 & 1.4243560 & 1.6434386 &
0.0814738\tabularnewline
\bottomrule
\end{longtable}

In the above, the complexity of \(\mathbf{A}\) and \(\mathbf{M}\), and
the mean \(C\), are also shown. For similar magnitudes of complexity as
in random networks of \(\sigma\sqrt{SC} \gtrapprox 1.26\), variation in
\(\gamma\) typically results in more stable than unstable systems.

Scale-free networks are also not saturated, but are defined by an
interaction frequency distribution that follows a power law. In other
words, a small number of components interact with many other components,
while most components interact with only a small number of other
components. Scale-free networks can be built by adding new components,
one by one, to an existing system, with each newly added component
interacting with a randomly selected subset of \(m\) existing
components\textsuperscript{\protect\hyperlink{ref-Albert2002}{5}}. The
network on the left below shows an example of a scale-free network in
which \(m = 3\). The histogram on the right shows the number of other
components with which each component interacts.

\begin{center}\includegraphics{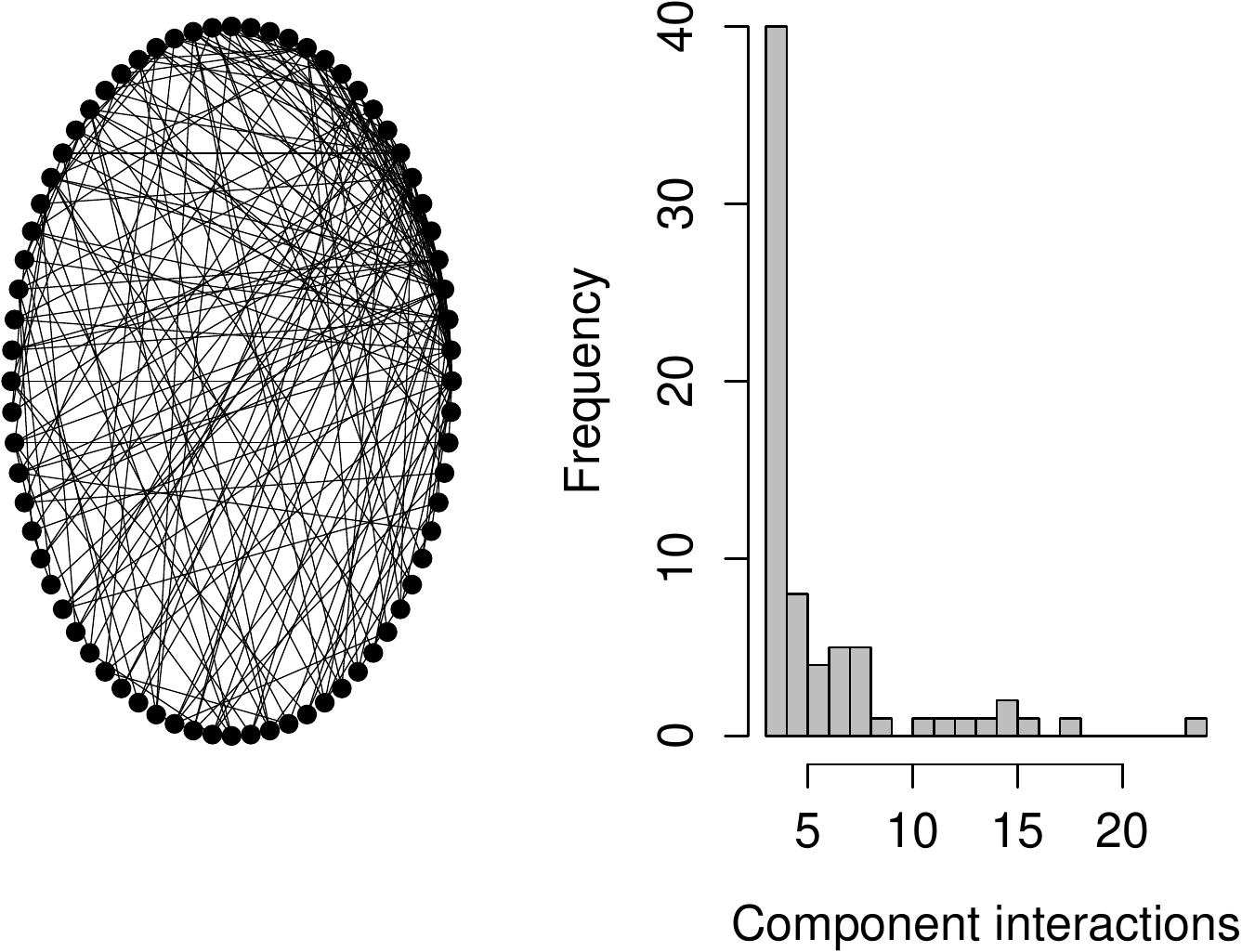} \end{center}

The table below shows how \(\sigma^{2}_\gamma\) affects stability across
different scale-free networks with different \(S\) and \(m\) values.

\begin{longtable}[]{@{}rrrrrrrrr@{}}
\toprule
m & S & A\_unstable & A\_stable & M\_unstable & M\_stable & complex\_A &
complex\_M & C\tabularnewline
\midrule
\endhead
2 & 24 & 152791 & 847209 & 156034 & 843966 & 0.7891257 & 0.9034663 &
0.1648551\tabularnewline
3 & 24 & 320481 & 679519 & 326351 & 673649 & 0.9566487 & 1.0967499 &
0.2409420\tabularnewline
4 & 24 & 504433 & 495567 & 504826 & 495174 & 1.0922870 & 1.2532761 &
0.3134058\tabularnewline
5 & 24 & 670676 & 329324 & 660426 & 339574 & 1.2073054 & 1.3857169 &
0.3822464\tabularnewline
6 & 24 & 798637 & 201363 & 779345 & 220655 & 1.3067095 & 1.5004508 &
0.4474638\tabularnewline
7 & 24 & 884082 & 115918 & 862215 & 137785 & 1.3942577 & 1.6013368 &
0.5090580\tabularnewline
8 & 24 & 936190 & 63810 & 915630 & 84370 & 1.4722315 & 1.6908563 &
0.5670290\tabularnewline
9 & 24 & 964868 & 35132 & 948297 & 51703 & 1.5414455 & 1.7707292 &
0.6213768\tabularnewline
10 & 24 & 981460 & 18540 & 967911 & 32089 & 1.6030044 & 1.8417459 &
0.6721014\tabularnewline
11 & 24 & 989838 & 10162 & 980232 & 19768 & 1.6586511 & 1.9059313 &
0.7192029\tabularnewline
12 & 24 & 994393 & 5607 & 987436 & 12564 & 1.7081503 & 1.9628898 &
0.7626812\tabularnewline
2 & 48 & 303963 & 696037 & 310053 & 689947 & 0.7946875 & 0.9132519 &
0.0828901\tabularnewline
3 & 48 & 577855 & 422145 & 579996 & 420004 & 0.9685494 & 1.1141445 &
0.1227837\tabularnewline
4 & 48 & 810001 & 189999 & 799132 & 200868 & 1.1122992 & 1.2799335 &
0.1617908\tabularnewline
5 & 48 & 938004 & 61996 & 924613 & 75387 & 1.2369960 & 1.4236817 &
0.1999113\tabularnewline
6 & 48 & 984975 & 15025 & 976433 & 23567 & 1.3478291 & 1.5514420 &
0.2371454\tabularnewline
7 & 48 & 997160 & 2840 & 994005 & 5995 & 1.4473792 & 1.6663763 &
0.2734929\tabularnewline
8 & 48 & 999584 & 416 & 998590 & 1410 & 1.5385445 & 1.7716359 &
0.3089539\tabularnewline
9 & 48 & 999955 & 45 & 999707 & 293 & 1.6227742 & 1.8687074 &
0.3435284\tabularnewline
10 & 48 & 999992 & 8 & 999939 & 61 & 1.7006157 & 1.9583879 &
0.3772163\tabularnewline
11 & 48 & 999999 & 1 & 999990 & 10 & 1.7731759 & 2.0420990 &
0.4100177\tabularnewline
12 & 48 & 1000000 & 0 & 999999 & 1 & 1.8410402 & 2.1203112 &
0.4419326\tabularnewline
2 & 72 & 427243 & 572757 & 434600 & 565400 & 0.7964226 & 0.9166566 &
0.0553599\tabularnewline
3 & 72 & 741345 & 258655 & 739020 & 260980 & 0.9723446 & 1.1195788 &
0.0823552\tabularnewline
4 & 72 & 931043 & 68957 & 921145 & 78855 & 1.1188220 & 1.2888100 &
0.1089593\tabularnewline
5 & 72 & 989644 & 10356 & 984372 & 15628 & 1.2466268 & 1.4361875 &
0.1351721\tabularnewline
6 & 72 & 999131 & 869 & 997914 & 2086 & 1.3604666 & 1.5674966 &
0.1609937\tabularnewline
7 & 72 & 999946 & 54 & 999804 & 196 & 1.4642496 & 1.6872501 &
0.1864241\tabularnewline
8 & 72 & 999999 & 1 & 999988 & 12 & 1.5596340 & 1.7974044 &
0.2114632\tabularnewline
9 & 72 & 1000000 & 0 & 999999 & 1 & 1.6482181 & 1.8994441 &
0.2361111\tabularnewline
10 & 72 & 1000000 & 0 & 1000000 & 0 & 1.7307859 & 1.9947150 &
0.2603678\tabularnewline
11 & 72 & 1000000 & 0 & 1000000 & 0 & 1.8086766 & 2.0847262 &
0.2842332\tabularnewline
12 & 72 & 1000000 & 0 & 1000000 & 0 & 1.8817533 & 2.1689764 &
0.3077074\tabularnewline
2 & 96 & 527633 & 472367 & 535188 & 464812 & 0.7974024 & 0.9183557 &
0.0415570\tabularnewline
3 & 96 & 842274 & 157726 & 837756 & 162244 & 0.9741293 & 1.1224709 &
0.0619518\tabularnewline
4 & 96 & 975834 & 24166 & 969478 & 30522 & 1.1220115 & 1.2931371 &
0.0821272\tabularnewline
5 & 96 & 998391 & 1609 & 996991 & 3009 & 1.2511287 & 1.4422331 &
0.1020833\tabularnewline
6 & 96 & 999955 & 45 & 999838 & 162 & 1.3669903 & 1.5757699 &
0.1218202\tabularnewline
7 & 96 & 999999 & 1 & 999996 & 4 & 1.4725862 & 1.6977057 &
0.1413377\tabularnewline
8 & 96 & 1000000 & 0 & 1000000 & 0 & 1.5699145 & 1.8099762 &
0.1606360\tabularnewline
9 & 96 & 1000000 & 0 & 1000000 & 0 & 1.6606162 & 1.9146804 &
0.1797149\tabularnewline
10 & 96 & 1000000 & 0 & 1000000 & 0 & 1.7457971 & 2.0129344 &
0.1985746\tabularnewline
11 & 96 & 1000000 & 0 & 1000000 & 0 & 1.8260368 & 2.1055559 &
0.2172149\tabularnewline
12 & 96 & 1000000 & 0 & 1000000 & 0 & 1.9018608 & 2.1929362 &
0.2356360\tabularnewline
2 & 120 & 609563 & 390437 & 616036 & 383964 & 0.7979355 & 0.9194404 &
0.0332633\tabularnewline
3 & 120 & 904064 & 95936 & 899040 & 100960 & 0.9753815 & 1.1243251 &
0.0496499\tabularnewline
4 & 120 & 991710 & 8290 & 988410 & 11590 & 1.1239922 & 1.2957520 &
0.0658964\tabularnewline
5 & 120 & 999781 & 219 & 999477 & 523 & 1.2539362 & 1.4458518 &
0.0820028\tabularnewline
6 & 120 & 999999 & 1 & 999981 & 19 & 1.3707937 & 1.5806987 &
0.0979692\tabularnewline
7 & 120 & 1000000 & 0 & 999999 & 1 & 1.4775366 & 1.7038860 &
0.1137955\tabularnewline
8 & 120 & 1000000 & 0 & 1000000 & 0 & 1.5762636 & 1.8177236 &
0.1294818\tabularnewline
9 & 120 & 1000000 & 0 & 1000000 & 0 & 1.6680647 & 1.9238257 &
0.1450280\tabularnewline
10 & 120 & 1000000 & 0 & 1000000 & 0 & 1.7545110 & 2.0233838 &
0.1604342\tabularnewline
11 & 120 & 1000000 & 0 & 1000000 & 0 & 1.8363882 & 2.1178385 &
0.1757003\tabularnewline
12 & 120 & 1000000 & 0 & 1000000 & 0 & 1.9135798 & 2.2069806 &
0.1908263\tabularnewline
\bottomrule
\end{longtable}

As in small-world networks, the mean \(C\) is shown, along with the mean
complexities of \(\mathbf{A}\) and \(\mathbf{M}\). Like all other
networks, \(\sigma^{2}_\gamma\) increases the stability of scale-free
networks given sufficiently high complexity.

Cascade food webs are saturated, and similar to predator-prey random
networks. What distinguishes them from predator-prey networks is that
cascade food webs are also defined by intactions in which components are
ranked such that if the rank of \(i > j\), then \(A_{ij} < 0\) and
\(A_{ji} > 0\)\textsuperscript{\protect\hyperlink{ref-Solow1998}{6},\protect\hyperlink{ref-Williams2000}{7}}.
In other words, if interpreting components as ecological species,
species can only feed off of a species of lower rank. The table below
shows how \(\sigma^{2}_\gamma\) affects stability across system sizes in
cascade food webs.

\begin{longtable}[]{@{}rrrrrrr@{}}
\toprule
S & A\_unstable & A\_stable & M\_unstable & M\_stable & complex\_A &
complex\_M\tabularnewline
\midrule
\endhead
2 & 0 & 1000000 & 0 & 1000000 & 0.6378839 & 0.6381485\tabularnewline
3 & 1 & 999999 & 1 & 999999 & 0.7055449 & 0.7525143\tabularnewline
4 & 2 & 999998 & 2 & 999998 & 0.8060500 & 0.8826100\tabularnewline
5 & 17 & 999983 & 17 & 999983 & 0.8974749 & 0.9967594\tabularnewline
6 & 42 & 999958 & 43 & 999957 & 0.9821323 & 1.0999762\tabularnewline
7 & 124 & 999876 & 124 & 999876 & 1.0600906 & 1.1938910\tabularnewline
8 & 303 & 999697 & 309 & 999691 & 1.1329713 & 1.2807302\tabularnewline
9 & 653 & 999347 & 661 & 999339 & 1.2009135 & 1.3616372\tabularnewline
10 & 1401 & 998599 & 1413 & 998587 & 1.2661142 &
1.4387567\tabularnewline
11 & 2534 & 997466 & 2566 & 997434 & 1.3276636 &
1.5113096\tabularnewline
12 & 4514 & 995486 & 4597 & 995403 & 1.3865754 &
1.5804005\tabularnewline
13 & 7570 & 992430 & 7722 & 992278 & 1.4424479 &
1.6462780\tabularnewline
14 & 12223 & 987777 & 12502 & 987498 & 1.4970134 &
1.7102322\tabularnewline
15 & 18433 & 981567 & 18879 & 981121 & 1.5498812 &
1.7719564\tabularnewline
16 & 26973 & 973027 & 27712 & 972288 & 1.6002970 &
1.8310447\tabularnewline
17 & 38272 & 961728 & 39499 & 960501 & 1.6494195 &
1.8884211\tabularnewline
18 & 52397 & 947603 & 54099 & 945901 & 1.6975099 &
1.9443860\tabularnewline
19 & 69986 & 930014 & 72342 & 927658 & 1.7439233 &
1.9987398\tabularnewline
20 & 92851 & 907149 & 95776 & 904224 & 1.7893524 &
2.0514394\tabularnewline
21 & 117487 & 882513 & 121095 & 878905 & 1.8335974 &
2.1030121\tabularnewline
22 & 147852 & 852148 & 151989 & 848011 & 1.8761874 &
2.1527108\tabularnewline
23 & 183501 & 816499 & 187888 & 812112 & 1.9186092 &
2.2019827\tabularnewline
24 & 222592 & 777408 & 226021 & 773979 & 1.9591518 &
2.2491948\tabularnewline
25 & 267691 & 732309 & 269822 & 730178 & 1.9999089 &
2.2963949\tabularnewline
26 & 316090 & 683910 & 316371 & 683629 & 2.0396325 &
2.3427211\tabularnewline
27 & 369830 & 630170 & 366550 & 633450 & 2.0785319 &
2.3879356\tabularnewline
28 & 426407 & 573593 & 419136 & 580864 & 2.1169703 &
2.4324407\tabularnewline
29 & 485068 & 514932 & 473666 & 526334 & 2.1545265 &
2.4759539\tabularnewline
30 & 544300 & 455700 & 527568 & 472432 & 2.1912376 &
2.5187795\tabularnewline
31 & 605803 & 394197 & 584385 & 415615 & 2.2271037 &
2.5603818\tabularnewline
32 & 664689 & 335311 & 638047 & 361953 & 2.2626270 &
2.6016360\tabularnewline
33 & 718848 & 281152 & 689172 & 310828 & 2.2979241 &
2.6424881\tabularnewline
34 & 770790 & 229210 & 737639 & 262361 & 2.3327303 &
2.6828460\tabularnewline
35 & 817531 & 182469 & 783112 & 216888 & 2.3666720 &
2.7221952\tabularnewline
36 & 858750 & 141250 & 823548 & 176452 & 2.3998286 &
2.7608037\tabularnewline
37 & 893017 & 106983 & 859194 & 140806 & 2.4332806 &
2.7994470\tabularnewline
38 & 921268 & 78732 & 890177 & 109823 & 2.4658414 &
2.8372307\tabularnewline
39 & 943551 & 56449 & 915655 & 84345 & 2.4974678 &
2.8741350\tabularnewline
40 & 961088 & 38912 & 936883 & 63117 & 2.5301278 &
2.9116114\tabularnewline
41 & 973664 & 26336 & 953645 & 46355 & 2.5616210 &
2.9481298\tabularnewline
42 & 982829 & 17171 & 967044 & 32956 & 2.5925309 &
2.9841081\tabularnewline
43 & 989464 & 10536 & 977033 & 22967 & 2.6228949 &
3.0191690\tabularnewline
44 & 993622 & 6378 & 984470 & 15530 & 2.6534626 &
3.0548439\tabularnewline
45 & 996221 & 3779 & 989678 & 10322 & 2.6832092 &
3.0890543\tabularnewline
46 & 997963 & 2037 & 993318 & 6682 & 2.7130588 &
3.1236201\tabularnewline
47 & 998818 & 1182 & 995957 & 4043 & 2.7423480 &
3.1575904\tabularnewline
48 & 999422 & 578 & 997446 & 2554 & 2.7714223 & 3.1912463\tabularnewline
49 & 999746 & 254 & 998532 & 1468 & 2.7999596 & 3.2244020\tabularnewline
50 & 999864 & 136 & 999132 & 868 & 2.8285547 & 3.2574510\tabularnewline
51 & 999934 & 66 & 999561 & 439 & 2.8566907 & 3.2900943\tabularnewline
52 & 999970 & 30 & 999761 & 239 & 2.8844703 & 3.3222721\tabularnewline
53 & 999985 & 15 & 999873 & 127 & 2.9122645 & 3.3544290\tabularnewline
54 & 999999 & 1 & 999935 & 65 & 2.9395400 & 3.3859103\tabularnewline
55 & 1000000 & 0 & 999971 & 29 & 2.9665996 & 3.4173273\tabularnewline
56 & 999999 & 1 & 999988 & 12 & 2.9936263 & 3.4486027\tabularnewline
57 & 1000000 & 0 & 999989 & 11 & 3.0199283 & 3.4789408\tabularnewline
58 & 1000000 & 0 & 999998 & 2 & 3.0460952 & 3.5094530\tabularnewline
59 & 1000000 & 0 & 999999 & 1 & 3.0728115 & 3.5401634\tabularnewline
60 & 1000000 & 0 & 1000000 & 0 & 3.0983367 & 3.5698067\tabularnewline
\bottomrule
\end{longtable}

Cascade food webs are more likely to be stable than small-world or
scale-free networks at equivalent magnitudes of complexity (note 
\(C = 1\) for all above rows). A higher number of stable \(\mathbf{M}\)
than \(\mathbf{A}\) was found given \(S \geq 27\).

\hypertarget{Feasibility}{\section{Feasibility of complex
systems}\label{Feasibility}}

When feasibility was evaluated with and without variation in \(\gamma\),
there was no increase in stability for \(\mathbf{M}\) where \(\gamma\)
varied as compared to where \(\gamma = 1\). Results below illustrate
this result, which was general to all other simulations performed.

\begin{longtable}[]{@{}rrrrrrr@{}}
\toprule
S & A\_infeasible & A\_feasible & M\_infeasible & M\_feasible &
A\_made\_feasible & A\_made\_infeasible\tabularnewline
\midrule
\endhead
2 & 749978 & 250022 & 749942 & 250058 & 35552 & 35516\tabularnewline
3 & 874519 & 125481 & 874296 & 125704 & 36803 & 36580\tabularnewline
4 & 937192 & 62808 & 937215 & 62785 & 26440 & 26463\tabularnewline
5 & 968776 & 31224 & 968639 & 31361 & 16319 & 16182\tabularnewline
6 & 984313 & 15687 & 984463 & 15537 & 9006 & 9156\tabularnewline
7 & 992149 & 7851 & 992161 & 7839 & 4991 & 5003\tabularnewline
8 & 996124 & 3876 & 996103 & 3897 & 2644 & 2623\tabularnewline
9 & 998014 & 1986 & 998027 & 1973 & 1361 & 1374\tabularnewline
10 & 999031 & 969 & 999040 & 960 & 698 & 707\tabularnewline
11 & 999546 & 454 & 999514 & 486 & 377 & 345\tabularnewline
12 & 999764 & 236 & 999792 & 208 & 160 & 188\tabularnewline
13 & 999883 & 117 & 999865 & 135 & 105 & 87\tabularnewline
14 & 999938 & 62 & 999945 & 55 & 40 & 47\tabularnewline
15 & 999971 & 29 & 999964 & 36 & 31 & 24\tabularnewline
16 & 999988 & 12 & 999991 & 9 & 8 & 11\tabularnewline
17 & 999996 & 4 & 999991 & 9 & 8 & 3\tabularnewline
18 & 999997 & 3 & 999999 & 1 & 1 & 3\tabularnewline
19 & 999998 & 2 & 999997 & 3 & 3 & 2\tabularnewline
20 & 1000000 & 0 & 999999 & 1 & 1 & 0\tabularnewline
21 & 1000000 & 0 & 1000000 & 0 & 0 & 0\tabularnewline
22 & 999999 & 1 & 1000000 & 0 & 0 & 1\tabularnewline
23 & 1000000 & 0 & 1000000 & 0 & 0 & 0\tabularnewline
24 & 1000000 & 0 & 1000000 & 0 & 0 & 0\tabularnewline
25 & 1000000 & 0 & 1000000 & 0 & 0 & 0\tabularnewline
26 & 1000000 & 0 & 1000000 & 0 & 0 & 0\tabularnewline
27 & 1000000 & 0 & 1000000 & 0 & 0 & 0\tabularnewline
28 & 1000000 & 0 & 1000000 & 0 & 0 & 0\tabularnewline
29 & 1000000 & 0 & 1000000 & 0 & 0 & 0\tabularnewline
30 & 1000000 & 0 & 1000000 & 0 & 0 & 0\tabularnewline
31 & 1000000 & 0 & 1000000 & 0 & 0 & 0\tabularnewline
32 & 1000000 & 0 & 1000000 & 0 & 0 & 0\tabularnewline
33 & 1000000 & 0 & 1000000 & 0 & 0 & 0\tabularnewline
34 & 1000000 & 0 & 1000000 & 0 & 0 & 0\tabularnewline
35 & 1000000 & 0 & 1000000 & 0 & 0 & 0\tabularnewline
36 & 1000000 & 0 & 1000000 & 0 & 0 & 0\tabularnewline
37 & 1000000 & 0 & 1000000 & 0 & 0 & 0\tabularnewline
38 & 1000000 & 0 & 1000000 & 0 & 0 & 0\tabularnewline
39 & 1000000 & 0 & 1000000 & 0 & 0 & 0\tabularnewline
40 & 1000000 & 0 & 1000000 & 0 & 0 & 0\tabularnewline
41 & 1000000 & 0 & 1000000 & 0 & 0 & 0\tabularnewline
42 & 1000000 & 0 & 1000000 & 0 & 0 & 0\tabularnewline
43 & 1000000 & 0 & 1000000 & 0 & 0 & 0\tabularnewline
44 & 1000000 & 0 & 1000000 & 0 & 0 & 0\tabularnewline
45 & 1000000 & 0 & 1000000 & 0 & 0 & 0\tabularnewline
46 & 1000000 & 0 & 1000000 & 0 & 0 & 0\tabularnewline
47 & 1000000 & 0 & 1000000 & 0 & 0 & 0\tabularnewline
48 & 1000000 & 0 & 1000000 & 0 & 0 & 0\tabularnewline
49 & 1000000 & 0 & 1000000 & 0 & 0 & 0\tabularnewline
50 & 1000000 & 0 & 1000000 & 0 & 0 & 0\tabularnewline
\bottomrule
\end{longtable}

Hence, in general, \(\sigma^{2}_{\gamma}\) does not appear to affect
feasibility in pure species interaction
networks\textsuperscript{\protect\hyperlink{ref-Servan2018}{8}}.

\hypertarget{ga}{\section{\texorpdfstring{Stability given targeted
manipulation of \(\gamma\) (genetic
algorithm)}{Stability given targeted manipulation of \textbackslash{}gamma (genetic algorithm)}}\label{ga}}

The figure below compares the stability of large complex systems given
\(\gamma = 1\) versus targeted manipulation of \(\gamma\) elements. For
each \(S\), 100000 complex systems are randomly generated. Stability of
each complex system is tested given variation in \(\gamma\) using a
genetic algorithm to maximise the effect of \(\gamma\) values on
increasing stability, as compared to stability in an otherwise identical
system in which \(\gamma\) is the same for all components. Blue bars
show the number of stable systems in the absence of component response
rate variation, while red bars show the number of stable systems that
can be generated if component response rate is varied to maximise system
stability. The black line shows the proportion of systems that are
stable when component response rate is targeted to increase stability,
but would not be stable if \(\sigma^{2}_{\gamma} = 0\). The y-axis shows
the \(\ln\) number of systems that are stable across
\(S = \{1, 2, ..., 39, 40\}\) for \(C = 1\), and the proportion of
systems wherein a targeted search of \(\gamma\) values successfully
resulted in system stability.

\includegraphics{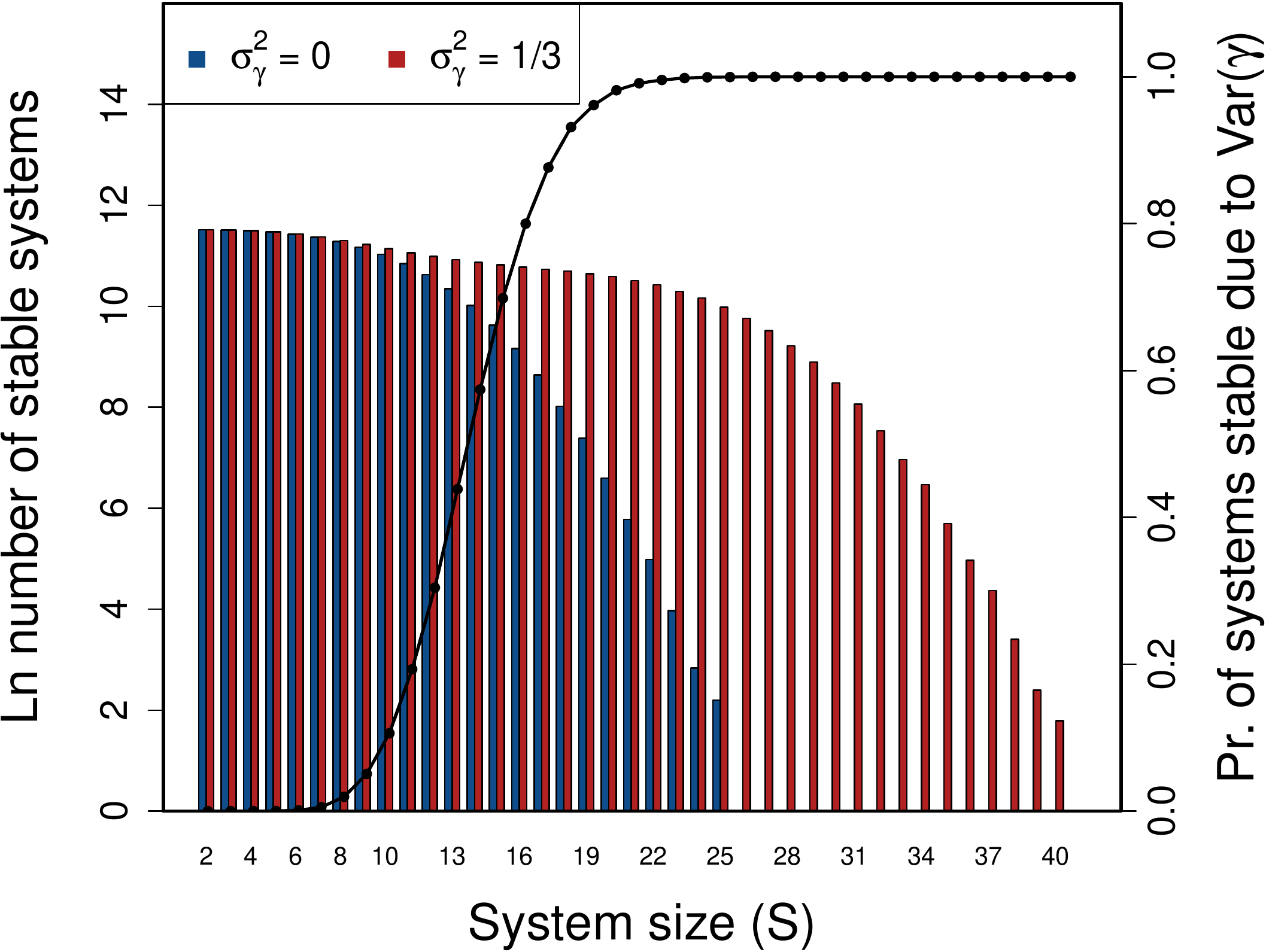}

Stability results are also shown in the table below. Results for
\texttt{A} indicate systems in which \(\gamma = 1\), while \texttt{M}
refers to systems in which the genetic algorithm searched for a set of
\(\gamma\) values that stabilised the system.

\begin{longtable}[]{@{}rrrrrrr@{}}
\toprule
S & A\_unstable & A\_stable & M\_unstable & M\_stable & A\_stabilised &
A\_destabilised\tabularnewline
\midrule
\endhead
2 & 26 & 99974 & 26 & 99974 & 0 & 0\tabularnewline
3 & 358 & 99642 & 358 & 99642 & 0 & 0\tabularnewline
4 & 1505 & 98495 & 1505 & 98495 & 0 & 0\tabularnewline
5 & 3995 & 96005 & 3982 & 96018 & 13 & 0\tabularnewline
6 & 8060 & 91940 & 7956 & 92044 & 104 & 0\tabularnewline
7 & 13420 & 86580 & 12953 & 87047 & 468 & 1\tabularnewline
8 & 20518 & 79482 & 18940 & 81060 & 1578 & 0\tabularnewline
9 & 28939 & 71061 & 25148 & 74852 & 3793 & 2\tabularnewline
10 & 38241 & 61759 & 30915 & 69085 & 7327 & 1\tabularnewline
11 & 48682 & 51318 & 36398 & 63602 & 12286 & 2\tabularnewline
12 & 58752 & 41248 & 40710 & 59290 & 18043 & 1\tabularnewline
13 & 68888 & 31112 & 44600 & 55400 & 24289 & 1\tabularnewline
14 & 77651 & 22349 & 47528 & 52472 & 30124 & 1\tabularnewline
15 & 84912 & 15088 & 49971 & 50029 & 34942 & 1\tabularnewline
16 & 90451 & 9549 & 52274 & 47726 & 38178 & 1\tabularnewline
17 & 94332 & 5668 & 54124 & 45876 & 40209 & 1\tabularnewline
18 & 96968 & 3032 & 55831 & 44169 & 41139 & 2\tabularnewline
19 & 98384 & 1616 & 58079 & 41921 & 40305 & 0\tabularnewline
20 & 99269 & 731 & 60181 & 39819 & 39088 & 0\tabularnewline
21 & 99677 & 323 & 63338 & 36662 & 36339 & 0\tabularnewline
22 & 99854 & 146 & 66350 & 33650 & 33504 & 0\tabularnewline
23 & 99947 & 53 & 70478 & 29522 & 29469 & 0\tabularnewline
24 & 99983 & 17 & 74121 & 25879 & 25862 & 0\tabularnewline
25 & 99991 & 9 & 78364 & 21636 & 21627 & 0\tabularnewline
26 & 99999 & 1 & 82635 & 17365 & 17364 & 0\tabularnewline
27 & 100000 & 0 & 86433 & 13567 & 13567 & 0\tabularnewline
28 & 100000 & 0 & 89951 & 10049 & 10049 & 0\tabularnewline
29 & 100000 & 0 & 92716 & 7284 & 7284 & 0\tabularnewline
30 & 100000 & 0 & 95171 & 4829 & 4829 & 0\tabularnewline
31 & 100000 & 0 & 96844 & 3156 & 3156 & 0\tabularnewline
32 & 100000 & 0 & 98128 & 1872 & 1872 & 0\tabularnewline
33 & 100000 & 0 & 98941 & 1059 & 1059 & 0\tabularnewline
34 & 100000 & 0 & 99358 & 642 & 642 & 0\tabularnewline
35 & 100000 & 0 & 99702 & 298 & 298 & 0\tabularnewline
36 & 100000 & 0 & 99856 & 144 & 144 & 0\tabularnewline
37 & 100000 & 0 & 99921 & 79 & 79 & 0\tabularnewline
38 & 100000 & 0 & 99970 & 30 & 30 & 0\tabularnewline
39 & 100000 & 0 & 99989 & 11 & 11 & 0\tabularnewline
40 & 100000 & 0 & 99994 & 6 & 6 & 0\tabularnewline
\bottomrule
\end{longtable}

The distributions of nine \(\gamma\) vectors from the highest \(S\)
values are shown below. This comparison shows the high number of stable
\(\mathbf{M}\) that can be produced through a targeted search of
\(\gamma\) values, and suggests that many otherwise unstable systems
could potentially be stabilised by an informed manipulation of their
component response times. Such a possibility might conceivably reduce
the dimensionality of problems involving stability in social-ecological
or economic systems.

\includegraphics{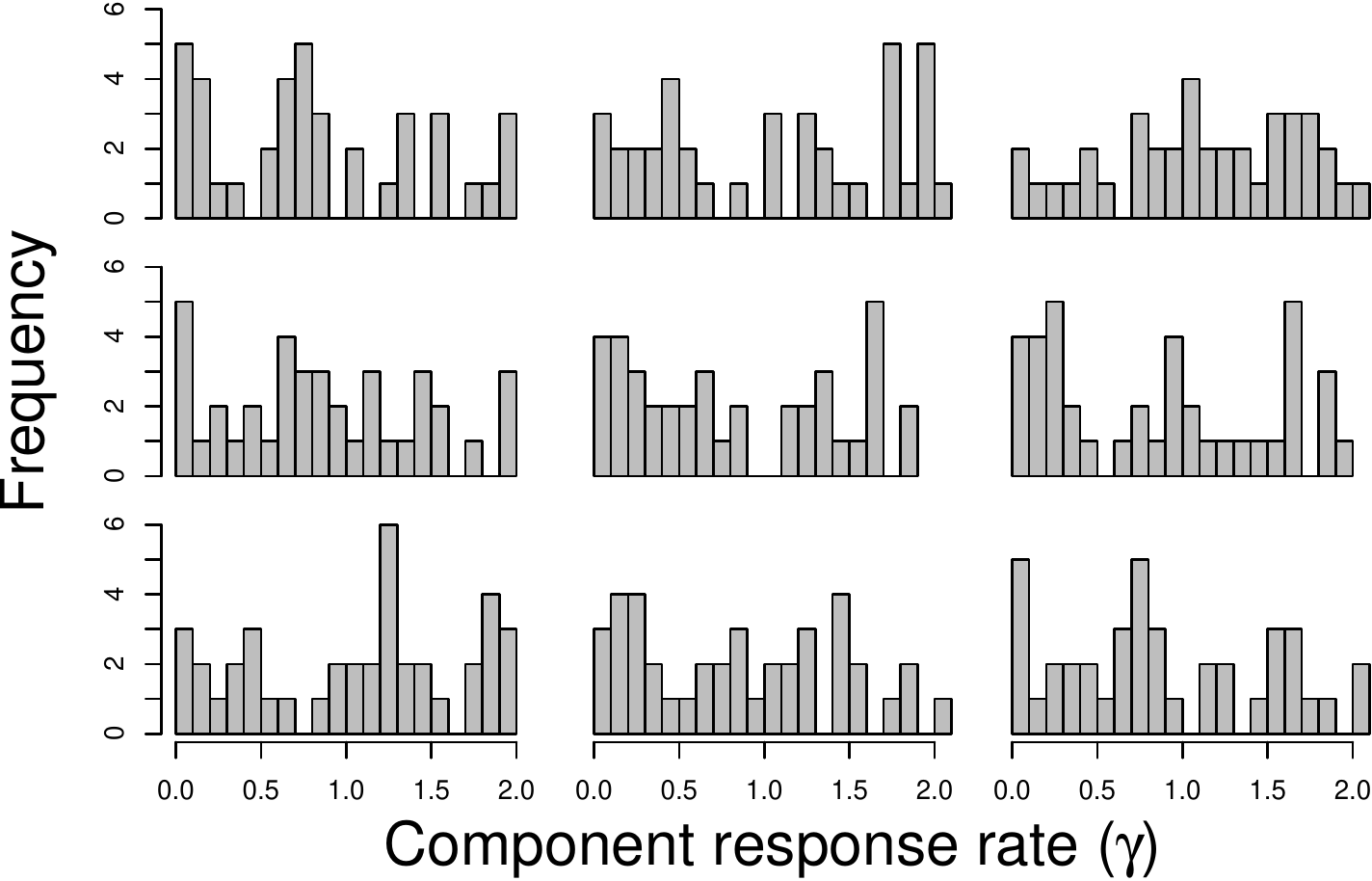}

The distribution of \(\gamma\) values found by the genetic algorithm is
uniform. A uniform distribution was used to initialise \(\gamma\)
values, so there is therefore no evidence that a particular distribution
of \(\gamma\) is likely to be found to stabilise a matrix
\(\mathbf{M}\).

\hypertarget{Gibbs}{\section{Consistency with Gibbs et al.
(2018)}\label{Gibbs}}

The question that I address in the main text is distinct from that of
Gibbs et al.\textsuperscript{\protect\hyperlink{ref-Gibbs2017}{9}}, who
focused instead on the effect of a diagonal matrix of biological species
densities \(\mathbf{X}\) on a community matrix \(\mathbf{M}\) given a
species interaction matrix \(\mathbf{A}\). This is modelled as below,

\[\mathbf{M} = \mathbf{XA}.\]

Mathematically, the above is identical to my model in the main text
where the system \(\mathbf{M}\) is defined by component interaction
strengths \(\mathbf{A}\) and individual component response rates
\(\boldsymbol{\gamma}\),

\[\mathbf{M} = \mathbf{\gamma A}.\]

I focused on the probability of observing a stable versus unstable
system given variation in \(\mathbf{\gamma}\) as system complexity
(\(\sigma\sqrt{SC}\)) increased. I increased system complexity by
holding \(C\) and \(\sigma\) constant and incrementally increasing \(S\)
to obtain numerical results. In contrast, Gibbs et
al.\textsuperscript{\protect\hyperlink{ref-Gibbs2017}{9}} applied
analytical techniques to instead focus on a different question
concerning the effect of \(\mathbf{\gamma}\) on the stability of
\(\mathbf{M}\) given \(\mathbf{A}\) as \(S \to \infty\), with \(\sigma\)
scaled so that \(\sigma = 1/\sqrt{S}\). Under such scaling, Gibbs et
al.\textsuperscript{\protect\hyperlink{ref-Gibbs2017}{9}} showed that
the effect of \(\gamma\) on stability should decrease exponentially as
\(S\) increases, which I demonstrate below by running simulations in
which \(\sigma = 1/\sqrt{S}\).

\begin{longtable}[]{@{}rrrrrrr@{}}
\toprule
S & A\_unstable & A\_stable & M\_unstable & M\_stable & A\_stabilised &
A\_destabilised\tabularnewline
\midrule
\endhead
2 & 3111 & 96889 & 3111 & 96889 & 0 & 0\tabularnewline
3 & 5203 & 94797 & 5237 & 94763 & 1 & 35\tabularnewline
4 & 6743 & 93257 & 6818 & 93182 & 6 & 81\tabularnewline
5 & 7889 & 92111 & 8005 & 91995 & 20 & 136\tabularnewline
6 & 8834 & 91166 & 8991 & 91009 & 55 & 212\tabularnewline
7 & 9885 & 90115 & 10072 & 89928 & 81 & 268\tabularnewline
8 & 10516 & 89484 & 10764 & 89236 & 108 & 356\tabularnewline
9 & 11135 & 88865 & 11383 & 88617 & 145 & 393\tabularnewline
10 & 11819 & 88181 & 12095 & 87905 & 181 & 457\tabularnewline
11 & 12414 & 87586 & 12700 & 87300 & 213 & 499\tabularnewline
12 & 12865 & 87135 & 13136 & 86864 & 283 & 554\tabularnewline
13 & 13530 & 86470 & 13836 & 86164 & 324 & 630\tabularnewline
14 & 13745 & 86255 & 14042 & 85958 & 362 & 659\tabularnewline
15 & 14401 & 85599 & 14720 & 85280 & 387 & 706\tabularnewline
16 & 14793 & 85207 & 15123 & 84877 & 428 & 758\tabularnewline
17 & 15004 & 84996 & 15356 & 84644 & 444 & 796\tabularnewline
18 & 15361 & 84639 & 15735 & 84265 & 472 & 846\tabularnewline
19 & 16062 & 83938 & 16303 & 83697 & 592 & 833\tabularnewline
20 & 15814 & 84186 & 16184 & 83816 & 566 & 936\tabularnewline
21 & 16171 & 83829 & 16492 & 83508 & 640 & 961\tabularnewline
22 & 16671 & 83329 & 17049 & 82951 & 641 & 1019\tabularnewline
23 & 17000 & 83000 & 17291 & 82709 & 718 & 1009\tabularnewline
24 & 17411 & 82589 & 17666 & 82334 & 765 & 1020\tabularnewline
25 & 17414 & 82586 & 17742 & 82258 & 783 & 1111\tabularnewline
26 & 17697 & 82303 & 18027 & 81973 & 806 & 1136\tabularnewline
27 & 18010 & 81990 & 18316 & 81684 & 880 & 1186\tabularnewline
28 & 18584 & 81416 & 18735 & 81265 & 1008 & 1159\tabularnewline
29 & 18401 & 81599 & 18572 & 81428 & 942 & 1113\tabularnewline
30 & 18497 & 81503 & 18754 & 81246 & 952 & 1209\tabularnewline
31 & 18744 & 81256 & 18942 & 81058 & 991 & 1189\tabularnewline
32 & 18936 & 81064 & 19194 & 80806 & 1022 & 1280\tabularnewline
33 & 19174 & 80826 & 19346 & 80654 & 1113 & 1285\tabularnewline
34 & 19477 & 80523 & 19632 & 80368 & 1120 & 1275\tabularnewline
35 & 19659 & 80341 & 19777 & 80223 & 1206 & 1324\tabularnewline
36 & 19883 & 80117 & 19929 & 80071 & 1275 & 1321\tabularnewline
37 & 20275 & 79725 & 20348 & 79652 & 1308 & 1381\tabularnewline
38 & 20067 & 79933 & 20190 & 79810 & 1275 & 1398\tabularnewline
39 & 20416 & 79584 & 20516 & 79484 & 1340 & 1440\tabularnewline
40 & 20370 & 79630 & 20489 & 79511 & 1359 & 1478\tabularnewline
41 & 20295 & 79705 & 20430 & 79570 & 1382 & 1517\tabularnewline
42 & 20767 & 79233 & 20839 & 79161 & 1418 & 1490\tabularnewline
43 & 20688 & 79312 & 20705 & 79295 & 1471 & 1488\tabularnewline
44 & 21049 & 78951 & 21028 & 78972 & 1555 & 1534\tabularnewline
45 & 21114 & 78886 & 21034 & 78966 & 1572 & 1492\tabularnewline
46 & 21163 & 78837 & 21195 & 78805 & 1463 & 1495\tabularnewline
47 & 21373 & 78627 & 21353 & 78647 & 1535 & 1515\tabularnewline
48 & 21338 & 78662 & 21285 & 78715 & 1632 & 1579\tabularnewline
49 & 21547 & 78453 & 21566 & 78434 & 1575 & 1594\tabularnewline
50 & 21738 & 78262 & 21633 & 78367 & 1636 & 1531\tabularnewline
51 & 21967 & 78033 & 21892 & 78108 & 1698 & 1623\tabularnewline
\bottomrule
\end{longtable}

Above table results can be compared to those of the
\protect\hyperlink{IncrS}{main results}. Note that 100000 (not 1
million), simulations are run to confirm consistency with Gibbs et
al.\textsuperscript{\protect\hyperlink{ref-Gibbs2017}{9}}. The
difference between my model and Gibbs et
al.\textsuperscript{\protect\hyperlink{ref-Gibbs2017}{9}} is that in the
latter, \(\sigma\sqrt{SC} = 1\) remains constant with increasing \(S\).
In the former, \(\sigma\sqrt{SC}\) increases with \(S\), so the expected
complexity of the system also increases accordingly. Consequently, for
the scaled \(\sigma\) in the table above, systems are not more likely to
be stabilised by \(\gamma\) as \(S\) increases, consistent with Gibbs et
al.\textsuperscript{\protect\hyperlink{ref-Gibbs2017}{9}}. Note that
overall stability does decrease with increasing \(S\) due to the
increased density of eigenvalues (see below).

\begin{center}
\includegraphics{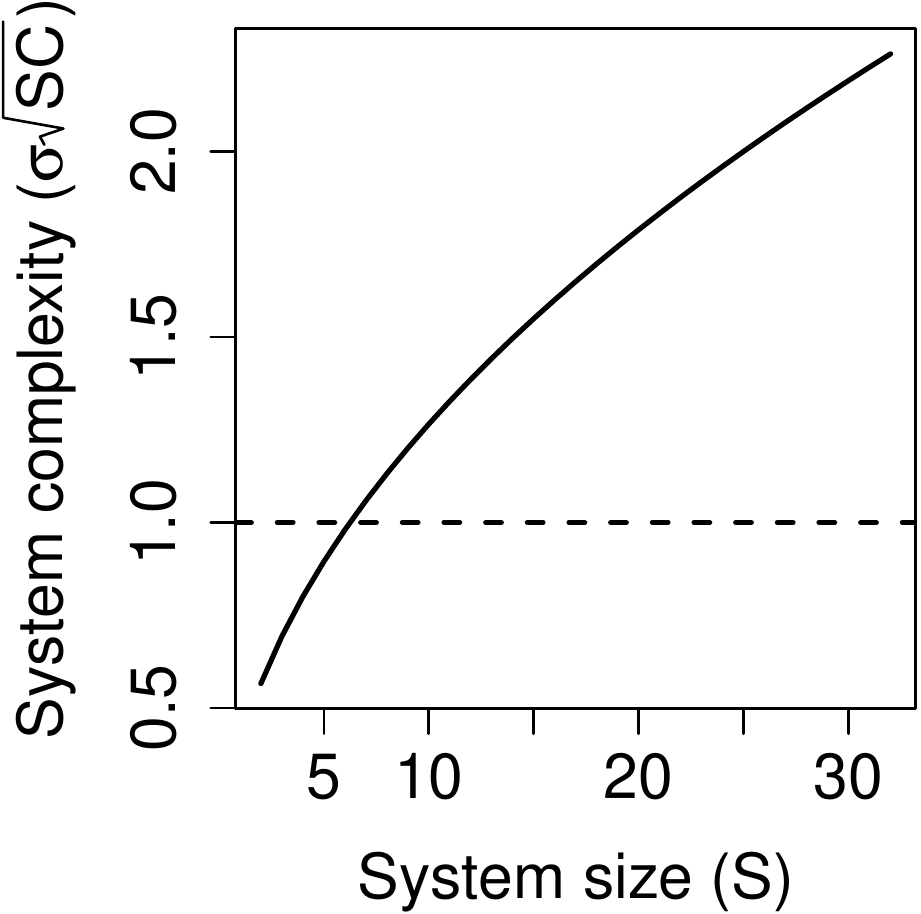} 
\end{center}

\textbf{Complexity as a function of \(S\) in the main text (solid)
versus in Gibbs et
al.\textsuperscript{\protect\hyperlink{ref-Gibbs2017}{9}} (dashed).}

When the complexity is scaled to \(\sigma\sqrt{SC} = 1\), an increase in
\(S\) increases the eigenvalue density within a circle with a unit
radius centred at \((-1, 0)\) on the complex plane. As \(S \to \infty\),
this circle becomes increasingly saturated. Gibbs et
al.\textsuperscript{\protect\hyperlink{ref-Gibbs2017}{9}} showed that a
diagonal matrix \(\mathbf{\gamma}\) will have an exponentially
decreasing effect on stability with increasing \(S\). Increasing \(S\)
is visualised below, first with a system size \(S = 100\).

\includegraphics{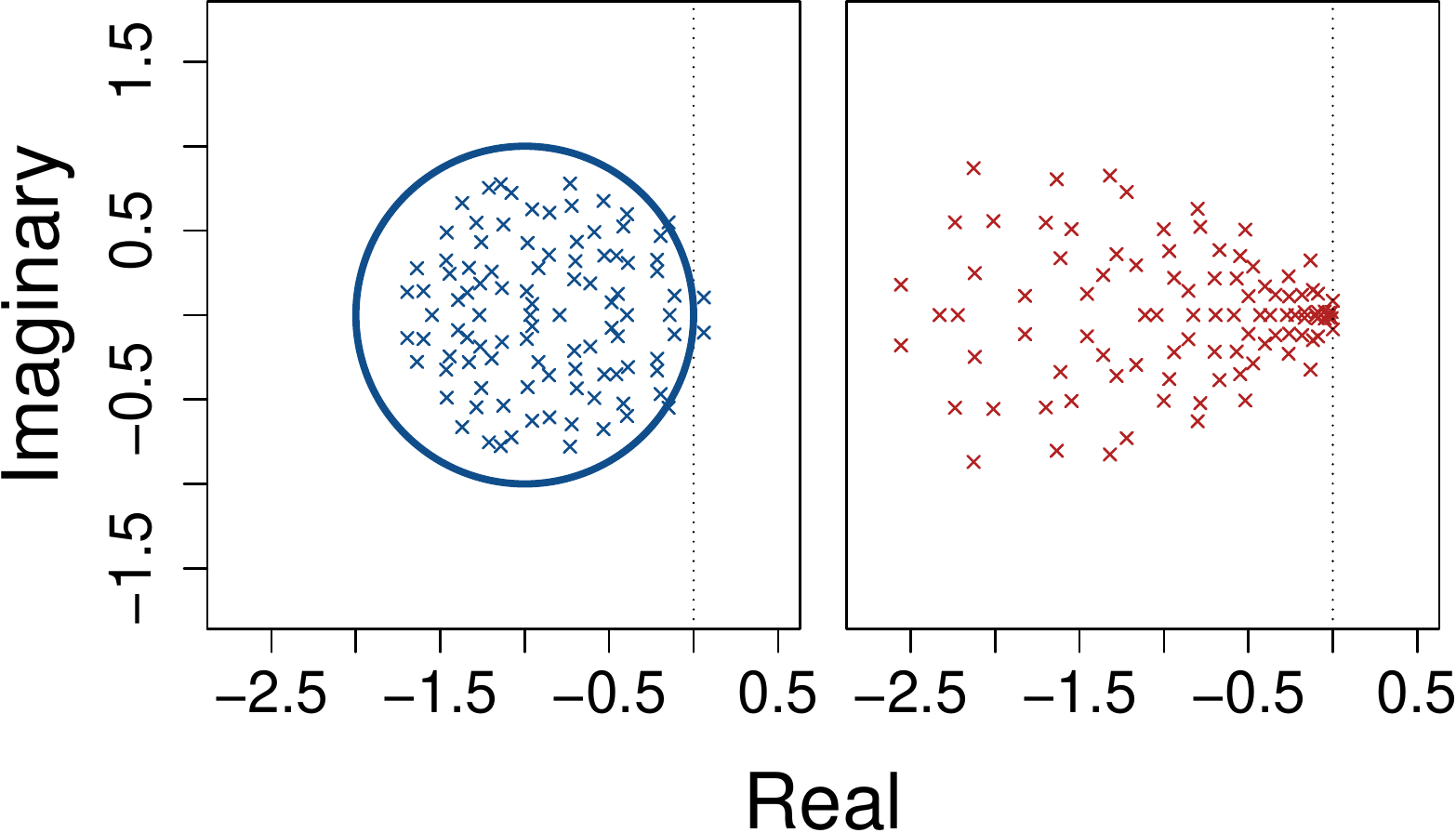}

The left panel above shows the distribution of eigenvalues; the blue
ellipse shows the unit radius within which eigenvalues are expected to
be contained. The right panel shows how eigenvalue distributions change
given \(\gamma \sim \mathcal{U}(0,2)\). The vertical dotted line shows
the threshold of stability, \(\Re = 0\). Increasing to \(S = 200\), the
scaling \(\sigma = 1 / \sqrt{S}\) maintains the expected distribution of
eigenvalues but increases eigenvalue density.

\includegraphics{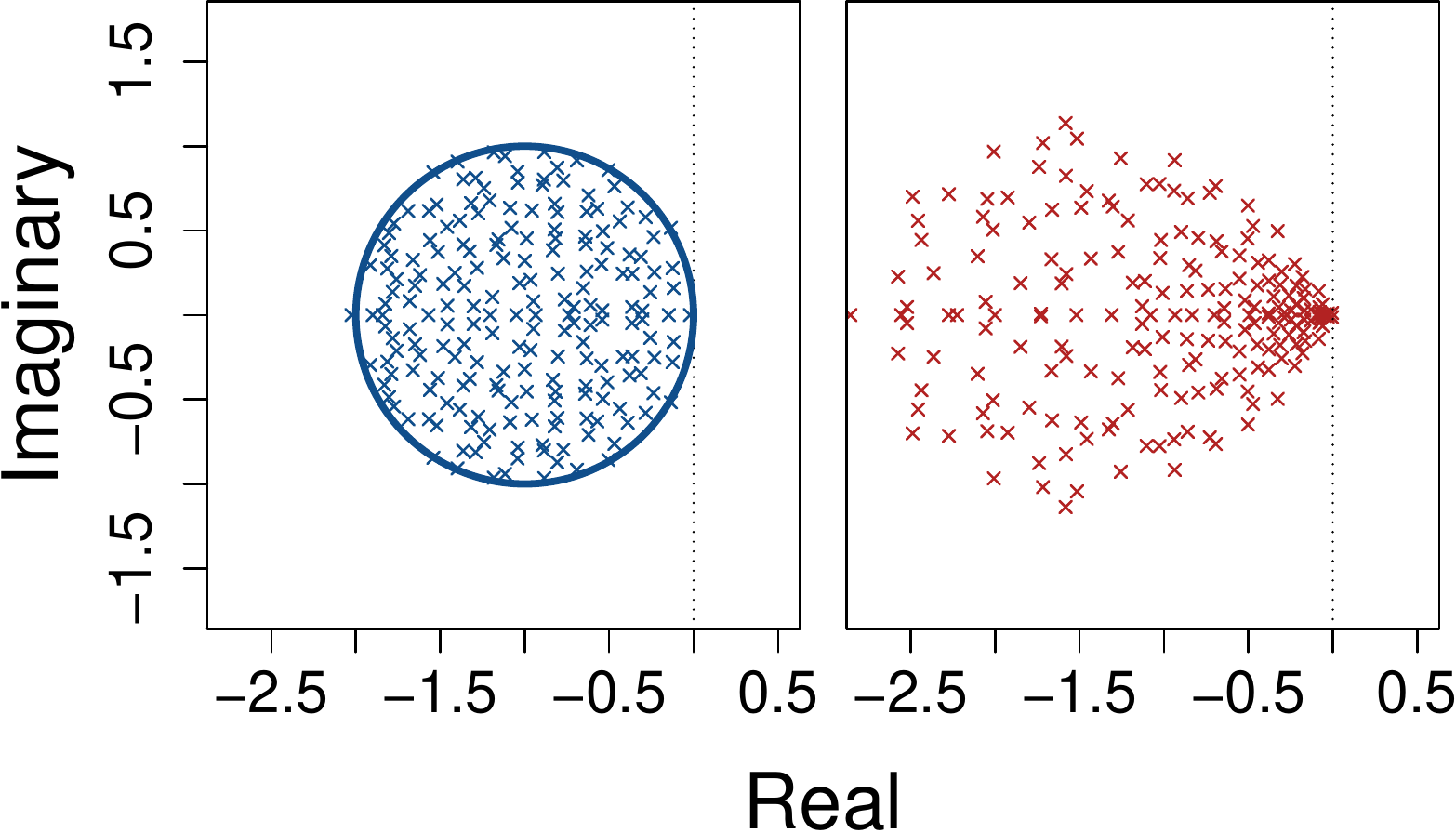}

We can increase the system size to \(S = 500\) and see the corresponding
increase in eigenvalue density.

\includegraphics{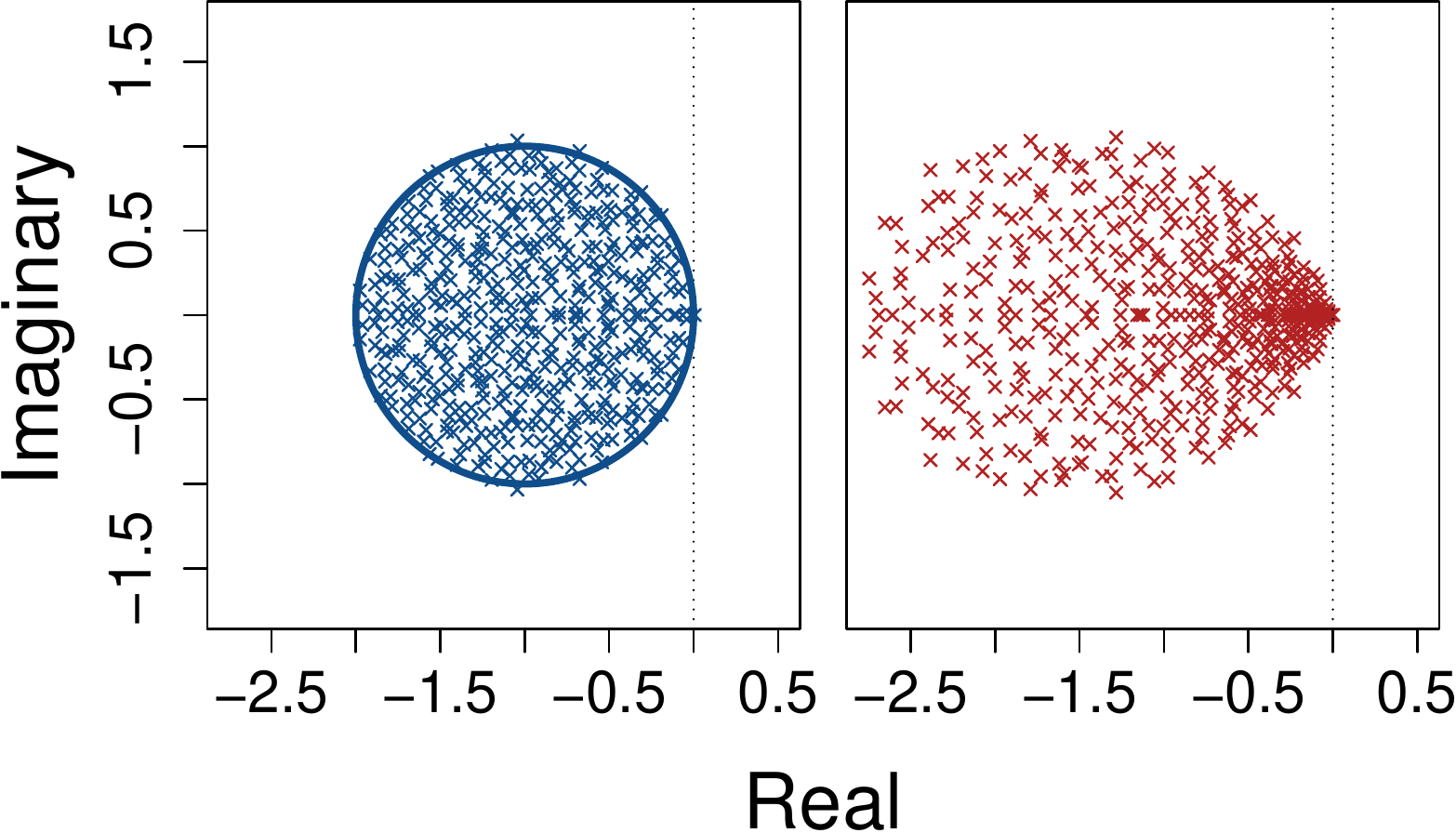}

Finally, below shows a increase in system size to \(S = 1000\).

\includegraphics{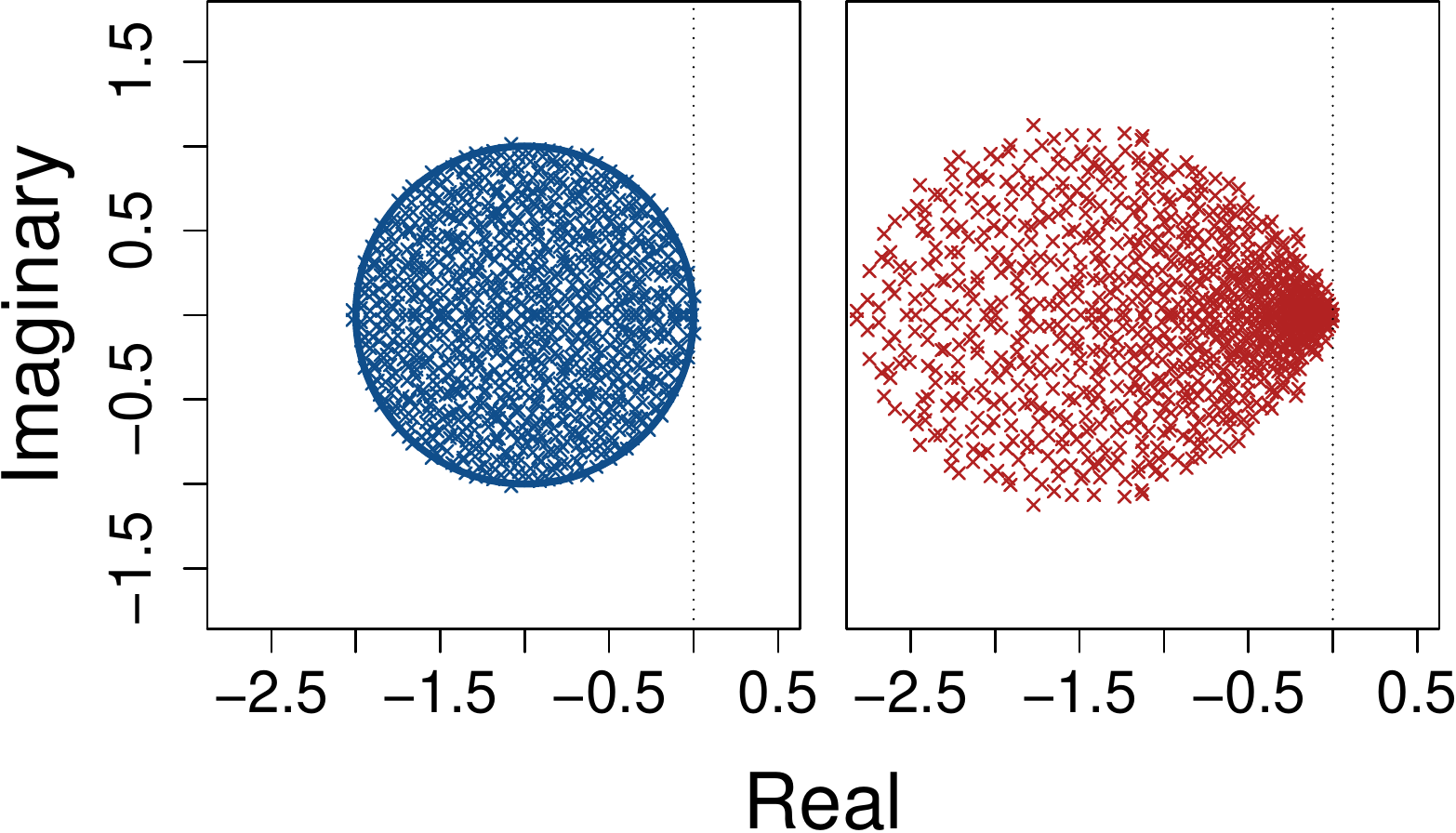}

In contrast, in the model of the main text, the complexity of system is
not scaled to \(\sigma\sqrt{SC} = 1\). Rather, the density of
eigenvalues within a circle centred at \((-1, 0)\) with a radius
\(\sigma\sqrt{SC}\) is held constant such that there are
\(S / \pi(\sigma\sqrt{SC})^2\) eigenvalues per unit area of the circle.
As \(S\) increases, so does the expected complexity of the system, but
the density of eigenvalues remains finite causing error around this
expectation. Below shows a system where \(S = 100\), \(C = 0.0625\), and
\(\sigma = 0.4\), where \(\sigma \sqrt{SC} = 1\) (identical to the first
example distribution above in which \(S = 100\) and
\(\sigma = 1/\sqrt{S}\)).

\includegraphics{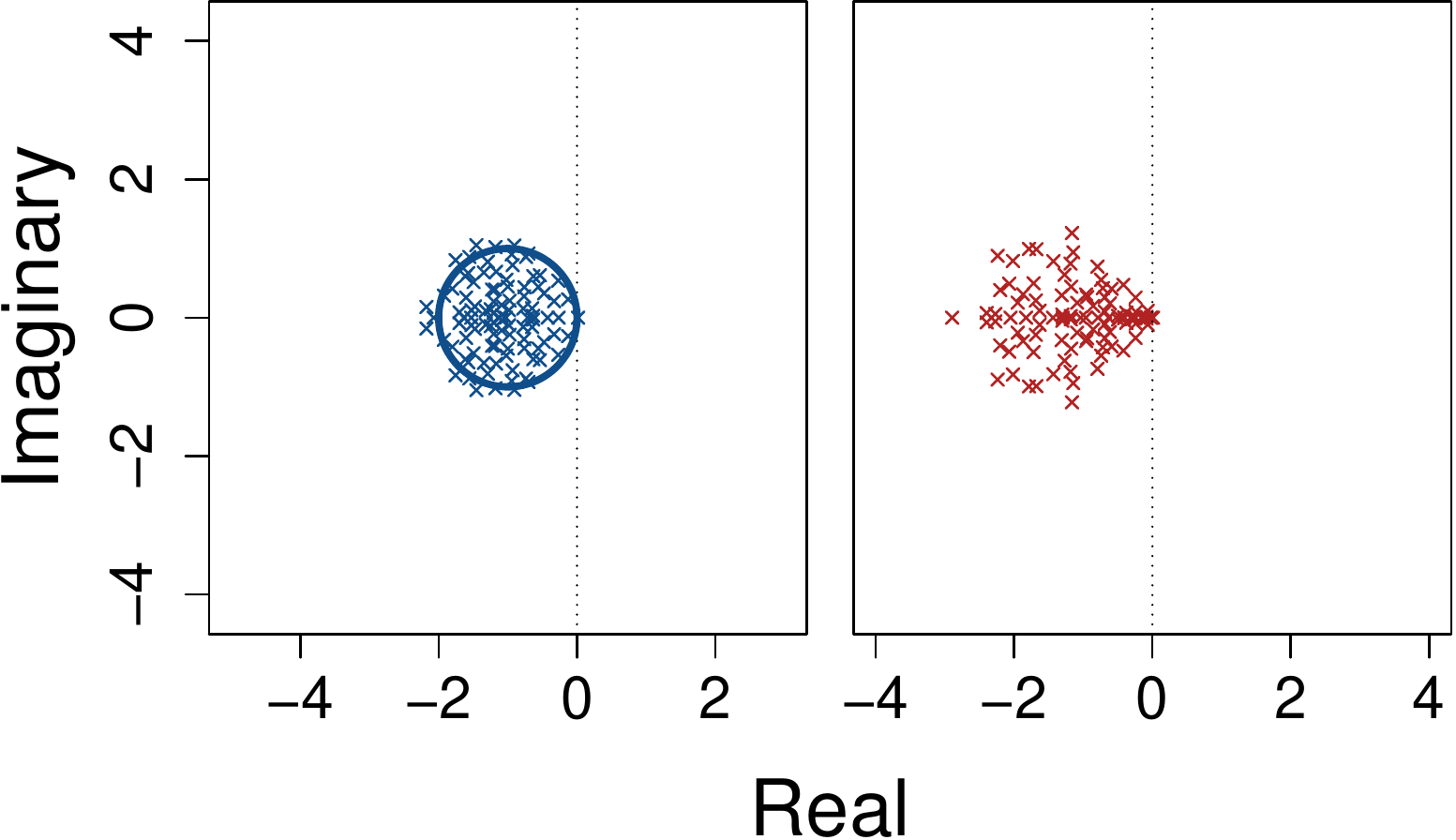}

Now when \(S\) is increased to \(200\) while keeping \(C = 0.0625\) and
\(\sigma = 0.4\), the area of the circle within which eigenvalues are
contained increases to keep the density of eigenvalues constant.

\includegraphics{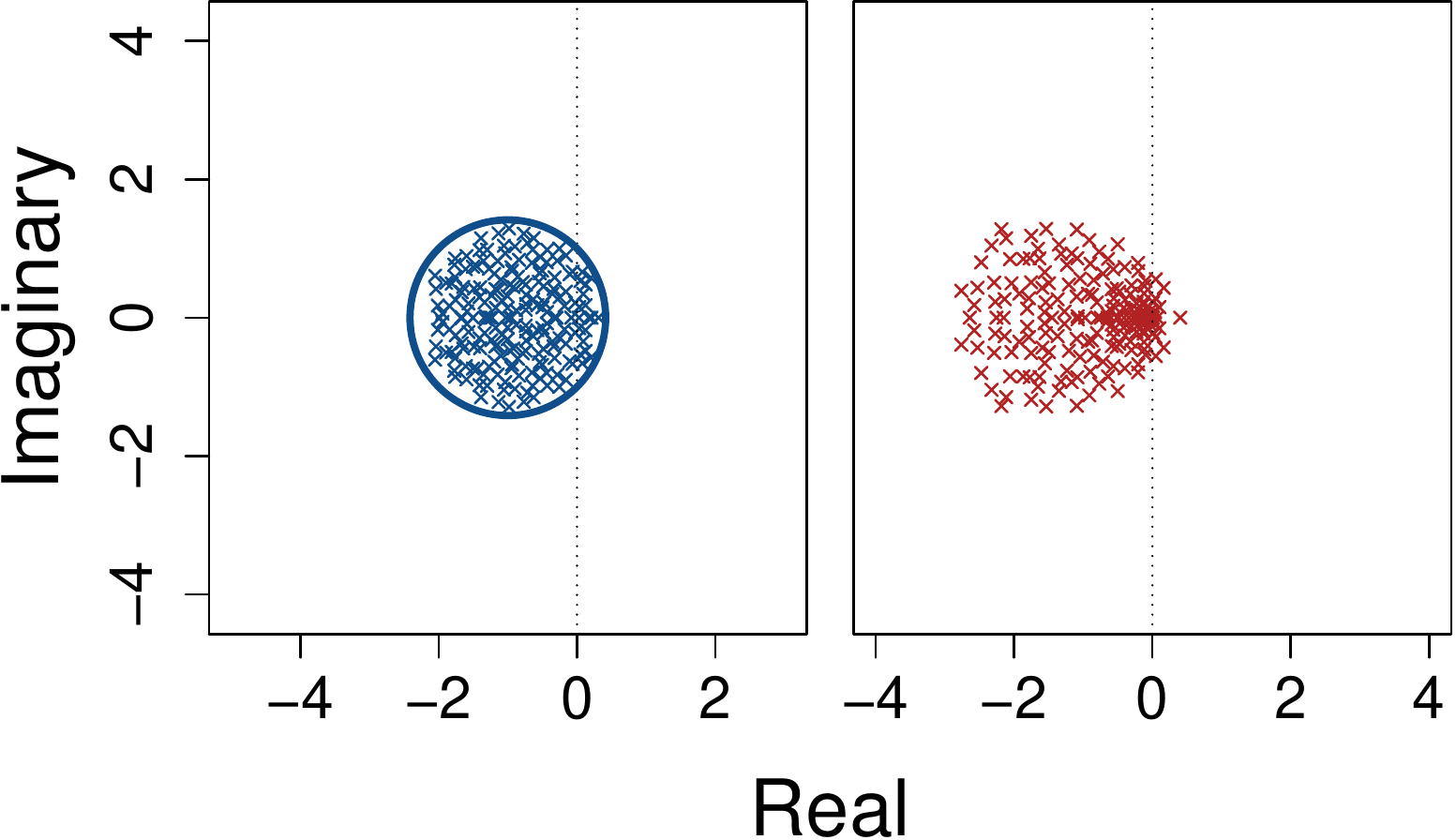}

Note that the expected distribution of eigenvalues increases so that the
threshold \(\Re = 0\) is exceeded. Below, system size is increased to
\(S = 500\).

\includegraphics{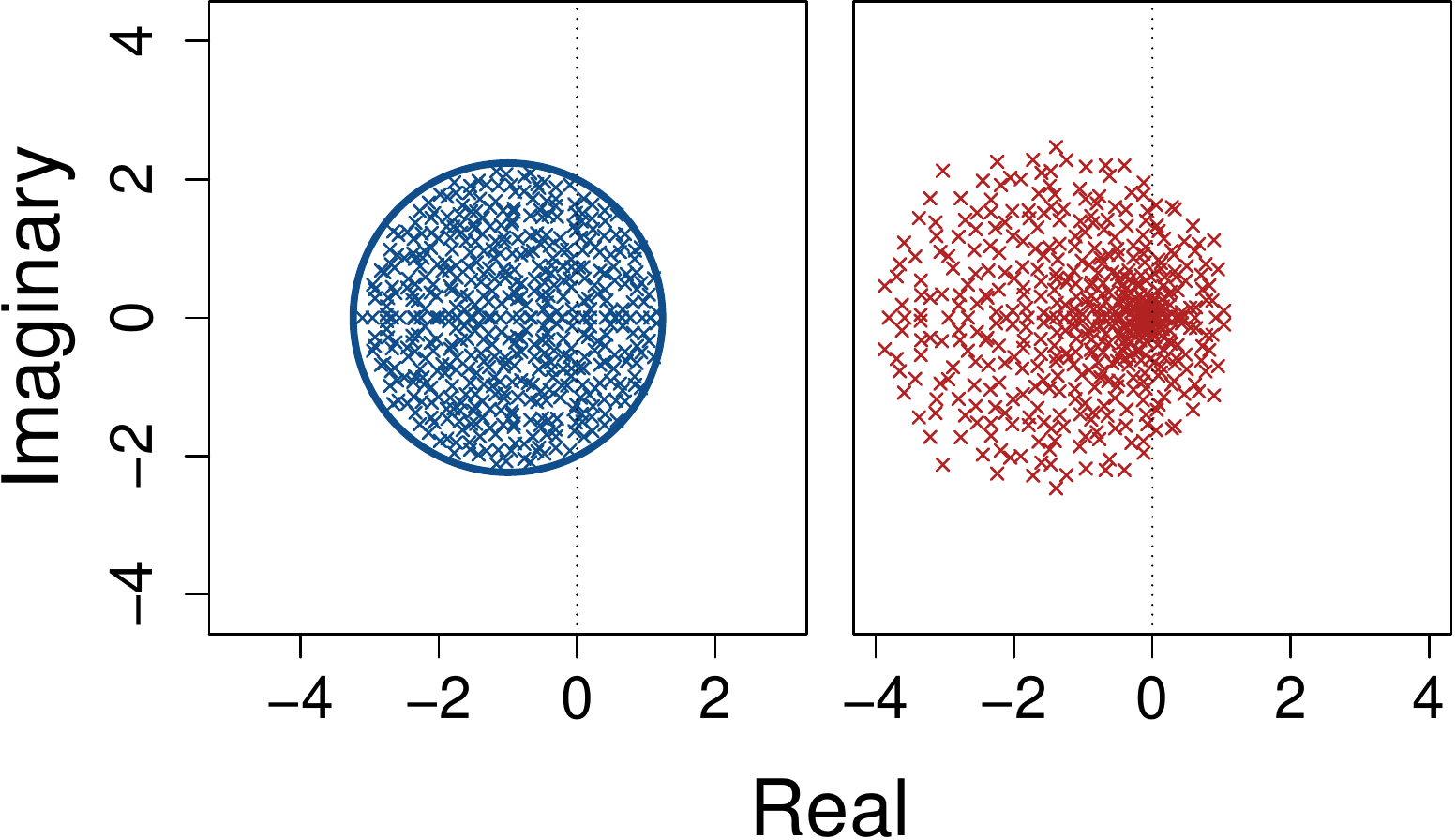}

Finally, \(S = 1000\) is shown below. Again, the density of eigenvalues
per unit remains constant at ca 2, but the system has increased in
complexity such that some real components of eigenvalues are almost
assured to be greater than zero.

\includegraphics{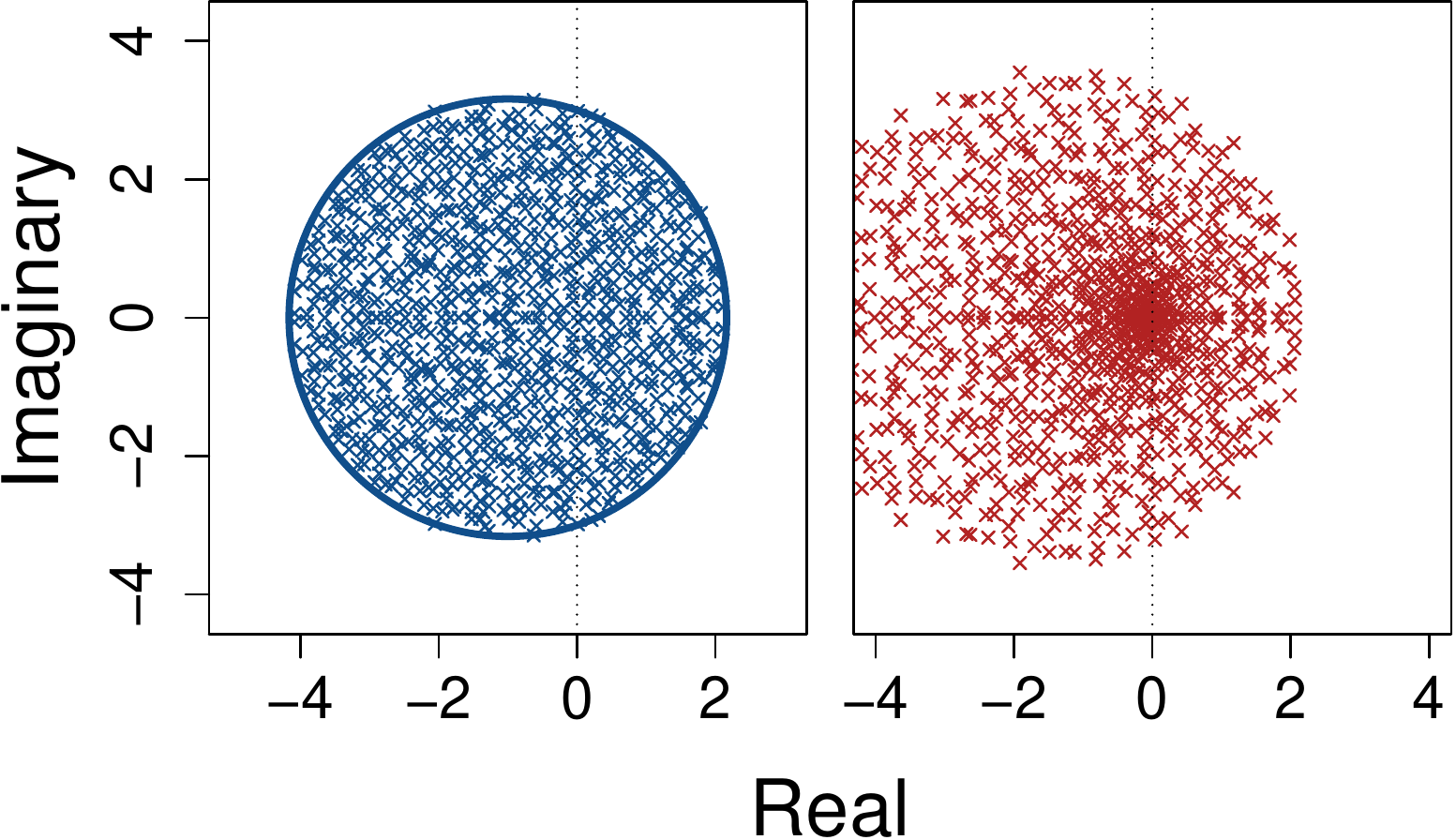}

\hypertarget{repr}{\section{Reproducing simulation results}\label{repr}}

All results in the main text and the literature cited can be reproduced
using the
\href{https://github.com/bradduthie/RandomMatrixStability}{RandomMatrixStability}
R package, which can be downloaded as instructed at the beginning of
this Supplemental Information document. The most relevant R functions
for reproducing simulations include the following:

\begin{enumerate}
\def\labelenumi{\arabic{enumi}.}
\tightlist
\item
  \texttt{rand\_gen\_var}: Simulates random complex systems and cascade
  food webs
\item
  \texttt{rand\_rho\_var}: Simulates random complex systems across a
  fixed correlation of \(\rho = cor(A_{ij}, A_{ji})\)
\item
  \texttt{rand\_gen\_swn}: Simulates randomly generated small-world
  networks
\item
  \texttt{rand\_gen\_sfn}: Simulates randomly generated scale-free
  networks
\item
  \texttt{Evo\_rand\_gen\_var}: Use a genetic algorithm to find stable
  random complex systems
\end{enumerate}

For the functions 1-4 above, R output will be a table of results. Below
describes the headers of this table to more clearly explain what is
being reported.

\footnotesize

\begin{longtable}[]{@{}llll@{}}
\toprule
Header & Description & Header\_cont. & Description\_cont.\tabularnewline
\midrule
\endhead
S & The system size & A\_rho & Corr. between elements A{[}ij{]} and
A{[}ji{]}\tabularnewline
A\_unstable & No. of A that were unstable & M\_rho & Corr. between
elements M{[}ij{]} and M{[}ji{]}\tabularnewline
A\_stable & No. of A that were stable & rho\_diff & Diff. between A and
M rho values\tabularnewline
M\_unstable & No. of M that were unstable & rho\_abs & Diff. between A
and M rho magnitudes\tabularnewline
M\_stable & No. of M that were stable & complex\_A & Complexity of
A\tabularnewline
A\_stabilised & No. of A stabilised by gamma & complex\_M & Complexity
of M\tabularnewline
A\_destabilised & No. of A destabilised by gamma & A\_eig & Expected
real part of leading A eigenvalue\tabularnewline
A\_infeasible & No. of A that were infeasible & M\_eig & Expected real
part of leading M eigenvalue\tabularnewline
A\_feasible & No. of A that were feasible & LR\_A & Lowest obs. real
part of leading A eigenvalue\tabularnewline
M\_infeasible & No. of M that were infeasible & UR\_A & Highest obs.
real part of leading A eigenvalue\tabularnewline
M\_feasible & No. of M that were feasible & LR\_M & Lowest obs. real
part of leading M eigenvalue\tabularnewline
A\_made\_feasible & No. of A made feasible by gamma & UR\_M & Highest
obs. real part of leading M eigenvalue\tabularnewline
A\_made\_infeasible & No. of A made infeasible by gamma & C & Obs.
network connectance\tabularnewline
\bottomrule
\end{longtable}

\normalsize

Note that output from \texttt{Evo\_rand\_gen\_var} only includes the
first seven rows of the table above, and \texttt{rand\_gen\_var} does
not include \(C\) (which can be defined as an argument). All results
presented here and in the main text are available in the
\href{https://github.com/bradduthie/RandomMatrixStability/tree/master/inst/extdata}{inst/extdata}
folder of the
\href{https://github.com/bradduthie/RandomMatrixStability}{RandomMatrixStability}
R package.

\hypertarget{ref}{\section*{Literature cited}\label{ref}}
\addcontentsline{toc}{section}{Literature cited}

\hypertarget{refs}{}
\hypertarget{ref-May1972}{}
1. May, R. M. Will a large complex system be stable? \emph{Nature}
\textbf{238,} 413--414 (1972).

\hypertarget{ref-Allesina2012}{}
2. Allesina, S. \& Tang, S. Stability criteria for complex ecosystems.
\emph{Nature} \textbf{483,} 205--208 (2012).

\hypertarget{ref-Allesina2015}{}
3. Allesina, S. \emph{et al.} Predicting the stability of large
structured food webs. \emph{Nature Communications} \textbf{6,} 7842
(2015).

\hypertarget{ref-Watts1998}{}
4. Watts, D. J. \& Strogatz, S. H. Collective dynamics of 'small world'
networks. \emph{Nature} \textbf{393,} 440--442 (1998).

\hypertarget{ref-Albert2002}{}
5. Albert, R. \& Barabási, A. L. Statistical mechanics of complex
networks. \emph{Reviews of Modern Physics} \textbf{74,} 47--97 (2002).

\hypertarget{ref-Solow1998}{}
6. Solow, A. R. \& Beet, A. R. On lumping species in food webs.
\emph{Ecology} \textbf{79,} 2013--2018 (1998).

\hypertarget{ref-Williams2000}{}
7. Williams, R. J. \& Martinez, N. D. Simple rules yield complex food
webs. \emph{Nature} \textbf{404,} 180--183 (2000).

\hypertarget{ref-Servan2018}{}
8. Serván, C. A., Capitán, J. A., Grilli, J., Morrison, K. E. \&
Allesina, S. Coexistence of many species in random ecosystems.
\emph{Nature Ecology and Evolution} \textbf{2,} 1237--1242 (2018).

\hypertarget{ref-Gibbs2017}{}
9. Gibbs, T., Grilli, J., Rogers, T. \& Allesina, S. The effect of
population abundances on the stability of large random ecosystems.
\emph{Physical Review E - Statistical, Nonlinear, and Soft Matter
Physics} \textbf{98,} 022410 (2018).

\end{document}